\documentclass[a4paper, amsfonts, amssymb, amsmath, reprint, footinbib, twoside,superscriptaddress,floatfix,longbibliography]{revtex4-1}

\usepackage{graphicx} 
\usepackage{amsmath}
\usepackage{comment}
\usepackage{booktabs}
\usepackage{hyperref}
\usepackage{url}
\usepackage{adjustbox}
\usepackage{array}
\usepackage[english]{babel}
\usepackage{verbatim}
\usepackage{multirow}
\usepackage{hhline}
\usepackage{makecell}
\usepackage[table]{xcolor}

\newcommand{\figLabelCapt}[1]{\textbf{\MakeLowercase{{#1}}}}
\newcommand{\refSub}[2]{\hyperref[#2]{\ref{#2}\figLabelCapt{#1}}}

\newcommand{\br}[1]{\mathbf{r}}
\newcommand{\bk}[1]{\mathbf{k}}

\begin{document}

\title{Polarizable atomic multipoles for learning long-range electrostatics}

\author{Yoonjae Park}
\thanks{These authors contributed equally.}
\affiliation{Department of Chemistry, UC Berkeley, California 94720, United States}
\affiliation{Bakar Institute of Digital Materials for the Planet, UC Berkeley, California 94720, United States}

\author{Dongjin Kim}
\thanks{These authors contributed equally.}
\affiliation{Department of Chemistry, UC Berkeley, California 94720, United States}

\author{Daniel S. King}
\thanks{These authors contributed equally.}
\affiliation{Bakar Institute of Digital Materials for the Planet, UC Berkeley, California 94720, United States}

\author{Nam H. \DJào}
\affiliation{Department of Chemistry, UC Berkeley, California 94720, United States}

\author{Roya Savoj}
\affiliation{Department of Chemistry, UC Berkeley, California 94720, United States}

\author{Sebastien Hamel}
\affiliation{Lawrence Livermore National Laboratory, Livermore, California, 94550, United States}

\author{Xiaoyu Wang}
\affiliation{Department of Chemistry, UC Berkeley, California 94720, United States}

\author{Bingqing Cheng}
\email{bingqingcheng@berkeley.edu}
\affiliation{Department of Chemistry, UC Berkeley, California 94720, United States}
\affiliation{Bakar Institute of Digital Materials for the Planet, UC Berkeley, California 94720, United States}
\affiliation{Chemical Sciences Division, Lawrence Berkeley National Laboratory, Berkeley, California, 94720, United States}

\date{\today}

\begin{abstract}
Long-range electrostatics and polarization remain central obstacles to extending machine learning interatomic potentials (MLIPs) to ionic, polar, and interfacial systems. 
Here we introduce a semi-local framework for learning electrostatics from energies and forces using polarizable atomic multipoles.
Local equivariant descriptors predict environment-dependent latent monopoles, dipoles, and quadrupoles, while residual non-local charge transfer and polarization are captured by non-self-consistent linear response in induced charges and dipoles. 
Across four diverse benchmarks and four short-range MLIP architectures, the multipole hierarchy and response terms systematically improve potential energy surface accuracy, with the largest gains in systems where long-range effects are essential. 
More importantly, physically meaningful electrical responses emerge without direct supervision.
The learned latent multipoles yield accurate Born effective charge tensors and infrared spectra in close agreement with experiments.
The induced-dipole extension introduces new capabilities: it predicts polarizabilities and thereby enables semi-quantitative Raman spectra for bulk water and hybrid MAPbI$_3$ perovskite, as well as the essential features of the surface-specific vibrational sum-frequency generation spectrum at the water-air interface.
In ferroelectric HfO$_2$, the predicted electrical response also captures LO-TO splitting and polarization switching.
This systematically improvable, physically transparent framework enables MLIPs trained on standard energy and force labels to predict polarization-sensitive observables.
\end{abstract}

\maketitle

\section{Introduction}

Although machine learning interatomic potentials (MLIPs) are traditionally built on the locality assumption~\cite{keith2021combining, unke2021machine, batzner20223, batatia2022mace}, it is increasingly recognized that explicitly modeling long-range interactions is important, especially for ionic systems, interfaces, and polar materials.
In recent years, various strategies~\cite{Baldwin2026,grasselli2026long,Kim2026} have been proposed to capture long-range interactions inside MLIPs.

One broad class of methods is to communicate global geometrical information via model architecture or representations~\cite{kosmala2023ewald, Caruso2026Extending, grisafi2019incorporating, huguenin2023physics, Faller2024Densitybased,Monacelli2026Electrostatic, Guo2026Capturing, Rumiantsev2025Learning,frank2026machine}. 
This includes approaches based on global or reciprocal-space message passing~\cite{kosmala2023ewald, Rumiantsev2025Learning},
the use of virtual global nodes for communicating global information~\cite{Caruso2026Extending}, 
and those constructing explicit long-range descriptors based on the fields generated by all atoms~\cite{grisafi2019incorporating, huguenin2023physics} or global atomic density distributions~\cite{Faller2024Densitybased}. 

The second class exploits the explicit physical formula of electrostatics.
Deep potential long-range (DPLR) model~\cite{zhang2022deep} learns from the maximally localized Wannier centers (MLWCs) for insulating systems.
3G-HDNNP~\cite{Morawietz2012neural,ko2021fourth} learns density functional theory (DFT) partial charges~\cite{Morawietz2012neural}, 
PhysNet~\cite{unke2019physnet} includes DFT dipoles in the loss function and predicts partial charges,
while SpookyNet learns from both DFT partial charges and dipoles~\cite{unke2021spookynet}.
The Latent Ewald Summation (LES) method~\cite{Cheng2025Latent,King2025Machine,zhong2025machine} infers partial charges and resulting long-range electrostatics, just from energy and force data. 
These methods are ``local'' in the sense that the partial charges are solely determined from the descriptors of local atomic environments.
Intriguingly, SpookyNet and LES demonstrate excellent performance even in the cases with nonlocal charge transfer effects~\cite{unke2021spookynet,King2025Machine}.

Non-local determination of partial charges can be achieved via charge equilibration, either using the classical Qeq scheme~\cite{rappe1991charge} combined with learned electronegativities~\cite{Ghasemi2015Interatomic,ko2021fourth,fuchs2025Learninga},
or redistributing the global excess charge during the message
passing steps~\cite{gong_predictive_2025,Anstine2025,batatia2026mace}.
It is worth noting that the Qeq scheme suffers from fundamental deficiencies,  such as incorrect fractional charge separation
upon dissociation~\cite{jensen2023unifying} and unphysical
system-size scaling of polarizability~\cite{lee2008origin},
and its introduced gain on training accuracy is inconclusive~\cite{Kim2026}.
Non-locality can also be introduced using self-consistent procedures.
In the self-consistent field neural network (SCFNN)~\cite{gao2022self},
the MLWC positions are 
iteratively updated in response to the predicted electric field of the last round.
The newly introduced fixed point model in MACE~\cite{Baldwin2026} incorporates electrostatic potential features in a loop to predict coarse-grained atomic charge densities until convergence.

Despite these advances, a gap remains. 
The long-range information from the global architectures is often not directly interpretable as electrostatics. 
Conversely, the existing physically-motivated models provide a clear electrostatic picture, but are often limited to monopoles, require additional electronic structure labels, or introduce self-consistent global solves that increase complexity. 
There is therefore room for a framework that combines three attributes: capturing electrostatics beyond atomic charges, a treatment of non-local charge transfer and polarization without global equilibration, and compatibility with the standard energy and force training of short-range MLIPs.

In this work, we address this gap with a semi-local framework that combines environment-dependent atomic multipoles with non-self-consistent polarization.
The atomic multipoles are determined locally, implicitly capturing short-range charge redistribution, while residual non-local contributions from distant electrostatic potentials and fields are incorporated through linear response in induced charges and dipoles.
The framework can be viewed as a generalization of the LES method, and likewise requires only energies and forces as training labels. 
This generalization also introduced the new capability of emergent polarizability, which enables Raman spectrum and vibrational sum-frequency generation (VSFG) spectrum predictions.
We benchmark the framework with different levels of expansions across four diverse benchmark systems,
using different short-range MLIP architectures as baselines.
We then apply the framework to the bulk and interfacial water system, MAPbI$_3$ perovskite, and ferroelectric hafnium dioxide,
focusing on the predictions of electric response properties, infrared, Raman spectra, and VSFG spectrum of the interface.
Here, the aim of the spectral predictions is not to replace the directly-supervised models, but to show that physically meaningful electrical response properties can emerge from standard energy-and-force training alone.

\section{Theory}

\paragraph{Multipole expansion of charge density and electric fields}

Extending the Molecular Dynamics in Electronic Continuum (MDEC) theory~\cite{leontyev2003continuum,leontyev2014polarizable,jorge2024theoretically} that separates free atomic charges and background electrons, 
we explicitly treat the free charge density associated with each atom, while representing the background as a homogeneous dielectric medium with a relative permittivity $\varepsilon_e$.
The long-range electrostatic potential generated by the localized charge distribution of an atom $i$ at the origin is
\begin{equation}
\Phi_i(\mathbf{r}) =
\frac{1}{4\pi\varepsilon_0 \varepsilon_e}
\int
\phi(|\mathbf{r}-\mathbf{r}'|)
\rho_i(\mathbf{r}')
d^3\mathbf{r}',
\end{equation}
where the screened Coulomb kernel,
\begin{equation}
\phi(r) = \frac{\mathrm{erf}(r/\sqrt{2}\sigma)}{r},
\label{eq:erf}
\end{equation}
replaces the bare $1/r$ interaction, 
to range-separate the short-ranged Coulomb accounted for in the baseline MLIPs.
Such screening is also equivalent to convolving the raw atomic charge density by a Gaussian function with the standard deviation $\sigma$.
The resulting electrostatic potential can be expanded in atomic multipoles as
\begin{multline}
\Phi_i(\mathbf{r})
= \frac{1}{4\pi\varepsilon_0 \varepsilon_e}
\Bigg(
q_i \phi(r)
-
\mathbf{u}_i \cdot \nabla_{\mathbf{r}} \phi(r)
\\
+ \frac{1}{2}\sum_{\alpha,\beta}
Q_{i\alpha\beta}
\frac{\partial^2 \phi(r)}{\partial r_\alpha \partial r_\beta}
+ \cdots
\Bigg),
\label{eq:Phi_i}
\end{multline}
where the monopole, dipole, and the traceless quadrupole moments are defined as
\begin{equation}
q_i =
\int \rho_i(\mathbf{r}') d^3\mathbf{r}',
\label{eqn:multipole-q}
\end{equation}
\begin{equation}
\mathbf{u}_i =
\int \mathbf{r}' \rho_i(\mathbf{r}') d^3\mathbf{r}',
\label{eqn:multipole-u}
\end{equation}
\begin{equation}
Q_{i\alpha\beta} =\int 
\left(r'_\alpha r'_\beta - \dfrac{1}{3}r'^2 \delta_{\alpha\beta}\right)
\rho_i(\mathbf{r}')
d^3\mathbf{r}'.
\label{eqn:multipole-Q}
\end{equation}

For an atom $j$ located at $\mathbf{r}_j$, the external potential at its vicinity $\mathbf{r}_j + \mathbf{r}'$ can be expanded as
\begin{equation}
\Phi(\mathbf{r}) =
\Phi(\mathbf{r}_j)
+ \mathbf{r}'\cdot\nabla\Phi(\mathbf{r}_j)
+ \frac{1}{2} r_\alpha' r_\beta'
\frac{\partial^2\Phi(\mathbf{r}_j)}{\partial r_\alpha \partial r_\beta}
+ \cdots ,
\end{equation}
where
\begin{equation}
\Phi(\mathbf{r}_j)=\sum_{i \ne j}\Phi_i(\mathbf{r}_j).
\end{equation}

The electrostatic energy of atom $j$ interacting with this external field is
\begin{equation}
U_j^{\mathrm{elec}}
=
\int
\rho_j(\mathbf{r})
\Phi(\mathbf{r})\,
d\mathbf{r} .
\end{equation}
Using the multipole definitions in
Eqns.~\eqref{eqn:multipole-q}–\eqref{eqn:multipole-Q}
and $\mathbf E=-\nabla\Phi$, we obtain
\begin{equation}
U_j^{\mathrm{elec}}
=
q_j\Phi(\mathbf{r}_j)
-
\mathbf{u}_j\cdot\mathbf{E}(\mathbf{r}_j)
-
\frac{1}{2}
\sum_{\alpha,\beta}
Q_{j\alpha\beta}
\frac{\partial E_\beta}{\partial r_\alpha}(\mathbf{r}_j)
+
\cdots .
\label{eq:W_j}
\end{equation}

Equations~\eqref{eq:Phi_i} and~\eqref{eq:W_j} show that the electrostatics generated by atomic charge density distributions can be hierarchically approximated using atomic multipoles ($q_i$, $\mathbf{u}_i$, $\mathbf{Q}_i$, \ldots).
As each successive term asymptotically decays with an additional factor of $1/r$, we therefore truncate the expansion at the dipole or the quadrupole level, while noting that higher-order multipoles can be added within the same framework.

\paragraph{Learning multipoles}

The atomic multipoles are learned from interatomic electrostatic interactions included in total energies and forces.
The learned LES charges correspond to scaled physical charges~\cite{zhong2025machine},
\begin{equation}
q_i^{\mathrm{les}} = \frac{q_i}{\sqrt{\varepsilon_e}},
\label{eq:scale-q}
\end{equation}
and analogously for the dipoles and the quadrupoles,
\begin{equation}
\mathbf{u}_i^{\mathrm{les}} =
\frac{\mathbf{u}_i}{\sqrt{\varepsilon_e}},
\qquad
\mathbf{Q}_i^{\mathrm{les}} =
\frac{\mathbf{Q}_i}{\sqrt{\varepsilon_e}}.
\end{equation}
The electrostatic potential and field are scaled consistently as
\begin{equation}
\Phi^{\mathrm{les}} = \Phi \sqrt{\varepsilon_e}, 
\qquad 
\mathbf{E}^{\mathrm{les}} = \mathbf{E} \sqrt{\varepsilon_e}.
\label{eq:scale-Phi}
\end{equation}
Under these transformations, the dielectric factor $\varepsilon_e$ is absorbed into the latent multipoles and fields, so that electrostatic interactions can be evaluated using the vacuum prefactor $1/(4\pi\varepsilon_0)$ only.

The latent charges $q_i^{\mathrm{les}}$ are predicted using a neural network acting on local invariant descriptors $B_i$, while the latent dipoles $\mathbf{u}_i^{\mathrm{les}}$ and quadrupole $\mathbf{Q}_i^{\mathrm{les}}$ are predicted using local equivariant features $A_i$.
As in the original LES method~\cite{Cheng2025Latent}, charge neutrality is not explicitly enforced; any residual net charge behaves as a uniform compensating background under the tinfoil boundary condition.

For periodic systems, the long-range electrostatic energy is evaluated using Ewald summation,
\begin{equation}
U^\mathrm{elec}
=
\frac{1}{2}
\sum_{i=1}^N
U_i^{\mathrm{elec}}
=
\frac{1}{2\varepsilon_0 V}
\sum_{0<k<k_c}
\frac{e^{-\sigma^2 k^2/2}}{k^2}
|S(\mathbf{k})|^2 ,
\label{eq:e_lr}
\end{equation}
where
\begin{equation}
S(\mathbf{k}) =
\sum_{i=1}^N
\left(
q_i^{\mathrm{les}}
+ i\mathbf{k}\cdot\mathbf{u}_i^{\mathrm{les}}
- \frac{1}{2} \mathbf{k} \cdot \mathbf{Q}_i^{\mathrm{les}} \cdot \mathbf{k}
\right)
e^{i\mathbf{k}\cdot\mathbf r_i}.
\label{eq:sfactor}
\end{equation}
Here $\mathbf{k}$ is a reciprocal lattice vector and $V$ is the cell volume.
Eqn.~\eqref{eq:e_lr} only includes reciprocal-space contribution, while the baseline MLIP adsorbs the short-range electrostatics.
Similarly, the reciprocal-space electrostatics can also be computed using particle–mesh methods~\cite{loche2025fast}.
Equivalent real-space expressions are given in the Methods and correspond to summing over pairwise charge--dipole--quadrupole interactions.

\paragraph{Non-local response}
We assume that the local atomic multipoles already capture the dominant electrostatic interactions, and
non-local effects due to distant atoms or external perturbations only introduce small corrections.
This assumption is justified by the good accuracy of the original monopole LES models across diverse systems~\cite{King2025Machine,zhong2025machine,Kim2025Universalb}.
To capture the non-local response without employing global charge equilibration schemes such as Qeq~\cite{rappe1991charge,Ghasemi2015Interatomic} or self-consistent polarizable force fields~\cite{Sala2010,Riera2023}, we introduce non-self-consistent induced charge and induced dipole terms.
Instead of iterative equilibration, the response is evaluated only once, using the field generated by fixed multipoles.

Neglecting the interactions between induced multipoles,
the energy associated with an induced charge $\Delta q_i$ on atom $i$ is
\begin{equation}
U_i^\mathrm{iq}
= \Phi(\mathbf{r}_i)\Delta q_i
+ \frac{1}{2}\kappa_i^{-1}\Delta q_i^2 ,
\end{equation}
where $\kappa_i^{-1}$ is the environment-dependent hardness parameter of atom $i$.
This quantity parallels the atomic hardness concept in conceptual
density functional theory~\cite{cardenas2011fukui} 
and in Qeq models~\cite{rappe1991charge,Ghasemi2015Interatomic}.
Resembling Qeq, the induced charge scheme essentially treats the system as a conductor.
Minimizing the energy with respect to $\Delta q_i$ yields
\begin{equation}
\Delta q_i = - \kappa_i \Phi(\mathbf{r}_i),
\label{eq:iq}
\end{equation}
and
\begin{equation}
U_i^\mathrm{iq} =
-\frac{1}{2} \kappa_i \Phi^2(\mathbf{r}_i).
\label{eq:U_i}
\end{equation}

Following the polarizable force field construction~\cite{Sala2010,leontyev2014polarizable,Riera2023}, the energy associated with induced dipole polarization is
\begin{equation}
U_i^\mathrm{iu}
= -\mathbf{E}(\mathbf{r}_i)\cdot\Delta\mathbf u_i
+ \frac{1}{2}
\Delta\mathbf u_i
\cdot \boldsymbol{\alpha}_i^{-1} \cdot
\Delta\mathbf u_i ,
\end{equation}
where $\Delta\mathbf u_i$ is the induced dipole and $\boldsymbol{\alpha}_i$ is the environment-dependent polarizability tensor of atom $i$.
If isotropic polarizability is assumed, this tensor can be reduced to a scalar, $\alpha_i = \mathrm{Tr}(\boldsymbol{\alpha}_i)/3$.
Minimization gives
\begin{equation}
\Delta\mathbf u_i =
\boldsymbol{\alpha}_i
\cdot 
\mathbf E(\mathbf{r}_i),
\label{eq:induced_u}
\end{equation}
and
\begin{equation}
U_i^\mathrm{iu}
= - \frac{1}{2}
\mathbf E(\mathbf{r}_i)
\cdot \boldsymbol{\alpha}_i \cdot
\mathbf E(\mathbf{r}_i) .
\end{equation}

Both the hardness and the atomic polarizability parameters are determined by the local atomic environments:
The latent $\kappa_i^{\mathrm{les}}$ is predicted using a neural network acting on local invariant descriptors $B_i$ of atom $i$.
Likewise, one can predict $\boldsymbol{\alpha}_i^\mathrm{les}$ using equivariant features $A_i$, or the isotropic $\alpha_i^\mathrm{les}$ based on $B_i$.
These learned parameters are related to the physical values by
$\kappa^\mathrm{les} = \kappa / \varepsilon_e$
and
$\boldsymbol{\alpha}^\mathrm{les} = \boldsymbol{\alpha} / \varepsilon_e$.
The scaling factor arises because the $q^\mathrm{les}$, $\mathbf{u}^\mathrm{les}$, $\mathbf{Q}^\mathrm{les}$, $\Phi^\mathrm{les}$, and $\mathbf E^\mathrm{les}$ that are used to infer $\alpha^\mathrm{les}$ and $\kappa^\mathrm{les}$ are themselves scaled according to Eqns.~\eqref{eq:scale-q} and ~\eqref{eq:scale-Phi}.

Finally, the total potential energy of the system can be assembled as 
\begin{equation}
    U = U^\mathrm{sr} +  U^\mathrm{elec}
    + \sum_{i=1}^N U_i^\mathrm{iq}
    + \sum_{i=1}^N U_i^\mathrm{iu},
    \label{eq:totE}
\end{equation}
where $U^\mathrm{sr}$ is from the baseline short-range MLIP.
In practice, the framework allows flexible truncation of the multipole and response terms.
For insulators, the induced charge term (-iq) may not be needed as it metalizes the system like the Qeq approach.
The training uses a loss function that contains energy loss, force loss, and sometimes stress loss.

\paragraph{Scaling factor and relative permittivity}

For systems in vacuum, such as isolated molecules, the effective dielectric constant entering the long-range electrostatics $\varepsilon_e = 1$.
For homogeneous bulk systems,
if no induced terms are included, the formulation reduces to the original LES framework~\cite{zhong2025machine}: $\varepsilon_e = \varepsilon_\infty$, the high-frequency (electronic) dielectric constant.
$\varepsilon_\infty$ can be obtained experimentally, such as from the square of the optical refractive index, or computed using density functional perturbation theory (DFPT) with frozen nuclei~\cite{farahvash2018dynamic}.

When induced dipoles are included, they provide an additional contribution to electric screening through the susceptibility:
\begin{equation}
\chi
=
\frac{1}{\varepsilon_0 V}
\sum_{i=1}^N \alpha_i
=
\frac{\varepsilon_e}{\varepsilon_0 V}
\sum_{i=1}^N \alpha_i^{\mathrm{les}}.
\label{eq:chi_les}
\end{equation}
The total electronic screening is then given by the combined contribution of the homogeneous background dielectric response and the induced-dipole susceptibility.
Requiring consistency with the overall high-frequency dielectric constant,  $\varepsilon_e + \chi = \varepsilon_\infty$, leads to
\begin{equation}
\varepsilon_e =
\frac{\varepsilon_\infty}{1 + \chi^{\mathrm{les}}},
\label{eq:eps_e}
\end{equation}
where $\varepsilon_\infty$ acts as a scaling constant, and $\chi^{\mathrm{les}}= \frac{1}{\varepsilon_0 V}
\sum_{i=1}^N \alpha_i^{\mathrm{les}}$
dependents on the atomic configuration.

Note that the configuration-dependent effective electronic dielectric constant $\varepsilon_e$ is naturally absorbed into all the latent variables and does not require consideration when training MLIPs or using them for energy and force predictions, 
and only the vacuum permittivity enters the Ewald summation in Eqn.~\eqref{eq:e_lr}.
$\varepsilon_e$ only enters during the rescaling procedure in recovering physical multipoles and electrical response properties as discussed below.
We also note that, the spatially-independent scalar $\varepsilon_e$ assumed here is reasonable for homogeneous bulk insulators, but it is not generally justified for heterogeneous systems such as interfaces, where electronic screening can be spatially varying and anisotropic.

\paragraph{Inferring electrical response properties}

Like LES~\cite{zhong2025machine},
the latent variables, learned without direct supervision by charges, BECs, or polarizabilities, can predict these electrical response properties.
For a finite system, the total polarization (dipole moment) is
\begin{equation}
\mathbf{P} = \mathbf{P}^q + \mathbf{P}^u =
\sum_{i=1}^{N}
(q_i+\Delta q_i)\mathbf r_i
+ \sum_{i=1}^{N}
\mathbf (\mathbf{u}_i + \Delta \mathbf{u}_i).
\end{equation}
The BEC tensor of atom $i$ can be computed using
\begin{equation}
Z^*_{i\alpha\beta}
= \frac{\partial P_\alpha}{\partial r_{i\beta}}.
\label{eq:z-finite}
\end{equation}
For a homogeneous periodic system, the dipole of the cell cannot be uniquely determined, based on the modern theory of polarization~\cite{resta2007theory}.
To determine the BECs, we followed the approach in the original LES~\cite{zhong2025machine},
\begin{equation}
Z^*_{i\alpha\beta}
=
\frac{\partial P_\alpha^{u}}{\partial r_{i\beta}}
+
\lim_{k\rightarrow0}
\Re
\left[
e^{-ik r_{i\alpha}}
\frac{\partial P_\alpha^q(k)}{\partial r_{i\beta}}
\right],
\label{eq:z-pbc}
\end{equation}
where $k$ is the reciprocal lattice vector along direction $\alpha$,
\begin{equation}
P_\alpha^q(k)
= \sum_{i=1}^{N}
\frac{1}{ik} q_i
e^{ik r_{i\alpha}}.
\end{equation}
Here $P_\alpha^q(k)$ is a complex-valued quantity that is not comparable to DFT polarization, and is only used as a device for obtaining BECs.

\paragraph{Polarization response}

For an insulating system, the polarizability tensor $\boldsymbol{\alpha}$ relates the induced polarization (or the Berry-phase polarization for periodic systems~\cite{resta2007theory}) to a uniform external electric field.
In the present framework, the local atomic multipoles ($q^\mathrm{les}$, $\mathbf{u}^\mathrm{les}$, $\mathbf{Q}^\mathrm{les}$) do not respond to an external field.
The polarization response thus arises from the induced dipole ($\Delta\mathbf u_i$) terms, whereas the induced charge contribution is assumed to be insignificant for insulators.
Under such, the total polarizability of the system is given by
\begin{equation}
\boldsymbol{\alpha}
=
\sum_{i=1}^N \boldsymbol{\alpha}_i.
\label{eq:alpha-u}
\end{equation}

\paragraph{Design space}
The proposed framework defines a flexible design space for incorporating physics-based long-range electrostatics into MLIPs.
The simplest configuration corresponds to the original LES model (-les), where only local environment-dependent atomic monopoles are included.
Higher-order multipoles, such as dipoles (-u) and quadrupoles (-Q), can be incorporated.
To capture charge transfer effects particularly in conductors, induced charge (-iq) can be added. The induced dipole (-iu) terms can be used to capture non-local polarization.
As shown in the benchmarks below, including additional terms leads to general improvements in model accuracy.
Moreover, the inclusion of the -iu term enables emergent polarizability predictions, which further enables spectrum calculations that utilize time-dependent polarizabilities.

For the induced-response parameters $\kappa$ and $\boldsymbol{\alpha}$, one may impose positive-definiteness constraints to ensure physical consistency.
For example, one can parameterize $\boldsymbol{\alpha} \leftarrow \boldsymbol{\alpha}^T \boldsymbol{\alpha}$, or $\alpha \leftarrow \alpha^2$ under the isotropic approximation.
In practice, however, we find that these parameters are typically learned to be positive without explicitly enforcing such constraints.

In the present formulation, the induced terms are evaluated in a non-self-consistent manner: induced charges and dipoles respond only to the field generated by the fixed multipoles, and not to each other.
A self-consistent scheme can be constructed by iteratively updating the fields to include contributions from the induced terms.
In our test cases, we find that while this procedure can occasionally yield marginal improvements in the accuracy, it may also introduce instability during training.
The limited benefit of self-consistency can be attributed to the relatively small magnitude of the induced terms compared to the fixed multipoles.
However, for cases involving strong charge transfer such as redox reactions, self-consistency may be required.

Although the long-range augmented MLIP can be trained only using standard energy and force labels,
additional observables such as dipoles, BECs, and potentially polarizabilities could be incorporated into the training loss.
Extensions such as global charge or spin-state embeddings can also be included.
Furthermore, the explicit atomic multipole representation enables seamless integration with other modeling frameworks, such as the Siepmann-Sprik metal model~\cite{siepmann1995influence, Wang2025Ionmodulateda}, hybrid quantum mechanics/machine learning (QM/ML) approaches~\cite{Semelak2025Advancing}, and classical force fields with fixed charges (ML/MM)~\cite{morado2025enhancing}.

\section{Implementation and benchmarks}

\begin{figure*}
    \centering
    \includegraphics[width=0.95\linewidth]{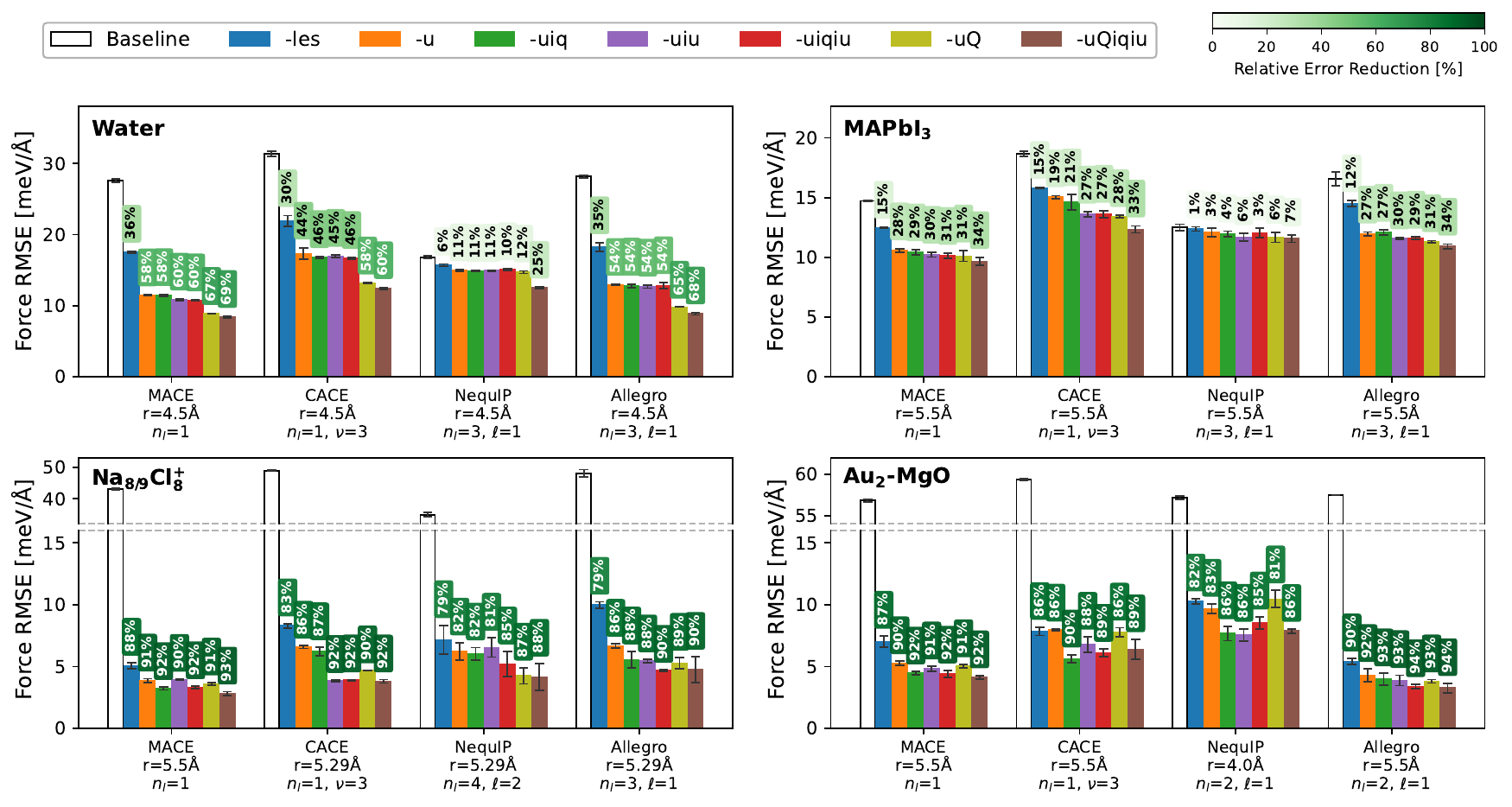}
    \caption{Benchmark results of baseline and long-range-augmented machine learning interatomic potentials (MLIPs) across four systems: \figLabelCapt{a}: Bulk water, \figLabelCapt{b}: MAPbI$_3$ (Methylammonium lead iodide), \figLabelCapt{c}: Na$_{8/9}$Cl$_8^{+}$, and \figLabelCapt{d}: Au$_2$ on MgO(001). 
    The test force root mean square errors (RMSEs) are compared between the baseline short-range models (hollow bars) and their long-range-augmented variants (solid bars) for different architectures (MACE~\cite{batatia2022mace}, CACE~\cite{cheng2024cartesian}, NequIP~\cite{batzner20223}, and Allegro~\cite{musaelian2023learning}). The colored bars represent the inclusion of long-range interactions:  monopoles (-les, blue), monopoles and dipoles (-u, orange), monopoles, dipoles, and induced charges (-uiq, green), monopoles, dipoles, and induced dipoles (-uiu, purple), monopoles, dipoles, induced charges, and induced dipoles (-uiqiu, red), monopoles, dipoles, and quadrupoles (-uQ, olive), and monopoles, dipoles, quadrupoles, induced charges, and induced dipoles (-uQiqiu, brown). Each bar reports the mean force RMSE from three independent training runs with different random training seeds, and the error bars indicate one standard deviation across the three runs. The color-coded badges above the bars indicate the relative force RMSE reduction (\%) of each augmented model compared to its corresponding baseline, with the background color intensity reflecting the magnitude of improvement. The y-axes in \figLabelCapt{c} and \figLabelCapt{d} are broken for visualization.
    Key model hyperparameters, including the cutoff $r$, the number of layers $n_l$, the order of irreducible representations (rotation order) $\ell$, and body order $\nu$, are specified for each MLIP architecture.}
    \label{fig:benchmark}
\end{figure*}

As all learned quantities (atomic multipoles, hardness parameters, and polarizabilities) are local, the framework remains fully compatible with virtually any existing short-range MLIPs.
In addition to scalar latent charges predicted from invariant features, as in the original LES method, the extended model also predicts higher-order multipoles from equivariant features.
The hardness is obtained from invariant features, while the polarizability can be predicted either as a $3\times3$ tensor from equivariant features or, under the isotropic approximation, as a scalar from invariant features.

We have integrated the extended LES with CACE~\cite{cheng2024cartesian}, MACE~\cite{batatia2022mace}, NequIP~\cite{batzner20223, Tan2026Highperformance}, and Allegro~\cite{musaelian2023learning, Tan2026Highperformance}.
Implementation details are provided in the Methods.
All four models support predictions of atomic charges $q$, dipoles $\mathbf{u}$, quadrupoles $\mathbf{Q}$, hardness $\kappa$, and isotropic polarizability or anisotropic polarizability tensors $\boldsymbol{\alpha}$.

We benchmark the accuracy of different levels of the long-range augmentation on the four architectures (MACE, CACE, NequIP, and Allegro), across four systems: (a) bulk liquid water~\cite{Schmiedmayer2024}, a dielectric fluid; (b) methylammonium lead iodide (MAPbI$_3$) hybrid halide perovskite~\cite{Schmiedmayer2024} as a polar and ionic material; (c) the ionic Na${_{8/9}}$Cl$_8^{+}$ cluster set~\cite{ko2021fourth} with non-local charge transfer; and (d) the Au$_2$-MgO(001)~\cite{ko2021fourth}, where Al dopants under the surface ($\approx$10~\AA{}) strongly influence the adsorption energetics of the gold dimer.
Further dataset descriptions, training details, and parameter counts are provided in the Methods section.
All models were trained using energies and forces. 

Fig.~\ref{fig:benchmark} reports mean root-mean-square errors (RMSEs) of atomic forces over three independent training runs with different random training seeds, which sensitively reflect the accuracy of the learned potential energy surface.
Energy errors are uniformly small across all long-range variants, in some cases approaching numerical precision.
The augmented models (solid bars), truncated at different levels of Eqn.~\eqref{eq:totE}, are compared to the baseline short-range models (hollow bars), with relative force error reductions indicated. The corresponding numerical values are tabulated in Table~\ref{tab:benchmark-error}. Across most comparisons, the variation across the three random-seed runs is smaller than the differences in mean force RMSE among the long-range variants, indicating that the overall trends are robust to training variability.

Long-range augmentation consistently reduces force errors across all systems and architectures, demonstrating both robustness and broad compatibility with short-range MLIPs.
The magnitude of improvement, however, depends on the baseline model.
Architectures with smaller receptive fields, such as single-layer ($n_l=1$) MACE and CACE, benefit the most.
Similarly, the strictly local Allegro architecture shows substantial gains, where $n_l$ denotes the body-order expansion rather than message-passing depth (e.g., $n_l=2$ corresponds to four-body features).
In comparison, NequIP models that employ multiple message-passing layers exhibit only modest accuracy gains.

Accuracy improves generally as additional terms, including monopoles (-les), dipoles (-u), quadrupoles (-Q), induced charges (-iq), and induced dipoles (-iu), are included.
The largest gain arises from introducing monopoles (-les) relative to the short-range baseline.
Higher-order multipoles (-u, -uQ) yield further, but diminishing, improvements.
Response terms (-iq, -iu) do not guarantee monotonic gains in every case, but are generally beneficial.
This trend reflects the hierarchical nature of the multipole expansion, as well as the dominant contribution of local multipole terms over induced responses.

The magnitude of error reduction from long-range augmentation is strongly system-dependent.
For bulk water (Fig.~\ref{fig:benchmark}a), force errors are reduced by up to 69\% (MACELES-uQiqiu), while for MAPbI$_3$ (Fig.~\ref{fig:benchmark}b), improvements remain substantial, reaching up to 34\% (MACELES-uQiqiu and Allegro-uQiqiu).
More dramatic gains are observed for the Na$_{8/9}$Cl$_8^{+}$ and Au$_2$-MgO(001) benchmarks, where long-range effects are central by construction~\cite{ko2021fourth}.
In the Na$_{8/9}$Cl$_8^{+}$ clusters, removing an atom induces global charge redistribution.
LES models that include induced-charge terms (-iq, -iqiu) capture this effect effectively, achieving up to 93\% error reduction (MACELES-uQiqiu).
For Au$_2$-MgO(001), short-range models generally show large errors, and
the most accurate model here (AllegroLES-uiqiu/uQiqiu) achieves a drastic, 94\% reduction in force error.

Furthermore, we benchmark the computational cost of different levels of long-range augmentation.
Fig.~\ref{fig:timing} (Methods) shows that, although additional terms introduce incremental overhead, the overall increase in evaluation cost relative to the baseline model remains modest, confirming the efficiency and scalability of the approach.
In addition, Table~\ref{tab:benchmark-error} shows that the MACE-based, CACE-based, and NequIP-based augmented models (-u, -uiq, -uiu) in general have very similar parameter counts compared to the monopole (-les) models, but can achieve more than 30\% in force error reduction. 
This indicates that the increased model capacity does not just come from the increase in the parameter count, but rather comes from the expanded electrostatic physics.
The parameter-controlled ablations in the Methods (Table~\ref{tab:ablation}) furthur prove this observation.

\section{Applications}

\subsection{Bulk and interfacial water}

\begin{figure*}
    \centering
    \includegraphics[width=0.95\linewidth]{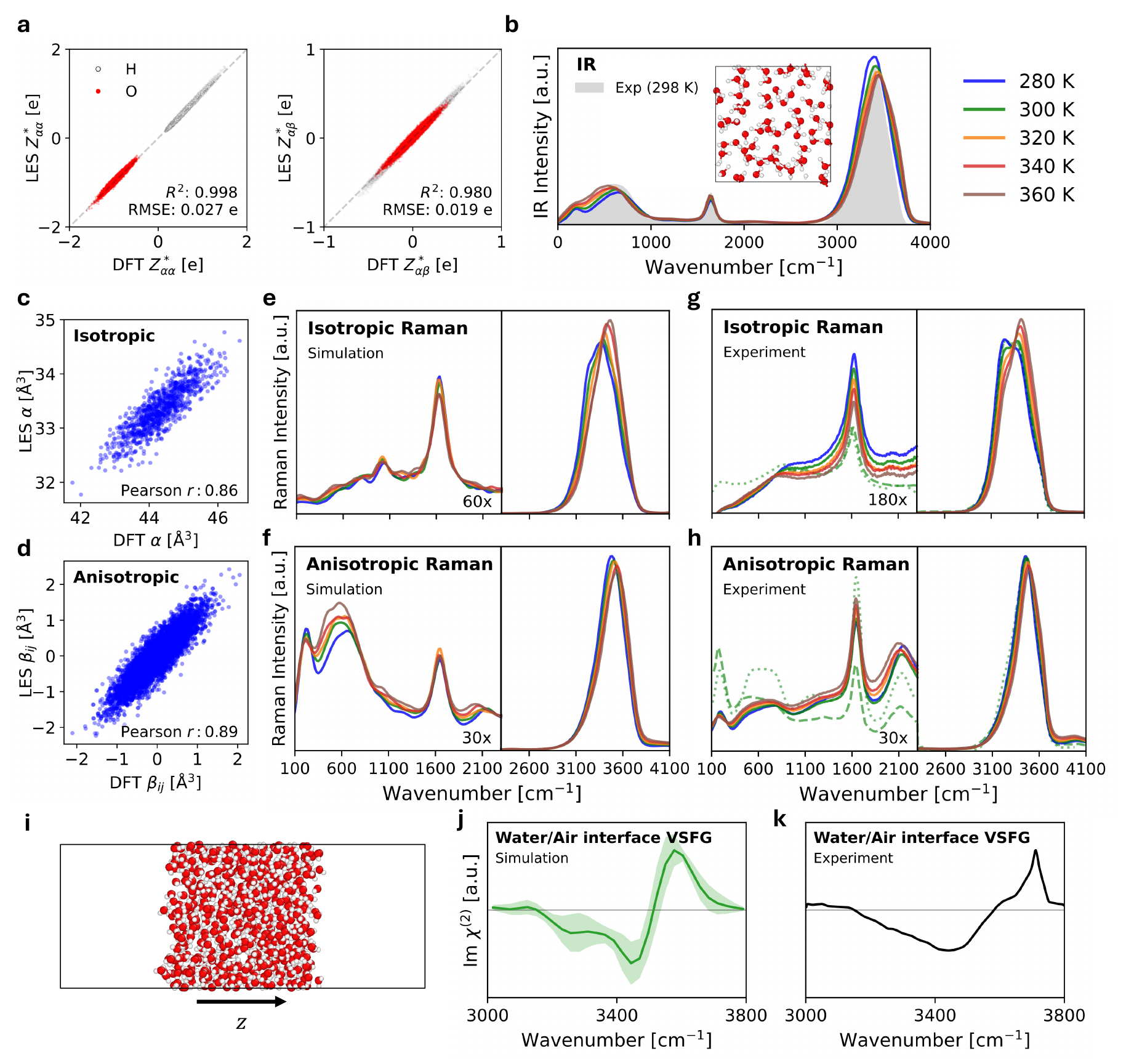}
    \caption{Bulk water and water-air interface properties predicted by the MACE augmented with long-range electrostatics with latent charges, dipoles, and induced dipoles (MACELES-uiu) with anisotropic polarizability tensors, compared with DFT reference data and experiments.
    \figLabelCapt{a}: Parity plots of the Born effective charge (BEC) tensors predicted by the MACELES-uiu model ($r = 4.5$~\AA{}, $n_l = 1$) relative to RPBE-D3 DFT reference data for 100 bulk water configurations~\cite{Schmiedmayer2024}. The left panel compares the diagonal components ($Z^*_{\alpha\alpha}$), while the right panel shows the off-diagonal components ($Z^*_{\alpha\beta}$ with $\alpha \neq \beta$). 
    The RMSE and $R^2$ values are reported in each panel.
    \figLabelCapt{b}: Infrared (IR) absorption spectra of bulk water at different temperatures. The experimental IR spectrum~\cite{Bertie1996Infrared} is included for comparison.
    \figLabelCapt{c, d}: Parity plots comparing predicted (c) isotropic polarizability $\alpha=\mathrm{Tr}(\boldsymbol{\alpha})/3$ and (d) anisotropic component $\boldsymbol{\beta}=\boldsymbol{\alpha}-{\alpha}\boldsymbol{I}$ with reference DFT values for 1000 bulk water configurations~\cite{grisafi2018symmetry}. The Pearson correlation coefficient $r$ is reported in each panel.
    \figLabelCapt{e, f}: Simulated (e) isotropic and (f) anisotropic Raman spectra in the reduced representation of bulk water at different temperatures.
    \figLabelCapt{g, h}: Experimental (g) isotropic and (h) anisotropic reduced Raman spectra of bulk water at different temperatures. The experimental reduced Raman spectra are reproduced from Refs.~\cite{Pattenaude2018Temperaturea, Morawietz2018Interplay}. Additional experimental reduced Raman spectra at 300~K (green dashed line) taken from Ref.~\cite{marsalek2017quantum} and at 298~K (green dotted line) from Ref.~\cite{Brooker1989Ramana} are included for comparison.
    For visual clarity, the low-frequency region is scaled by the factor indicated in each panel; the dashed-line spectrum is scaled by a factor of 50 and 10 for the isotropic and anisotropic cases, respectively, while the scaling factor for the dotted line is not explicitly specified in Ref.~\cite{Brooker1989Ramana}. All spectra are truncated below 100~$\mathrm{cm}^{-1}$, the lowest frequency for which reliable experimental intensities are available~\cite{Pattenaude2018Temperaturea}. 
    \figLabelCapt{i}: A representative snapshot of the water slab with the surface-normal ($z$) direction indicated by the arrow.
    \figLabelCapt{j, k}: Simulated (j) and experimental (k) vibrational sum-frequency generation (VSFG) spectrum of the water-air interface. The experimental VSFG spectrum is taken from Ref.~\cite{tang2020Molecular}, which was obtained by reanalyzing the experimental spectra reported in Ref.~\cite{sun2016Phase}.
    All spectra are normalized to unit area and reported in arbitrary units (a.u.).
    }
    \label{fig:water_panel}
\end{figure*}

Electrical response properties, including polarizability $\boldsymbol{\alpha}$ and BECs, naturally emerge from the extended LES framework. 
Here, we explore whether such learned quantities are physically meaningful and if they are predictive of the Raman and infrared (IR) spectra of bulk water, as well as the vibrational sum-frequency generation (VSFG) spectrum of the water-air interface.
Vibrational Raman and IR spectroscopy provide complementary probes of the structure and dynamics of liquid water, as the two follow different selection rules while spanning similar frequency ranges~\cite{Bertie1996Infrared, Brooker1989Ramana}. Meanwhile, the VSFG signal is a surface-specific second-order optical technique because centrosymmetric bulk contribution vanishes in the second-order susceptibility, $\chi^{(2)}$, further enabling direct information on the molecular orientation at the interface~\cite{tang2020Molecular, Litman2023Fully}. Accurate theoretical prediction of these spectra remains an active area of research~\cite{Morawietz2018Interplay, Sommers2020Raman, wang2024n, medders2015Infrareda, Schmiedmayer2024, cassone2019ab, wan2013Raman, LaCour2023Predicting, Litman2023Fully, tang2020Molecular, Kapil2024Firstprinciples, moberg2018Temperature}.

We trained a MACELES-uiu model ($r=4.5$~\AA{}, $n_l=1$) with anisotropic latent polarizability tensors using the same RPBE-D3 bulk water dataset~\cite{Schmiedmayer2024} employed in the benchmark section.
The $\varepsilon_e$ used to determine BECs and polarizabilities was calculated using Eqn.~\eqref{eq:chi_les} and Eqn.~\eqref{eq:eps_e}, with water experimental $\varepsilon_\infty=1.78$. 
Fig.~\ref{fig:water_panel}a compares both the diagonal and off-diagonal elements of the BEC tensors predicted via Eqn.~\eqref{eq:z-pbc} against DFT reference values for bulk water~\cite{Schmiedmayer2024}, which exhibit excellent agreement with an RMSE of 0.022~e, even smaller than the bespoke ML model directly trained on the BECs (0.032~e)~\cite{Schmiedmayer2024}.
The accurate BECs predictions are also shown for other long-range augmented variants in Fig.~\ref{fig:water_bec} of Methods, with systematic improvement when more multipole and induced terms are included.

Figs.~\ref{fig:water_panel}c and d further compare the isotropic ($\alpha = \text{Tr}(\boldsymbol{\alpha})/3$) and the anisotropic components ($\boldsymbol{\beta} = \boldsymbol{\alpha} - \alpha\boldsymbol{I}$) of the predicted polarizabilities with DFT reference calculated at the PBE level~\cite{grisafi2018symmetry},
achieving Pearson coefficients of $r=$ 0.86 and 0.89, respectively.
The isotropic component exhibits a systematic offset of 11.0~\AA{}$^3$ (24.7\%) relative to the DFT values, while the anisotropic components are centered around zero and therefore do not exhibit a systematic offset.

We performed equilibrium NVT molecular dynamics simulations of bulk water across a temperature range of 280~K to 360~K at the corresponding experimental densities.
Using the predicted BEC tensors and polarizabilities, we calculated IR and reduced Raman spectra as detailed in the Methods. 

Fig.~\ref{fig:water_panel}b (green curve) compares the calculated IR spectra at 300~K with the experimental result at 298~K~\cite{Bertie1996Infrared}. The model reproduces the major vibrational features with correct peak positions and relative intensities, including the hydrogen-bond translational stretching mode ($\sim$200~$\mathrm{cm}^{-1}$), the librational region (400--750~$\mathrm{cm}^{-1}$), the bending mode ($\sim$1640~$\mathrm{cm}^{-1}$), and the O--H stretching band (3200--3600~$\mathrm{cm}^{-1}$). 

The learned $\boldsymbol{\alpha}$ enabled Raman spectra calculations: 
the isotropic $\alpha$ is used to compute the isotropic Raman, and the anisotropic $\boldsymbol{\beta}$ is used to compute the anisotropic spectra, as detailed in the Methods.
The green curves in Figs.~\ref{fig:water_panel}e and f present the simulated reduced isotropic and anisotropic Raman spectra at 300~K, compared to different sets of experimental spectra (solid green, dashed green, and dotted green)~\cite{Pattenaude2018Temperaturea, Brooker1989Ramana, Morawietz2018Interplay, marsalek2017quantum} shown in Figs.~\ref{fig:water_panel}g and h. All the simulated and experimental Raman spectra are shown in reduced representation~\cite{Brooker1989Ramana, Murphy1989Further}, which removes the dependence on the incident laser frequency and is particularly useful for highlighting low-frequency features. Although different definitions of reduced Raman spectra involve different frequency-dependent scaling factors, primarily affecting relative intensities in the low-frequency region, the peak positions remain unchanged~\cite{Brooker1989Ramana, Pattenaude2018Temperaturea}.

The agreement between simulated and experimental Raman spectra is somewhat less quantitative compared to IR, particularly for the relative intensities of low-frequency peaks. 
Given the good agreement for IR spectra, which suggests that the main vibrational dynamics and charge-response trends are captured, these discrepancies are likely due to the limitations in the predicted polarizability and may be related to the frequency-dependent scaling factors assumed in the Raman measurements.
Nevertheless, the simulated Raman spectra reproduce the main features:
The isotropic spectra capture the positions and shapes of the bending ($\sim$1640~$\mathrm{cm}^{-1}$) and O--H stretching modes (3200--3600~$\mathrm{cm}^{-1}$), while the anisotropic spectra additionally reproduce the hydrogen-bond translational stretching ($\sim$200~$\mathrm{cm}^{-1}$) and librational bands (400--750~$\mathrm{cm}^{-1}$), with minor deviations in peak shape.
The anisotropic Raman spectra further reproduce subtle features that are weak or absent in IR spectra. These include the broad association band near 2100~$\mathrm{cm}^{-1}$, commonly assigned to a combination of libration and bending modes~\cite{McCoy2014Role, marechal2011molecular}, and the high-frequency combination band ($\sim$4100~$\mathrm{cm}^{-1}$), arising from the coupling between librational and O--H stretching modes~\cite{Walrafen2004Raman, Morawietz2018Interplay}. 

We further investigate the temperature dependence of the water vibrational spectra. For both the computed IR (Fig.~\ref{fig:water_panel}b) and anisotropic Raman (Fig.~\ref{fig:water_panel}f) spectra, a decrease in temperature causes the intramolecular O--H stretching band (3200--3600~$\mathrm{cm}^{-1}$) to exhibit a red shift. The red shift is associated with the strengthening of the hydrogen-bond network~\cite{ojha2018Hydrogen,wang2004Vibrational} with increasing tetrahedrality of the water structure as the temperature decreases~\cite{Morawietz2018Interplay}.
The intermolecular librational band (400--750~$\mathrm{cm}^{-1}$) of IR and anisotropic Raman spectra blue-shifts with decreasing temperature. 
This blue shift is attributed to increasing tetrahedrality~\cite{Morawietz2018Interplay}, which imposes greater rotational restrictions on the water molecules~\cite{cassone2019ab}. 
These shifts resemble the behavior observed in experimental IR~\cite{marechal2011molecular} and anisotropic Raman~\cite{Pattenaude2018Temperaturea, Scherer1974Raman} spectra. Our results are also consistent with previous theoretical studies using MB-pol models~\cite{Reddy2017Temperaturedependent} and by combining MLIP MD and DFPT for anisotropic Raman spectra~\cite{Morawietz2018Interplay}.
Interestingly, lowering the temperature and imposing the external electric fields (See Fig.~\ref{fig:water_e_field} of Methods) yield similar shifts in IR and Raman spectra, consistent with a previous \emph{ab initio} molecular dynamics (AIMD) result at the BLYP-D3(BJ) level~\cite{cassone2019ab}.

The computed isotropic Raman spectra in Fig.~\ref{fig:water_panel}e capture a notable blue shift of the O--H stretching band (3200--3600~$\mathrm{cm}^{-1}$) with increasing temperature, which aligns well with experimental isotropic Raman data~\cite{Pattenaude2018Temperaturea, Scherer1974Raman}. This blue shift is further consistent with theoretical isotropic Raman studies using MB-pol~\cite{Reddy2017Temperaturedependent}, MLIP MD coupled with a ML model trained separately on polarizabilities~\cite{Sommers2020Raman}, and by combining MLIP MD and DFPT~\cite{Morawietz2018Interplay}.
However, we note that a distinct bimodal profile for the O--H stretching band (3200--3600~$\mathrm{cm}^{-1}$) of isotropic Raman spectra from experiments~\cite{Brooker1989Ramana, Pattenaude2018Temperaturea} is absent. Such bimodal feature is also absent in previous theoretical studies, including those based on the MB-pol models~\cite{Reddy2017Temperaturedependent}, AIMD simulations using revPBE-D3 and revPBE0-D3 functionals with polarizabilities evaluated via DFPT~\cite{marsalek2017quantum}, and the MLIP MD coupled with directly-learned polarizabilities~\cite{Sommers2020Raman}.

Meanwhile, our model qualitatively captures the isosbestic point~\cite{Walrafen1986Raman, Geissler2013Water}, located at $\sim$3370~$\mathrm{cm}^{-1}$ for isotropic Raman and $\sim$3540~$\mathrm{cm}^{-1}$ for anisotropic Raman, where the spectral intensity remains approximately constant upon temperature change. 
These values are in good agreement with the corresponding experimental observations ($\sim$3330~$\mathrm{cm}^{-1}$ and $\sim$3490~$\mathrm{cm}^{-1}$ for isotropic and anisotropic spectra, respectively)~\cite{Pattenaude2018Temperaturea}. While the isosbestic point can be interpreted as a balance between two temperature-independent spectral components with different tetrahedral environments~\cite{Morawietz2018Interplay}, 
it does not necessarily imply the presence of discrete structural environments, 
as it can be explained by the continuity of equilibrium
distributions~\cite{Geissler2005Temperature, Geissler2013Water}.

While IR and Raman spectra discussed above probe the vibrational response of bulk water, VSFG spectroscopy selectively probes water molecules at the water-air interface, providing additional information on their molecular orientation.
Since the VSFG spectrum is computed from the predicted BEC tensors and polarizabilities, as detailed in the Methods, it provides a meaningful test of whether the electrical response learned by the LES framework remains physically meaningful at interfaces.

We trained a MACELES-uiu model with anisotropic polarizability tensors on a revPBE-D3 interfacial water dataset~\cite{niblett2021, Cheng2025Latent}, and then performed equilibrium NVT molecular dynamics simulations of the water slab (Fig.~\ref{fig:water_panel}i) at 300~K. 

Figs.~\ref{fig:water_panel}j and~\ref{fig:water_panel}k show the calculated $\mathrm{Im}\,\chi^{(2)}$ and the experimental spectrum. 
Despite the shortcomings of the underlying GGA level DFT functional and the neglect of nuclear quantum effects in our classical MD simulations, the simulated VSFG spectrum qualitatively reproduces the two characteristic interfacial features, a broad negative band spanning 3200--3500~$\mathrm{cm}^{-1}$ and a positive free-O--H peak centered at $\sim$3600~$\mathrm{cm}^{-1}$.
These features originate from hydrogen-bonded O--H groups of interfacial water molecules oriented toward the bulk~\cite{Ji2008Characterization, Bonn2015Molecular} and dangling O--H groups of the outermost interfacial water layer~\cite{Du1993Vibrational, Bonn2015Molecular}, respectively.
Overall, the agreement demonstrates that the MACELES-uiu potential successfully captures the tensorial electrical response required to reproduce the essential features in the water-air VSFG.

\subsection{Methylammonium lead iodide perovskite}

\begin{figure*}
    \centering
    \includegraphics[width=0.9\linewidth]{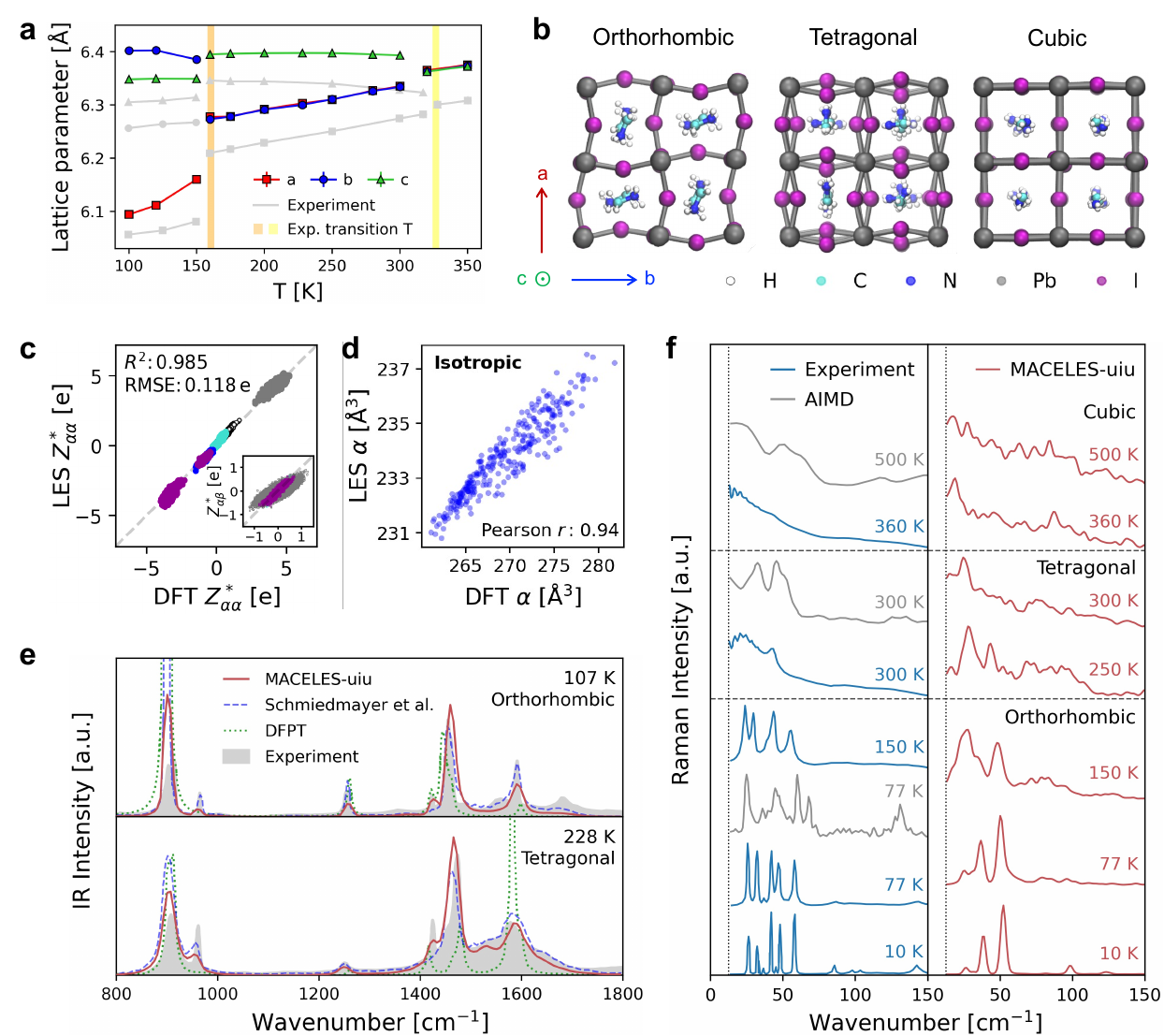}
    \caption{Phase diagram and electrical response properties of MAPbI$_3$ perovskite predicted by MACE augmented with long-range electrostatics with latent charges, dipoles, and induced dipoles (MACELES-uiu). 
    \figLabelCapt{a}: The pseudocubic lattice constants at different temperatures, compared to experiment~\cite{whitfield2016structures-b43}. The lattice parameters $a$, $b$, and $c$ correspond to the directions defined in panel \figLabelCapt{b}. Vertical orange and yellow bands mark the experimentally reported orthorhombic$-$tetragonal and tetragonal$-$cubic phase transition temperatures, respectively~\cite{whitfield2016structures-b43}. 
    \figLabelCapt{b}: Illustrations of the orthorhombic, tetragonal, and cubic MAPbI$_3$ phases. 
    \figLabelCapt{c}: Parity plot of Born effective charge (BEC) tensor $Z^*$ computed from MACELES-uiu model versus from DFT calculations. The main panel shows the diagonal components ($Z^*_{\alpha \alpha}$), while the inset displays the off-diagonal components ($Z^*_{\alpha \beta}$ with $\alpha \ne \beta$). 
    \figLabelCapt{d}: Parity plot of isotropic polarizability $\alpha$ predicted by the model compared to DFT values.
    \figLabelCapt{e}: Infrared spectra computed from the model (red) in the orthorhombic phase at 107~K and the tetragonal phase at 228~K. For comparison, experimental data (gray)~\cite{schuck2018infrared-342}, spectra using DFPT (green), and MLIP MD combined with direct learning of BECs from Schmiedmayer and Kresse~\cite{Schmiedmayer2024} (blue) are included. The computed vibrational frequencies are redshifted by 1.5\% to align with experiment, in accordance with Ref.~\cite{Schmiedmayer2024}.
    \figLabelCapt{f}: The Stokes lines from Raman spectra across a range of temperatures. Experimental unpolarized Raman intensities and AIMD results are taken from Ref.~\cite{sharma2020elucidating}. 
    Infrared and Raman intensities are reported in arbitrary units (a.u.). 
    }
    \label{fig:mapi}
\end{figure*}

The vibrational spectra of the organic-inorganic halide perovskite methylammonium lead iodide (MAPbI$_3$) have been extensively studied both experimentally and theoretically~\cite{perez2015vibrational}.
Experimental Raman measurements on MAPbI$_3$ are challenging because of its high light absorption coefficient and limited thermal stability~\cite{park2015resonance}, while theoretical predictions are complicated by its complex crystal structure, large unit cell, and strong anharmonicity~\cite{sharma2020elucidating,zhenbang2022}.
As a result, discrepancies between computational and experimental Raman spectra persist~\cite{quarti2013raman,brivio2015lattice,perez2018raman}, making MAPbI$_3$ a stringent test for models that aim to predict polarization-sensitive structural and vibrational observables.

Trained on the MAPbI$_3$ dataset~\cite{Schmiedmayer2024} based on SCAN DFT as described in the benchmark section, 
the MACE MLIPs with different levels of long-range augmentation all predict consistent phase diagrams and lattice constants, as shown in the Fig.~\ref{fig:mapi_lc} of the Methods.
This suggests that the long-range corrections do not significantly alter the thermodynamic behaviors, consistent with the previous consensus on the bulk material~\cite{yue2021short}.
Here we focus on the MACELES-uiu model with isotropic latent polarizability $\alpha$.
This model has test RMSEs of 0.3~meV/atom for energy and 10.0~meV/\AA{} for forces, lower than the previous MLIP fitted to the same dataset (0.4~meV/atom and 18~meV/\AA{})~\cite{Schmiedmayer2024}.

Fig.~\ref{fig:mapi}a shows the temperature-dependent pseudocubic lattice constants predicted by the MACELES-uiu model.
As temperature increases, the system undergoes transformations between orthorhombic, tetragonal, and cubic phases (atomic structures illustrated in Fig.~\ref{fig:mapi}b), 
with transition temperatures well-captured by the MLIP.
The orthorhombic-to-tetragonal transition shows a clear discontinuity at $\approx 160$ K, whereas transition at $\approx300$~K between tetragonal and cubic phases is more gradual due to the small free energy barrier~\cite{fransson2023revealing-ad5}.
These transition temperatures are quite consistent compared to the experimental measurements indicated by the vertical bands in Fig.~\ref{fig:mapi}a.
The computed lattice constants of the three phases are also quite consistent with a previous MLIP study based on the same SCAN functional~\cite{jinnouchi2019phase-507},
and the difference with experimental values may be due to the underlying DFT level of theory.

Fig.~\ref{fig:mapi}c compares the predicted BEC tensor components against PBE DFT reference values. The RMSE of $0.118$~e is larger than another ML model that was directly trained on the BECs (RMSE $=0.04$~e)~\cite{Schmiedmayer2024}. 
The larger error may be due to the different learning objectives (energies and forces versus BECs), and may also be affected by the difference in the underlying DFT functionals (SCAN versus PBE).
The BEC prediction accuracy of the MLIP is helped by the addition of the dipole (-u) and the induced dipole (-iu) terms to the monopoles, as shown in Fig.~\ref{fig:mapi_bec} of Methods.
As is characteristic of this polar and ionic perovskite, 
the diagonal BEC components massively exceed the nominal ionic charges, indicating strong coupling between atomic displacements and polarization.

The predicted BECs enabled IR spectra calculations, as shown in Fig.~\ref{fig:mapi}e for the orthorhombic phase at 107~K and the tetragonal phase at 228~K.
In addition to the experimental results (shaded gray areas)~\cite{schuck2018infrared-342}, 
we also compare to IR spectra calculated using DFPT (dotted green curves)~\cite{Schmiedmayer2024},
and another ML-based approach combining MLIP MD with a separate model directly predicting BECs (dashed blue curves)~\cite{Schmiedmayer2024}.
The DFPT spectra show a number of deficiencies, reflecting the lack of treatment for the anharmonicity.
Meanwhile, the MACELES-uiu spectra (red curves) show a small improvement over the direct BEC learning approach~\cite{Schmiedmayer2024}:
The peak heights associated with the rocking motion of MA cations at $\sim$900~$\mathrm{cm}^{-1}$ are in better agreement with the experiment,
and the peak near 1420 cm$^{-1}$ associated with CH$_3$ bending modes~\cite{perez2015vibrational} can be identified.

The isotropic polarizability $\alpha$ from the MACELES-uiu model is well-correlated with the PBE DFT reference values, although it exhibits a systematic offset of 35.7~\AA{}$^3$ (13.3\%) relative to the DFT reference,
as shown in Fig.~\ref{fig:mapi}d.
The Raman spectra in the low-frequency region based on the latent $\alpha$ for orthorhombic, tetragonal, and cubic phases across a range of temperatures are shown in Fig.~\ref{fig:mapi}f,
compared with experimental unpolarized spectra~\cite{sharma2020elucidating},
and AIMD simulations based on the PBE functional and the Tkatchenko-Scheffler dispersion combined with DFPT for computing polarizabilities~\cite{sharma2020elucidating}.
The frequency range probed here mainly corresponds to the vibrational modes for the PbI$_3$ cage and the libration/translation of the MA cations~\cite{perez2018raman}.

Here, both AIMD and MACELES-uiu Raman spectra do not quantitatively agree with experiments,
which is very typical for MAPbI$_3$~\cite{quarti2013raman,brivio2015lattice,perez2018raman} and reflects the challenges from both the simulation and the experimental sides.
From the MACELES-uiu side, one deficiency in the prediction is the lack of anisotropic components in the polarizabilities.
Nevertheless, both the AIMD and the MACELES-uiu reproduce the most important temperature-dependent trends observed in the experiment:
In the orthorhombic phase, the spectra at very low temperatures show sharp peaks, which become broadened with increasing temperature; 
the transition to the tetragonal phase introduces merging and broadening of peaks, and the spectra exhibit relatively broad features at low frequencies.
Entering the cubic phase,
the amplitudes for the very low-frequency mode further grow at higher temperatures, with a shape
resembling the Raman spectrum of a fluid~\cite{sharma2020elucidating}.

\subsection{Hafnium Dioxide}

\begin{figure*}
    \centering
    \includegraphics[width=0.9\linewidth]{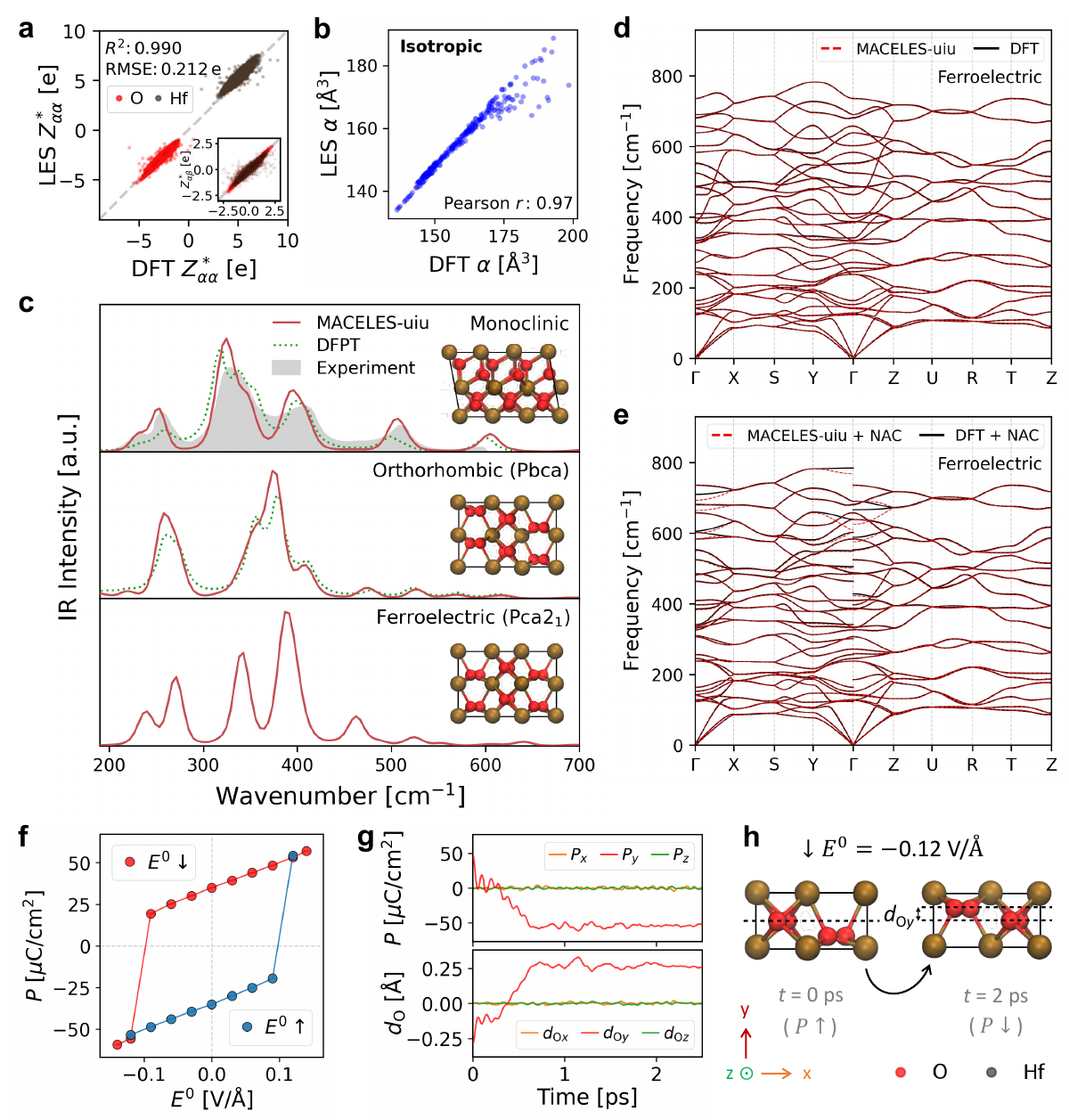}
    \caption{Electrical response properties of HfO$_2$ predicted by MACE augmented with long-range electrostatics with latent charges, dipoles, and induced dipoles (MACELES-uiu). 
    \figLabelCapt{a}: Parity plot of Born effective charge (BEC) tensor predicted by the MACELES-uiu model relative to the reference PBEsol DFT values. The diagonal components ($Z^*_{\alpha\alpha}$) are shown in the main panel, whereas the off-diagonal components ($Z^*_{\alpha\beta}$, $\alpha \ne \beta$) are presented in the inset. 
    \figLabelCapt{b}: Parity plot of isotropic polarizability $\alpha$ computed from the model versus from DFPT calculations. 
    \figLabelCapt{c}: Infrared spectra of the monoclinic, orthorhombic ($Pbca$), and ferroelectric ($Pca2_1$) phase of HfO$_2$ computed by the model (red) at room temperature, with insets illustrating the corresponding phase. The experimental IR spectrum (gray) \cite{toriumi2006doped-e52} and spectra using DFPT (green) \cite{zhou2014simulated-42d} are included for comparison. Infrared intensities are shown in arbitrary units (a.u.). 
    \figLabelCapt{d, e}: Phonon dispersions of ferroelectric ($Pca2_1$) phase HfO$_2$ with (e) and without (d) non-analytical correction (NAC). The red dotted and black solid lines are phonon dispersions calculated from the model and PBEsol DFT, respectively. 
    \figLabelCapt{f}: $P$-$E^0$ hysteresis loop of $Pca2_1$ HfO$_2$ at $T=300$ K computed by the model under different electric fields. 
    \figLabelCapt{g}: Time evolution of the polarization (top) and displacements of oxygen atoms (bottom) during the polarization-switching process at $E^0=-0.12 \, \mathrm{V/\AA}$.
    \figLabelCapt{h}: Schematics showing atomic displacement before and after polarization switching. }
    \label{fig:hfo2}
\end{figure*}

Ferroelectric hafnium dioxide (HfO$_2$) has attracted much attention as emerging memory devices owing to their spontaneous and switchable polarization~\cite{lee2023unveiled-8a5, schroeder2022fundamentals-ce3},
but the metastability of the ferroelectric phase makes both experimental and theorectial charecterization challenging~\cite{zhu2024progress-cd2, zhao2022microstructural-5dc, park2024atomic-scale-b11}.
HfO$_2$ is highly polymorphic even at ambient conditions: The monoclinic $P2_1/c$ is the thermodynamic ground state of the bulk,
the antipolar orthorhombic $Pbca$ phase is characterized by oppositely aligned local dipoles and zero net polarization,
the ferroelectric orthorhombic $Pca2_1$ is the phase that we primarily focus on here.

We recomputed the PBEsol DFT energies and forces for a HfO$_2$ dataset~\cite{chen2026electric-c45}, and trained different long-range augmented MACE variants (see Table~\ref{tab:mace-hfo2}).
Here we use a MACELES-uiu model, incorporating latent charges, dipoles, and induced dipoles with isotropic latent polarizability $\alpha$. 
This model uses a cutoff radius of $r=5.0$~\AA{}, 2 layers ($n_l=2$), and a total of 625,584 trainable parameters, yielding test RMSEs of 0.6 meV/atom for energy and 31.7 meV/$\mathrm{\AA}$ for forces.
For comparison, the previous short-range MACE MLIP with $r=6.0$~\AA{} and 1,882,904 trainable parameters and was constructed by finetuning the MACE-MP-0 foundation model~\cite{batatia2023foundation} on the same data set has larger errors: test mean-absolute errors (MAEs) of 2.14 meV/atom for energy and 30.2 meV/$\mathrm{\AA}$ for forces.

Fig.~\ref{fig:hfo2}a compares the BEC tensors predicted by the MACELES-uiu model with the reference PBEsol DFT values for 276 test configurations, covering six HfO$_2$ phases including monoclinic ($P2_1/c$), three orthorhombic ($Pca2_1$, $Pbca$, $Pnma$), tetragonal ($P4_2/nmc$), and cubic ($Fm\bar{3}m$).
The predicted values reproduce the anomalously large BECs in HfO$_2$ and show good agreement with DFT, yielding MAE of 0.125~e. 
For comparison, a pretrained BEC-prediction model produced a MAE of 0.065~e on six minimized structures of the six HfO$_2$ phases~\cite{chen2026electric-c45}. 
Our larger error may arise from the energy-and-force learning versus the direct BEC learning strategy, and may be due to the different, more challenging test configuration.
In addition, as shown in Fig.~\ref{fig:hfo2}b, the learned isotropic polarizability $\alpha$ from the MACELES-uiu model is also well-correlated with DFT reference values, yielding Pearson coefficient of $r= 0.97$, although it exhibits a systematic offset of 3.7~\AA{}$^3$ (2.3\%) relative to the DFT reference.

\begin{figure}
    \centering
    \includegraphics[width=0.85\linewidth]{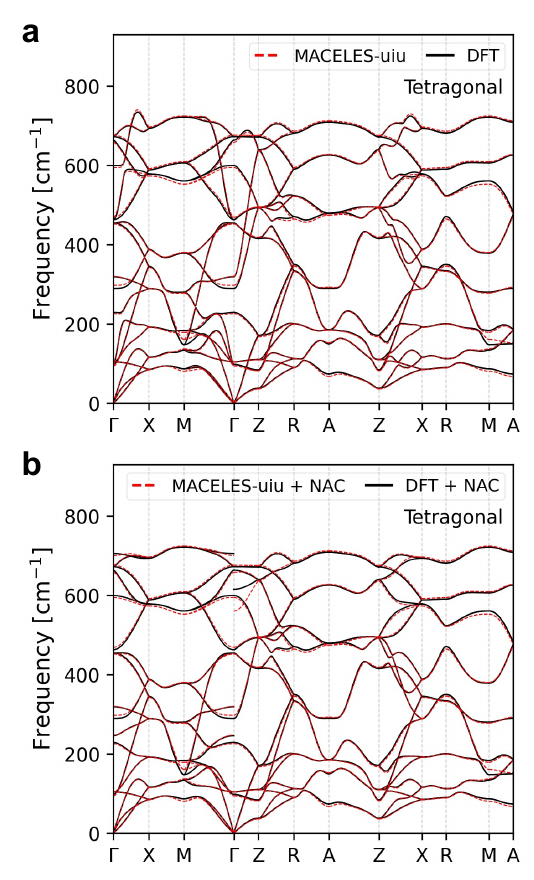}
    \caption{Phonon dispersions of tetragonal HfO$_2$ computed with (b) and without (a) the non-analytical correction (NAC). Red dotted and black solid lines denote the MACELES-uiu model and PBEsol DFT results, respectively.
    }
    \label{fig:hfo2_phonon}
\end{figure}

Long-wavelength optical phonons provide a stringent test of the learned electrical response because, in polar crystals like HfO$_2$, longitudinal optical (LO) and transverse optical (TO) modes split near the $\Gamma$ point through the macroscopic electric field generated by atomic displacements. 
This LO-TO splitting is governed by the magnitude of BECs and is therefore especially relevant for ferroelectric HfO$_2$. 
The BEC predictions enabled the inclusion of the non-analytical correction (NAC) in the phonon dispersion calculations, allowing the model to account for LO-TO splittings near $\Gamma$~\cite{gonze1997dynamical-202, zhong1994giant-6dc}; further details are given in the Methods. 
For the ferroelectric $Pca2_1$ phase of HfO$_2$, Fig.~\ref{fig:hfo2}d and Fig.~\ref{fig:hfo2}e show the phonon dispersions computed without and with NAC, respectively. In both cases, the model predictions, shown as dotted red lines, agree well with the PBEsol DFT results, shown as solid black lines. 
When NAC is included, the model also reproduces the large LO-TO splitting, demonstrating its ability to capture polarization-sensitive phonon behavior. Given the close agreement in the phonon dispersion without NAC, the remaining small deviations in the NAC-corrected dispersion are likely due to residual errors in the predicted BECs. To further assess robustness, phonon dispersions for the tetragonal phase are provided in Fig.~\ref{fig:hfo2_phonon}.

The IR spectra further provides complementary probes of 
finite-temperature vibrational dynamics and polarization response.
Fig.~\ref{fig:hfo2}c shows the calculated IR spectra (red curves) for three HfO$_2$ phases from NVE MD at 300~K: monoclinic, antipolar orthorhombic ($Pbca$), and ferroelectric orthorhombic ($Pca2_1$). 
The calculated IR for the monoclinic phase well reproduces the experimental peak positions and relative intensities~\cite{toriumi2006doped-e52} as well as the DFPT results~\cite{zhou2014simulated-42d}. 
The IR for the antipolar orthorhombic ($Pbca$) phase also agree well with DFPT~\cite{zhou2014simulated-42d}.
For the ferroelectric $Pca2_1$ phases, direct experimental or DFPT comparisons are not available,
but the signature frequencies agree with previous PBEsol DFT analysis on IR active modes reported by Ref.~\cite{fan2022vibrational-811}.

To characterize ferroelectric behavior of HfO$_2$, we computed polarization-electric field ($P$-$E^0$) hysteresis loop in NPT simulations at $T=300$~K, as shown in Fig.~\ref{fig:hfo2}f. 
With the absence of the external field, the predicted spontaneous polarization is approximately $35~\mu\mathrm{C/cm}^2$, comparable to the experimentally reported value of $30~\mu\mathrm{C/cm}^2$ \cite{shimizu2018ferroelectricity-7a5}.
The external electric field $E^0$ was then applied along or against the polarization direction (along the $y$-axis) of the equilibrated structure, with magnitudes ranging from $0.03$ to $0.14\, \mathrm{V/\AA^{}}$. 
The results shown in red and blue were obtained from two branches of simulations initiated from structures with opposite polarizations. 
The two simulation series exhibit polarization reversal at oppositely directed fields, forming a hysteresis loop and thereby confirming the ferroelectricity.  

Fig.~\ref{fig:hfo2}g shows the time evolution of polarization components (top) and the average oxygen displacement $d_{\mathrm{O}}$ (bottom) under an applied field of $E^0=-0.12\, \mathrm{V/\AA^{}}$. Here, $d_{\mathrm{O}}$ denotes the average displacements of the oxygen atoms relative to the non-polar centrosymmetric reference structure, as illustrated in Fig.~\ref{fig:hfo2}h. As the field is applied opposite to the initial polarization, $P_y$ rapidly reverses sign, accompanied by a simultaneous sign change in $d_{\mathrm{O}}$. This correlated reversal indicates that polarization switching is primarily associated with the collective displacement of oxygen atoms, consistent with previous first-principle studies based on PBE DFT combined with variable-cell nudged elastic band method on the switching mechanism~\cite{hu2025mapping-961, ma2023ultrahigh-8db}. 
Overall, the observed polarization switching demonstrates that the MACELES-uiu model captures the essential ferroelectric response in HfO$_2$.

\section{Discussion}

We introduce a physically transparent and hierarchical framework for learning long-range electrostatics in MLIPs. 
The model represents electrostatics through environment-dependent atomic multipoles predicted from local descriptors, while long-range charge transfer and polarization are captured through non-self-consistent response terms governed by local hardness parameters and polarizabilities. 
The framework can therefore be viewed as a semi-local approach to long-range electrostatics, in which dominant electrostatic interactions are encoded locally and residual non-local effects are incorporated through explicit linear response.

The framework retains the two key design principles of the LES method~\cite{Kim2026}:
(i) employing a physics-based representation with environment-dependent atomic multipoles and polarization to describe electrostatic interactions, and
(ii) avoiding explicit training from DFT partial charge labels, instead learning electrical response directly from energies and forces.

Compared to the original LES, the inclusion of higher-order multipoles and response terms yields general improvements in the accuracy of energies and forces.
Although the explicit atomic dipoles or quadrupole may not be essential in every system, it provides a systematic higher-order correction beyond monopoles and can thereby improve the model accuracy particularly for intermediate range electrostatic descriptions. Beyond these,
the accuracy of the predicted BECs is qualitatively and significantly enhanced compared to the monopole LES (Fig.~\ref{fig:water_bec} and Fig.~\ref{fig:mapi_bec}),
while the latent polarizability is a new capability.
These learned polarizabilities show strong correlation with physically computed values, and enable semi-quantitative modeling of VSFG and Raman spectra.

The present framework also offers a unifying perspective that conceptually connects many existing approaches.
The relationship between original LES and other long-range electrostatic models has been discussed in Ref.~\cite{Kim2026}.
The inclusion of fixed atomic multipoles beyond scalar charges, in particular equivariant dipole terms, brings the present framework closer to equivariant long-range message-passing approaches that incorporate Ewald summation, such as LOREM~\cite{Rumiantsev2025Learning} and Euclidean fast attention (EFA)~\cite{frank2026machine}.
On the other hand, the multipole expansion in 
Eqns.~\eqref{eqn:multipole-q}–\eqref{eqn:multipole-Q} 
can be viewed as a simplified representation of electronic density expansions in atomic orbitals~\cite{cuevas2020analytical,Baldwin2026}.

The induced charge component closely resembles charge equilibration schemes~\cite{rappe1991charge,Ghasemi2015Interatomic,ko2021fourth,fuchs2025Learninga}, while the induced dipole part follows the formulation behind polarizable force fields~\cite{Sala2010} and data-driven polarizable models such as MB-pol~\cite{Riera2023}.
A key distinction, however, is that the Qeq or polarizable force field approaches typically determine charges or induced dipoles through matrix inversion or iterative self-consistent procedures.
In contrast, the present framework separates electrostatic interactions into two contributions.
The atomic multipoles are determined by the local descriptors, which encode the positions and environments of neighboring atoms and thus effectively perform short-range charge equilibration.
The remaining, smaller, contribution arises from the electric potential and field generated by more distant atoms, and is treated non-self-consistently based on linear response.
This scheme has been tested in the diverse insulating systems in the paper. However, we caution that for systems where long-range charge transfer dominates, such as charged defects and redox reactions on metal-insulator interfaces, a global solve for charge redistribution may be needed.

Finally, our work clarifies a practical role for physics-informed modeling in increasingly flexible MLIPs.
Much of the current debate has focused on architectural choices, such as whether symmetry constraints (e.g., equivariance) remain essential as models scale~\cite{yuan2026foundation}, echoing the broader ``bitter lesson'' argument.
Our results point to a different use of physical structure: not to replace scale, but to expand what can be learned from standard labels.
Here, a compact and hierarchical electrostatic structure turns energies and forces into indirect supervision for measurable electrical response.
As a result, the model learns not only an accurate potential energy surface, but also latent variables that predict BECs, polarizabilities, and vibrational spectra.
This points to a route toward designing MLIPs whose internal representations are not only predictive, but physically interpretable and experimentally actionable.

\section{Methods}

\subsection{Screened electrostatics for finite systems}
We consider the total real-space electrostatic energy $U^\mathrm{elec} $ for a system with fixed charges, dipoles, and traceless quadrupoles,
based on the screened Coulomb kernel in Eqn.~\eqref{eq:erf}:
\begin{equation}
\phi(r) = \frac{\mathrm{erf}(a r)}{r}, 
\qquad g(ar) = e^{-a^2r^2},
\qquad a = \frac{1}{\sqrt{2}\sigma}.
\end{equation}
For simplicity, here we omit the Coulomb prefactor ($1/(4\pi\varepsilon_0)$) in all derived electric potential, field, and energy terms.
We define the displacement vector between a pair of atoms as $\mathbf{r}_{ij} = \mathbf{r}_j - \mathbf{r}_i = \mathbf{r}$, $r = |\mathbf{r}_{ij}|$, and $\hat{\mathbf{r}}= \mathbf{r}_{ij}/r_{ij}$,
and their interaction tensors are:
\begin{align}
f_{qq} \; \equiv\;& \phi(r), \\
f_{qu}^{a}      \;=\;& -\nabla_a \phi = s_1\ r \hat{\mathbf{r}}_a, \\
f_{uu}^{ab}     \;=\;& -\nabla_b f_{qu}^{a} = s_2\,\hat{\mathbf{r}}_a\hat{\mathbf{r}}_b - s_1\,\delta_{ab}, \\
f_{Qu}^{abc}    \;=\;& -\nabla_c f_{uu}^{ab} = s_3\,\hat{\mathbf{r}}_a\hat{\mathbf{r}}_b\hat{\mathbf{r}}_c - \tfrac{s_2}{r}\,[\delta\hat{\mathbf{r}}]^{(3)}_{abc}, \\
f_{QQ}^{abcd}   \;=\;& -\nabla_d f_{Qu}^{abc} = s_4\,\hat{\mathbf{r}}_a\hat{\mathbf{r}}_b\hat{\mathbf{r}}_c\hat{\mathbf{r}}_d \notag\\
  &- \tfrac{s_3}{r}\,[\delta\hat{\mathbf{r}}\hat{\mathbf{r}}]^{(6)}_{abcd} + \tfrac{s_2}{r^{2}}\,[\delta\delta]^{(3)}_{abcd}.
\end{align}
with the radial terms
\begin{align}
s_1 \;=\;& \frac{\operatorname{erf}(ar)}{r^{3}} - \frac{2a}{\sqrt{\pi}}\frac{g(ar)}{r^{2}}, \\
s_2 \;=\;& \frac{3\,\operatorname{erf}(ar)}{r^{3}} - \frac{6a}{\sqrt{\pi}}\frac{g(ar)}{r^{2}} - \frac{4a^{3}}{\sqrt{\pi}}g(ar), \\
s_3 \;=\;& \frac{15\,\operatorname{erf}(ar)}{r^{4}} - \frac{30a}{\sqrt{\pi}}\frac{g(ar)}{r^{3}} \notag\\
  &- \frac{20a^{3}}{\sqrt{\pi}}\frac{g(ar)}{r} - \frac{8a^{5}}{\sqrt{\pi}}g(ar)\,r, \\
s_4 \;=\;& \frac{105\,\operatorname{erf}(ar)}{r^{5}} - \frac{210a}{\sqrt{\pi}}\frac{g(ar)}{r^{4}} \notag\\
  &- \frac{140a^{3}}{\sqrt{\pi}}\frac{g(ar)}{r^{2}} - \frac{56a^{5}}{\sqrt{\pi}}g(ar) - \frac{16a^{7}}{\sqrt{\pi}}g(ar)\,r^{2}.
\end{align}
and completely symmetric tensors
\begin{align}
[\delta\hat{\mathbf{r}}]^{(3)}_{abc} \;\equiv\;& \delta_{ab}\hat{\mathbf{r}}_c + \delta_{ac}\hat{\mathbf{r}}_b + \delta_{bc}\hat{\mathbf{r}}_a, \\
[\delta\hat{\mathbf{r}}\hat{\mathbf{r}}]^{(6)}_{abcd} \;\equiv\;& \delta_{ab}\hat{\mathbf{r}}_c\hat{\mathbf{r}}_d + \delta_{ac}\hat{\mathbf{r}}_b\hat{\mathbf{r}}_d + \delta_{ad}\hat{\mathbf{r}}_b\hat{\mathbf{r}}_c \notag\\
   &+ \delta_{bc}\hat{\mathbf{r}}_a\hat{\mathbf{r}}_d + \delta_{bd}\hat{\mathbf{r}}_a\hat{\mathbf{r}}_c + \delta_{cd}\hat{\mathbf{r}}_a\hat{\mathbf{r}}_b, \\
[\delta\delta]^{(3)}_{abcd} \;\equiv\;& \delta_{ab}\delta_{cd} + \delta_{ac}\delta_{bd} + \delta_{ad}\delta_{bc}.
\end{align}

\noindent Given these interaction tensors, the electric potential at atom $j$ is given by
\begin{align}
\Phi_j \;=\; \sum_{i\neq j}\Bigl[
& \;q_i\,f_{qq}^{ij} + u_i^{a}\,f_{qu}^{ij,a} 
+ \tfrac{1}{2}\,Q_i^{ab}\,f_{uu}^{ij,ab} \Bigr],
\end{align}
where repeated Cartesian indices $(a,b,c)$ are summed over.
The $c$ component of the electric field at $j$ is
\begin{align}
E_{j}^c = -\nabla_c \phi_j = \sum_{i\neq j}\Bigl[
q_i\,f_{qu}^{ij,c} + u_i^{a}\,f_{uu}^{ij,ac}
+ \tfrac{1}{2}\,Q_i^{ab}\,f_{Qu}^{ij,abc}\Bigr],
\end{align}
and the gradient of the electric field component is
\begin{align}
\nabla_a E_{j}^b \;=\; -\sum_{i\neq j}\Bigl[
 \;q_i\,f_{uu}^{ij,ab} + u_i^{c}\,f_{Qu}^{ij,abc} 
+ \tfrac{1}{2}\,Q_i^{cd}\,f_{QQ}^{ij,abcd} \,\Bigr].
\end{align}

\noindent The total electrostatic energy is then given by

\begin{align}
U^\mathrm{elec}  \;=\; \tfrac{1}{2}\sum_{j}\Bigl[
& \Phi_j q_j - \mathbf{E}_j \cdot \mathbf{u}_j 
-\, \tfrac{1}{2}\,Q_j^{ab}\,\nabla_a E_j^b \,\Bigr].
\end{align}

\noindent The Ewald summation in Eqn.~\eqref{eq:e_lr} includes self-interaction terms that are excluded explicitly by the real-space summation via $\sum_{i\neq j}$. These self-interactions are given by
\begin{align}
\phi_j^{\mathrm{self}}   \;=\;& \frac{2a}{\sqrt{\pi}}\,q_j, \\
\mathbf{E}_j^{\mathrm{self}}  \;=\;& -\,\frac{4a^{3}}{3\sqrt{\pi}}\,\mathbf{u}_j, \\
(\nabla_a E_j^b)^{\mathrm{self}} \;=\;& -\,\frac{8a^{5}}{5\sqrt{\pi}}\,Q_j^{ab}.
\end{align}
which can be derived by taking the limit of the appropriate interaction tensors as $r \rightarrow 0$. In our implementation, these self-interaction terms are subtracted from the Ewald summation by default. 

\subsection{Implementation}
The LES package is implemented in PyTorch~\cite{Paszke2019PyTorch}, allowing seamless compatibility with other PyTorch-based MLIP frameworks.
For all architectures considered here (CACE, MACE, NequIP, and Allegro), invariant and equivariant features are generated from the same backbone within each architecture.

\paragraph{CACE}

Cartesian vectors and matrices are obtained from $\nu$-order tensor contractions \cite{wang2024n} over the CACE $A$ basis ~\cite{cheng2024cartesian}. For example, 
\begin{equation}
    A^{'l \neq 0, \nu=2}_{a_1...a_{x-z}b_1...b_{y-z}} = \bigoplus_{x+y-2z=l} A^{l_1=x}_{a_1...a_{x-z}c_1...c_z} A^{l_2=y}_{a_1...a_{y-z}c_1...c_z}
\end{equation}
The symmetrized CACE $B$-basis ~\cite{cheng2024cartesian} is maintained as the $l=0$ output. Given the equivariant $A'$ features, per-atom electrostatic multipoles and polarizabilities are obtained via a \texttt{TensorReadout} head that feeds the CACE $B$ features and norms of $A'$ features through a multi-layer perceptron (MLP) to obtain activations of linear combinations of the $A'$ features as well as point charges. Dipoles, quadrupoles, and polarizabilities are then read out as a linear combination of the activated equivariant features. Quadrupoles are made to be traceless before prediction. The CACE $B$-basis is additionally fed through an MLP to predict scalars, which are added to the diagonal of the predicted polarizability matrices.

\paragraph{MACE}

MACE~\cite{batatia2022mace} provides internal spherical representations, which are used to compute dipoles and polarizabilities. Dipoles are obtained as linear combinations of $1o$ features, and polarizabilities and quadrupoles are obtained via weighted outer products of the $1o$ features. Polarizabilities and quadrupoles are additionally obtained from linear combinations of the MACE $2e$ features if they exist in the hidden representation. Additionally, linear combinations are taken of the MACE $0e$ features to predict scalars, which are added to the diagonal of the predicted polarizability matrices. Quadrupoles are then made to be traceless.

\paragraph{NequIP and Allegro} 
For NequIP~\cite{batzner20223, Tan2026Highperformance} and Allegro~\cite{musaelian2023learning, Tan2026Highperformance}, we updated an NequIP extension package \texttt{NequIP-LES}~\cite{Kim2025Universalb}. The baseline architecture is selected via the \texttt{base\_model} key, and additional target quantities can be activated using \texttt{use\_dipole}, \texttt{use\_induced\_charge}, \texttt{use\_induced\_dipole}, \texttt{use\_quad}, and \texttt{use\_anisotropic\_polarizability} flags, without modifying the underlying training workflow.
Optional positive-definiteness constraints for $\alpha$ and $\kappa$ are also available.

In NequIP, the per-atom scalar quantities, namely $\alpha$, $\kappa$, and latent charges, are predicted by passing the $\ell=0$ node features from the final convolution layer block into an MLP with an output dimension of 1, analogous to how local atomic energies are computed. To predict atomic dipoles, quadrupoles, and anisotropic polarizabilities, the $\ell=1$ (odd parity, $p=-1$) node features are extracted just before the final convolution layer and passed through a single atom-wise linear layer. This linear layer projects the high-dimensional vector features down to a single three-dimensional vector representing the atomic dipole moment, and quadrupole and anisotropic polarizability tensors are constructed via outer products of three-dimensional vectors. When available, additional contributions to the tensorial quantities are additionally obtained from the $\ell=2$ (even parity, $p=1$) node features. For anisotropic polarizabilities, the scalar contributions $\alpha$ predicted from the $\ell=0$ node features are added to the diagonal components of the polarizability tensors. Quadrupoles are further constrained to be traceless.

In Allegro, which relies on local edge features, all per-atom quantities are derived by aggregating pairwise edge contributions. For scalar predictions (charges, $\kappa$, and $\alpha$), the $\ell=0$ edge features are processed through an edge-wise MLP with an output dimension of 1, and the resulting values are pooled across neighboring edges to form per-atom scalar values. For atomic dipoles, quadrupoles, and anisotropic polarizabilities, an edge-wise scalar weight from the $\ell=0$ edge features is computed in the same manner as other scalar quantities. This weight is then multiplied by the $\ell=1$ or $\ell=2$ component of the edge attributes ($Y_{l=1, p=-1}^{ij}$ or $Y_{l=2, p=1}^{ij}$, the spherical harmonic projection of the unit direction vector $\hat{\mathbf{r}}_{ij}$) to generate equivariant edge-wise vectors or tensors, which are then aggregated over all neighbors to yield the corresponding per-atom quantities. For quadrupoles and anisotropic polarizabilities, the quantities are treated analogously to the NequIP implementation. Quadrupoles are constrained to be traceless, while scalar contributions from the $\ell=0$ edge features are added to the diagonal components of the anisotropic polarizability tensors.

\paragraph{Timing}
\begin{figure}
\centering
    \includegraphics[width=0.9\linewidth]{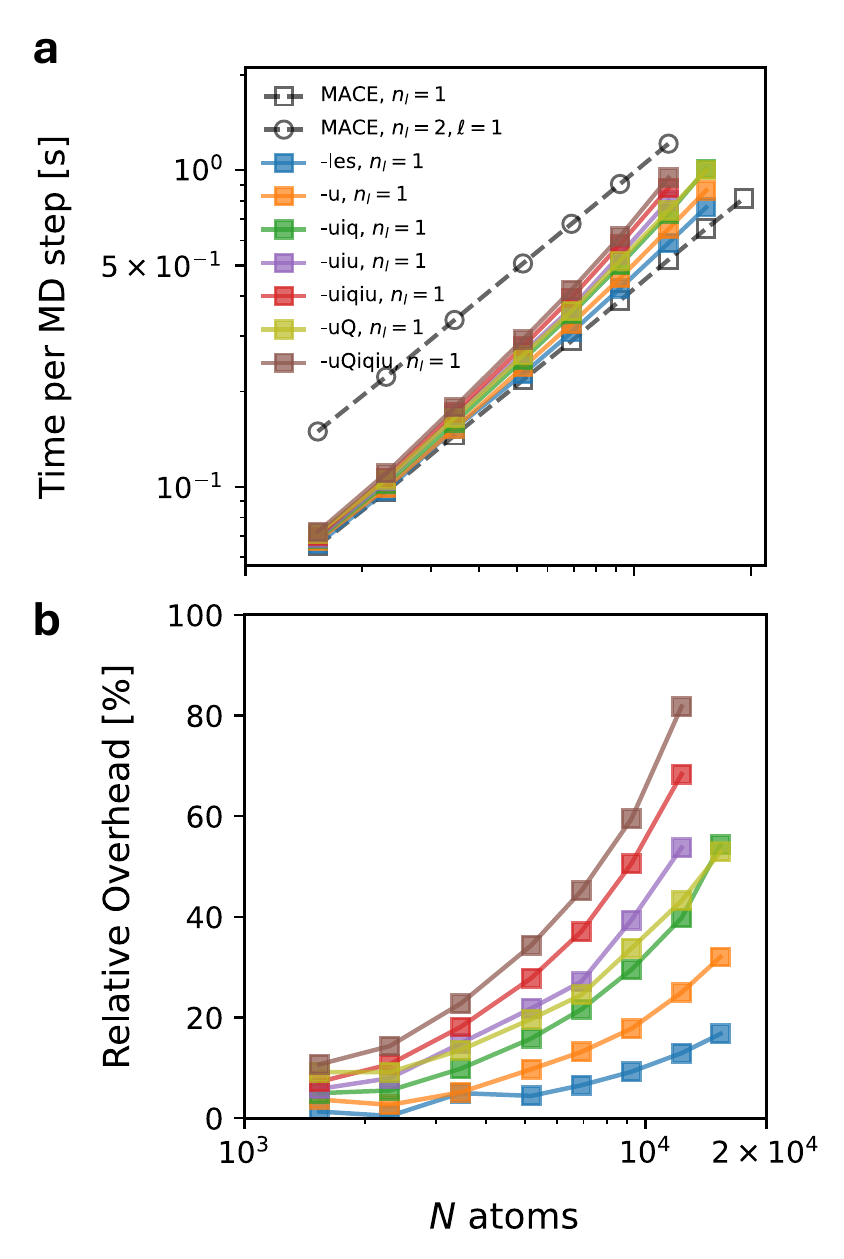}
     \caption{
     Computational performance benchmarks of molecular dynamics (MD) simulations using the baseline MACE models and their long-range augmented variants for bulk liquid water.
     \figLabelCapt{a}: Time per MD step for all models as a function of system size ($N$).
     \figLabelCapt{b}: Relative computational overhead of long-range augmented variants compared to the baseline MACE model ($n_l=1$).
     The benchmarks were performed on an NVIDIA L40S GPU (48 GB memory) using single-precision (float32) with ASE implementations.
     }
    \label{fig:timing}
\end{figure}

Fig.~\ref{fig:timing}a compares the inference speed of the short-range MACE~\cite{batatia2022mace} baseline with various long-range augmented variants used in Fig.~\ref{fig:benchmark}a for bulk liquid water. To more directly assess the cost of these augmentations, Fig.~\ref{fig:timing}b shows the relative computational overhead of each variant compared to the baseline MACE model ($n_l=1$). The results demonstrate that the additional computational overhead introduced by the long-range terms remains minimal relative to the baseline evaluation cost. While incorporating higher-order electrostatic terms, such as multipoles (-u, -Q), and response terms (-iq, -iu), increases the inference time, all variants maintain a similar favorable scaling with system size. Notably, even at the maximum system size supportable on a single GPU, the overhead of the most comprehensive model (MACELES-uQiqiu) remains lower than the cost of simply adding a message-passing interaction layer to the short-range baseline (e.g., $n_l=2$). 
In terms of memory consumption, the multipole-augmented models can simulate systems up to approximately 16,000 atoms on a single GPU, only slightly lower than 19,000 atoms of the baseline model.
These confirm that the long-range augmentation at any level of Eqn.~\eqref{eq:totE} provides an efficient route to incorporating long-range physics without compromising the high performance of underlying MLIP. 

\subsection{Datasets}
\paragraph{Bulk water}
The RPBE-D3 bulk water dataset~\cite{Schmiedmayer2024} consists of 604 train and 50 test configurations, each containing 64 water molecules.
The configurations were generated from MD trajectories spanning temperatures from 270~K to 420~K at the experimental water density under room temperature ($\approx 997$~kg/m$^3$), using an on-the-fly learning scheme. 
For the BEC inference test, we used an additional set of 100 bulk water snapshots obtained from an NVT simulation at room temperature and experimental density~\cite{Schmiedmayer2024}, with BEC values also computed at the RPBE-D3 DFT level.
For the polarizability inference test in Figs.~\ref{fig:water_panel}c and d, we used a separate dataset~\cite{grisafi2018symmetry} of 1000 bulk water snapshots containing 32 water molecules with polarizabilities computed at the PBE DFT level.
We assumed the experimental high-frequency permittivity ($\epsilon_\infty=1.78$) of water.

\paragraph{MAPbI$_3$}
The SCAN DFT methylammonium lead iodide (MAPbI$_3$) dataset~\cite{Schmiedmayer2024} contains 1414 configurations (95\% train/5\% test split) each with 96 atoms. The configurations were generated from MD trajectories using an on-the-fly learning scheme at temperatures ranging from 80~K to 430~K, covering the orthorhombic and tetragonal phases.
For the BEC inference test, we used an additional set of 300 configurations~\cite{Schmiedmayer2024}, spanning the orthorhombic, tetragonal, and cubic phases, computed using DFPT at the PBE level. 
The polarizabilities of the configurations in this dataset were calculated from the static dielectric tensor using the Clausius–Mossotti relation.
The constant high-frequency permittivity ($\epsilon_\infty=4.7$) of MAPbI$_3$ was obtained as the average value from DFPT calculations.

\paragraph{Na$_{8/9}$Cl$_{8}^{+}$}
The Na$_{8/9}$Cl$_{8}^{+}$ set from Ko et al.~\cite{ko2021fourth} consists of 5000 configurations (95\% train/5\% test split), including ionic Na$_{9}$Cl$_{8}^{+}$ clusters and Na$_{8}$Cl$_{8}^{+}$ obtained by removing a neutral Na atom. Since the overall positive charge of the cluster remains unchanged, this system provides an example of global charge transfer. 

\paragraph{Au$_{2}$ on MgO(001)}
The Au$_{2}$-MgO(001) dataset from Ko et al.~\cite{ko2021fourth} consists of 5000 configurations (90\% train/10\% test split) of a gold dimer adsorbed on an MgO(001) surface in upright non-wetting and parallel wetting geometries. In a subset of configurations, three Al dopant atoms are introduced in the fifth subsurface layer. Despite being more than 10~\AA{} away from the gold dimer, the dopants strongly influence the electronic structure and the relative stability of the adsorption geometries, highlighting the importance of long-range interactions.

\subsection{Details on benchmarks}

The training scripts, corresponding hyperparameters, and the trained models are available in the Supplemental Information (SI) repository. For the long-range augmentation, we utilized the default parameters across the three periodic systems (bulk water, MAPbI$_3$, and Au$_{2}$ on MgO(001)): $\sigma = 1$~\AA{}, $dl=2$~\AA{}, corresponding to a $\mathbf{k}$-point cutoff of $k_c = \pi$ in the Ewald summation.
The Na$_{8/9}$Cl$_{8}^{+}$ system used real-space summation.
It is also worth noting that, in our trials, while the training dynamics of NequIP were more sensitive and hyperparameter-dependent than those of other MLIPs, the architecture still exhibited an unambiguous trend of improvement across all cases upon incorporating long-range augmentation, as shown in Fig.~\ref{fig:benchmark} and Table~\ref{tab:benchmark-error}.

\begin{table*}
\centering
\caption{Energy root-mean-square errors (RMSEs) in meV/atom, force RMSEs in meV/\AA{}, and the number of trainable parameters ($N_{\mathrm{params}}$)
for various MLIP architectures and their long-range augmented variants compared to DFT reference data~\cite{Schmiedmayer2024, ko2021fourth}. The reported energy and force RMSEs are mean values from three independent training runs with different training seeds, and the sample standard deviation of the force RMSE is given in parentheses. The best-performing model (lowest force RMSE) within each architecture for each system is highlighted in bold.}
\label{tab:benchmark-error}
\footnotesize
\renewcommand{\arraystretch}{1.1}

\definecolor{tabblue}{HTML}{1F77B4}
\definecolor{taborange}{HTML}{FF7F0E}
\definecolor{tabgreen}{HTML}{2CA02C}
\definecolor{tabpurple}{HTML}{9467BD}
\definecolor{tabred}{HTML}{D62728}
\definecolor{tabolive}{HTML}{BCBD22}
\definecolor{tabbrown}{HTML}{8C564B}

\begin{tabular}{c|c||cccc}
\hline
\hline
\textbf{MLIP} & \textbf{Model} & \textbf{Water} & \textbf{MAPbI$_\mathbf{3}$} & \textbf{Na$_\mathbf{_{8/9}}$Cl$_\mathbf{_{8}}^+$} & \textbf{Au$_\mathbf{_2}$-MgO} \\
\hline
& & \multicolumn{4}{c}{E RMSE [meV/atom] $|$ F RMSE [meV/\AA{}] $|$ $N_{\mathrm{params}}$} \\
\hline

\multirow{8}{*}{\textbf{MACE}}
  & Baseline & 0.3 $|$ 27.57(0.21) $|$ 211,328 & 0.4 $|$ 14.73(0.06) $|$ 102,144 & 1.6 $|$ 43.13(0.29) $|$ 211,328 & 2.3 $|$ 56.83(0.15) $|$ 266,368 \\
& -les     & \cellcolor{tabblue!20} 0.2 $|$ 17.50(0.17) $|$ 211,456 & \cellcolor{tabblue!20} 0.3 $|$ 12.47(0.06) $|$ 102,208 & \cellcolor{tabblue!20} 0.1 $|$ 5.07(0.23) $|$ 211,456 & \cellcolor{tabblue!20} 0.1 $|$ 7.03(0.45) $|$ 266,496 \\
& -u       & \cellcolor{taborange!20} 0.1 $|$ 11.47(0.06) $|$ 211,584 & \cellcolor{taborange!20} 0.3 $|$ 10.57(0.15) $|$ 102,272 & \cellcolor{taborange!20} 0.1 $|$ 3.87(0.15) $|$ 211,584 & \cellcolor{taborange!20} 0.1 $|$ 5.27(0.15) $|$ 266,624 \\
& -uiq     & \cellcolor{tabgreen!20} 0.2 $|$ 11.43(0.15) $|$ 211,712 & \cellcolor{tabgreen!20} 0.3 $|$ 10.43(0.21) $|$ 102,336 & \cellcolor{tabgreen!20} 0.1 $|$ 3.22(0.11) $|$ 211,712 & \cellcolor{tabgreen!20} 0.1 $|$ 4.47(0.12) $|$ 266,752 \\
& -uiu     & \cellcolor{tabpurple!20} 0.2 $|$ 10.83(0.12) $|$ 211,712 & \cellcolor{tabpurple!20} 0.3 $|$ 10.23(0.21) $|$ 102,336 & \cellcolor{tabpurple!20} 0.1 $|$ 3.94(0.06) $|$ 211,712 & \cellcolor{tabpurple!20} 0.1 $|$ 4.80(0.20) $|$ 266,752 \\
& -uiqiu   & \cellcolor{tabred!20} 0.2 $|$ 10.77(0.06) $|$ 211,840 & \cellcolor{tabred!20} 0.3 $|$ 10.13(0.21) $|$ 102,400 & \cellcolor{tabred!20} 0.1 $|$ 3.30(0.10) $|$ 211,840 & \cellcolor{tabred!20} 0.1 $|$ 4.40(0.26) $|$ 266,880 \\
& -uQ      & \cellcolor{tabolive!20} 0.1 $|$ 8.87(0.06) $|$ 260,864 & \cellcolor{tabolive!20} 0.3 $|$ 10.10(0.44) $|$ 114,624 & \cellcolor{tabolive!20} 0.1 $|$ 3.60(0.10) $|$ 260,864 & \cellcolor{tabolive!20} 0.1 $|$ 5.03(0.15) $|$ 315,904 \\
& -uQiqiu  & \cellcolor{tabbrown!20} 0.1 $|$ \textbf{8.40(0.10)} $|$ 310,400 & \cellcolor{tabbrown!20} 0.3 $|$ \textbf{9.67(0.32)} $|$ 127,104 & \cellcolor{tabbrown!20} 0.1 $|$ \textbf{2.80(0.17)} $|$ 261,120 & \cellcolor{tabbrown!20} 0.1 $|$ \textbf{4.13(0.15)} $|$ 316,160 \\
\hline

\multirow{8}{*}{\textbf{CACE}}
  & Baseline & 0.3 $|$ 31.35(0.35) $|$ 11,248 & 0.4 $|$ 18.67(0.20) $|$ 24,590 & 1.7 $|$ 48.94(0.11) $|$ 7,696 & 2.3 $|$ 59.36(0.10) $|$ 43,240 \\
& -les     & \cellcolor{tabblue!20} 0.3 $|$ 21.89(0.78) $|$ 21,330 & \cellcolor{tabblue!20} 0.3 $|$ 15.81(0.05) $|$ 46,552 & \cellcolor{tabblue!20} 0.2 $|$ 8.29(0.19) $|$ 14,610 & \cellcolor{tabblue!20} 0.1 $|$ 7.85(0.34) $|$ 81,834 \\
& -u       & \cellcolor{taborange!20} 0.2 $|$ 17.33(0.79) $|$ 22,346 & \cellcolor{taborange!20} 0.3 $|$ 15.04(0.12) $|$ 48,828 & \cellcolor{taborange!20} 0.2 $|$ 6.57(0.11) $|$ 15,290 & \cellcolor{taborange!20} 0.1 $|$ 7.98(0.08) $|$ 85,874 \\
& -uiq     & \cellcolor{tabgreen!20} 0.2 $|$ 16.76(0.15) $|$ 22,346 & \cellcolor{tabgreen!20} 0.3 $|$ 14.61(0.64) $|$ 48,828 & \cellcolor{tabgreen!20} 0.2 $|$ 6.22(0.33) $|$ 15,290 & \cellcolor{tabgreen!20} 0.1 $|$ \textbf{5.63(0.33)} $|$ 85,874 \\
& -uiu     & \cellcolor{tabpurple!20} 0.2 $|$ 16.96(0.14) $|$ 23,169 & \cellcolor{tabpurple!20} 0.2 $|$ 13.58(0.21) $|$ 50,671 & \cellcolor{tabpurple!20} 0.1 $|$ 3.84(0.06) $|$ 15,841 & \cellcolor{tabpurple!20} 0.1 $|$ 6.77(0.63) $|$ 89,145 \\
& -uiqiu   & \cellcolor{tabred!20} 0.2 $|$ 16.69(0.11) $|$ 23,169 & \cellcolor{tabred!20} 0.2 $|$ 13.58(0.29) $|$ 50,671 & \cellcolor{tabred!20} 0.1 $|$ 3.91(0.04) $|$ 15,841 & \cellcolor{tabred!20} 0.1 $|$ 6.11(0.31) $|$ 89,145 \\
& -uQ      & \cellcolor{tabolive!20} 0.2 $|$ 13.14(0.07) $|$ 23,169 & \cellcolor{tabolive!20} 0.2 $|$ 13.40(0.10) $|$ 50,671 & \cellcolor{tabolive!20} 0.1 $|$ 4.67(0.02) $|$ 15,841 & \cellcolor{tabolive!20} 0.1 $|$ 7.78(0.36) $|$ 89,145 \\
& -uQiqiu  & \cellcolor{tabbrown!20} 0.1 $|$ \textbf{12.42(0.14)} $|$ 34,825 & \cellcolor{tabbrown!20} 0.2 $|$ \textbf{12.34(0.29)} $|$ 76,127 & \cellcolor{tabbrown!20} 0.1 $|$ \textbf{3.81(0.14)} $|$ 17,096 & \cellcolor{tabbrown!20} 0.1 $|$ 6.38(0.81) $|$ 96,480 \\
\hline

\multirow{8}{*}{\textbf{NequIP}}
  & Baseline & 0.2 $|$ 16.81(0.17) $|$ 177,768 & 0.5 $|$ 12.51(0.26) $|$ 177,864 & 1.6 $|$ 34.91(0.67) $|$ 489,576 & 2.4 $|$ 57.18(0.20) $|$ 58,536 \\
& -les     & \cellcolor{tabblue!20} 0.2 $|$ 15.74(0.13) $|$ 177,800 & \cellcolor{tabblue!20} 0.4 $|$ 12.37(0.18) $|$ 177,896 & \cellcolor{tabblue!20} 0.4 $|$ 7.17(1.17) $|$ 489,608 & \cellcolor{tabblue!20} 0.8 $|$ 10.28(0.21) $|$ 58,568 \\
& -u       & \cellcolor{taborange!20} 0.2 $|$ 14.93(0.14) $|$ 177,832 & \cellcolor{taborange!20} 0.4 $|$ 12.07(0.35) $|$ 177,928 & \cellcolor{taborange!20} 0.3 $|$ 6.23(0.70) $|$ 489,640 & \cellcolor{taborange!20} 0.7 $|$ 9.67(0.40) $|$ 58,600 \\
& -uiq     & \cellcolor{tabgreen!20} 0.2 $|$ 14.88(0.09) $|$ 177,864 & \cellcolor{tabgreen!20} 0.5 $|$ 11.96(0.23) $|$ 177,960 & \cellcolor{tabgreen!20} 0.7 $|$ 6.01(0.52) $|$ 489,672 & \cellcolor{tabgreen!20} 0.3 $|$ 7.67(0.58) $|$ 58,632 \\
& -uiu     & \cellcolor{tabpurple!20} 0.2 $|$ 14.93(0.07) $|$ 177,864 & \cellcolor{tabpurple!20} 0.5 $|$ 11.66(0.33) $|$ 177,960 & \cellcolor{tabpurple!20} 0.9 $|$ 6.54(0.78) $|$ 489,672 & \cellcolor{tabpurple!20} 1.0 $|$ \textbf{7.54(0.50)} $|$ 58,632 \\
& -uiqiu   & \cellcolor{tabred!20} 0.2 $|$ 15.07(0.14) $|$ 177,896 & \cellcolor{tabred!20} 0.5 $|$ 12.03(0.39) $|$ 177,992 & \cellcolor{tabred!20} 0.4 $|$ 5.19(1.02) $|$ 489,704 & \cellcolor{tabred!20} 0.2 $|$ 8.54(0.50) $|$ 58,664 \\
& -uQ      & \cellcolor{tabolive!20} 0.2 $|$ 14.71(0.18) $|$ 177,864 & \cellcolor{tabolive!20} 0.5 $|$ 11.67(0.42) $|$ 177,960 & \cellcolor{tabolive!20} 0.6 $|$ 4.25(0.65) $|$ 489,704 & \cellcolor{tabolive!20} 0.3 $|$ 10.45(0.70) $|$ 58,632 \\
& -uQiqiu  & \cellcolor{tabbrown!20} 0.2 $|$ \textbf{12.51(0.10)} $|$ 177,960 & \cellcolor{tabbrown!20} 0.5 $|$ \textbf{11.57(0.32)} $|$ 178,056 & \cellcolor{tabbrown!20} 0.3 $|$ \textbf{4.14(1.08)} $|$ 489,832 & \cellcolor{tabbrown!20} 0.3 $|$ 7.87(0.16) $|$ 58,728 \\
\hline

\multirow{8}{*}{\textbf{Allegro}}
  & Baseline & 0.3 $|$ 28.17(0.17) $|$ 308,992 & 0.7 $|$ 16.56(0.57) $|$ 309,376 & 1.8 $|$ 48.09(1.20) $|$ 308,992 & 2.3 $|$ 57.51(0.03) $|$ 214,848 \\
& -les     & \cellcolor{tabblue!20} 0.2 $|$ 18.24(0.63) $|$ 374,656 & \cellcolor{tabblue!20} 0.5 $|$ 14.50(0.24) $|$ 375,040 & \cellcolor{tabblue!20} 0.3 $|$ 9.97(0.26) $|$ 374,656 & \cellcolor{tabblue!20} 0.1 $|$ 5.40(0.25) $|$ 264,128 \\
& -u       & \cellcolor{taborange!20} 0.2 $|$ 12.95(0.06) $|$ 440,320 & \cellcolor{taborange!20} 0.5 $|$ 11.96(0.16) $|$ 440,704 & \cellcolor{taborange!20} 0.3 $|$ 6.67(0.16) $|$ 440,320 & \cellcolor{taborange!20} 0.1 $|$ 4.28(0.51) $|$ 313,408 \\
& -uiq     & \cellcolor{tabgreen!20} 0.2 $|$ 12.79(0.26) $|$ 505,984 & \cellcolor{tabgreen!20} 0.5 $|$ 12.07(0.20) $|$ 506,368 & \cellcolor{tabgreen!20} 0.8 $|$ 5.55(0.66) $|$ 505,984 & \cellcolor{tabgreen!20} 0.2 $|$ 3.96(0.48) $|$ 362,688 \\
& -uiu     & \cellcolor{tabpurple!20} 0.2 $|$ 12.68(0.22) $|$ 505,984 & \cellcolor{tabpurple!20} 0.6 $|$ 11.57(0.07) $|$ 506,368 & \cellcolor{tabpurple!20} 0.4 $|$ 5.43(0.14) $|$ 505,984 & \cellcolor{tabpurple!20} 0.2 $|$ 3.88(0.42) $|$ 362,688 \\
& -uiqiu   & \cellcolor{tabred!20} 0.2 $|$ 12.82(0.43) $|$ 571,648 & \cellcolor{tabred!20} 0.6 $|$ 11.61(0.09) $|$ 572,032 & \cellcolor{tabred!20} 0.4 $|$ \textbf{4.67(0.05)} $|$ 571,648 & \cellcolor{tabred!20} 0.2 $|$ 3.39(0.17) $|$ 411,968 \\
& -uQ      & \cellcolor{tabolive!20} 0.1 $|$ 9.79(0.01) $|$ 505,984 & \cellcolor{tabolive!20} 0.5 $|$ 11.30(0.10) $|$ 506,368 & \cellcolor{tabolive!20} 0.6 $|$ 5.26(0.44) $|$ 505,984 & \cellcolor{tabolive!20} 0.2 $|$ 3.81(0.14) $|$ 362,688 \\
& -uQiqiu  & \cellcolor{tabbrown!20} 0.1 $|$ \textbf{8.85(0.15)} $|$ 702,976 & \cellcolor{tabbrown!20} 0.5 $|$ \textbf{10.92(0.19)} $|$ 703,360 & \cellcolor{tabbrown!20} 0.4 $|$ 4.73(1.05) $|$ 702,976 & \cellcolor{tabbrown!20} 0.1 $|$ \textbf{3.27(0.38)} $|$ 510,528 \\
\hline
\hline
\end{tabular}
\end{table*}

Table~\ref{tab:benchmark-error} provides the mean energy and force RMSEs over three independent training runs with different random seeds, with the sample standard deviation of the force RMSE in parentheses, for all architectures and benchmark datasets across different levels of long-range augmentation (see Fig.~\ref{fig:benchmark}). Although the force RMSE generally decreases with increasing long-range augmentation, the trend is not strictly monotonic. In some cases, such as the NequIP models for the Na$_{8/9}$Cl$_8^+$ system, differences in force RMSE among some of the augmented models are comparable to the seed-to-seed variation and therefore cannot be unambiguously attributed to the added terms alone.

\begin{table}
\centering
\caption{Energy (meV/atom), force (meV/\AA{}) RMSEs, and the number of trainable parameters ($N_{\mathrm{params}}$)} for various CACE models ($r=4.5$~\AA{}, $n_l=1$, $\nu=3$) on the bulk water dataset~\cite{Schmiedmayer2024}. The benchmark includes different levels of long-range augmentation and modeling options, including the type of polarizability, the applied positive-definiteness constraints, and the number of self-consistent loops ($N_{\mathrm{sc}}$).
\label{tab:cace-water}
\small
\renewcommand{\arraystretch}{1.1}

\begin{tabular}{l|c|c|cccc}
\hline
\hline
\textbf{Model} & \textbf{E} & \textbf{F} & {Polarizability} & Positive & $N_{\mathrm{sc}}$ & $N_{\mathrm{params}}$\\
 &  &  & iso/aniso& constraint & \\
\hline
\hline
CACE & 0.31 & 31.0 & -- & -- & -- & 11,248\\
-les & 0.25 & 21.0 & -- & -- & -- & 21,330\\
-u & 0.20 & 17.6 & -- & -- & -- & 22,346\\
-uiq & 0.19 & 16.8 & -- & -- & -- & 22,346\\
\hline
-uiu & 0.18 & 16.8 & iso ($\alpha$) & -- & -- & 32,428\\
-uiu & 0.21 & 17.1 & aniso ($\boldsymbol{\alpha}$) & -- & -- & 23,169\\
-uiu & 0.19 & 16.6 & aniso ($\boldsymbol{\alpha}$) & $\boldsymbol{\alpha}$ & -- & 23,169\\
-uiu & 0.19 & 16.6 & aniso ($\boldsymbol{\alpha}$) & -- & 1  & 23,169\\
-uiu & 0.19 & 16.6 & aniso ($\boldsymbol{\alpha}$) & -- & 2  & 23,169\\
\hline
-uiqiu & 0.18 & 16.9 & iso ($\alpha$) & -- & -- & 32,428\\
-uiqiu & 0.20 & 16.8 & aniso ($\boldsymbol{\alpha}$) & -- & -- & 23,169\\
-uiqiu & 0.21 & 17.2 & aniso ($\boldsymbol{\alpha}$) & $\boldsymbol{\alpha}$ & -- & 23,169\\
-uiqiu & 0.20 & 16.9 & aniso ($\boldsymbol{\alpha}$) & $\boldsymbol{\alpha}, \kappa$ & -- & 23,169\\
-uiqiu & 0.19 & 16.8 & aniso ($\boldsymbol{\alpha}$) & -- & 1 & 23,169\\
-uiqiu & 0.21 & 16.6 & aniso ($\boldsymbol{\alpha}$) & -- & 2 & 23,169\\
\hline
-uQ & 0.15 & 13.2 & -- & -- & -- & 23,169\\
-uQiu & 0.10 & 11.7 & aniso ($\boldsymbol{\alpha}$) & -- & -- & 34,825\\
-uQiqiu & 0.10 & 12.3 & aniso ($\boldsymbol{\alpha}$) & -- & -- & 34,825\\
\hline
\hline
\end{tabular}
\end{table}

Table~\ref{tab:cace-water} summarizes a detailed error analysis for CACE models on bulk water across various long-range augmentation levels and model design space options. 
The direct comparison between the CACELES(-uiu/uiqiu) models with isotropic $\alpha$ and the ones with anisotropic tensors $\boldsymbol{\alpha}$ reveals no clear advantage for either choice in terms of energy or force RMSEs.
Enforcing positive-definiteness constraints on polarizability or hardness $\kappa$ also does not significantly alter performance, as unconstrained networks naturally learn to output positive values in the cases that we observed.
Introducing a self-consistent scheme, where the fields are iteratively updated to include contributions from induced terms, leads to negligible changes in accuracy.

\begin{table}[t]
\centering
\caption{
Energy (meV/atom) and force (meV/\AA{}) RMSEs and the number of trainable parameters ($N_{\mathrm{params}}$) for parameter-matched CACE models ($r=4.5$~\AA{}, $n_l=1$, $\nu=3$) on the bulk water dataset~\cite{Schmiedmayer2024}. The left and right columns correspond to two model families with approximately 11k and 35k trainable parameters, respectively.}
\label{tab:cace-param-matched}
\small
\renewcommand{\arraystretch}{1.2}
\begin{tabular}{l|ccc|ccc}
\hline
\hline
\textbf{Model} & \textbf{E} & \textbf{F} & \textbf{$N_{\mathrm{params}}$} & \textbf{E} & \textbf{F} & \textbf{$N_{\mathrm{params}}$} \\
\hline
\hline
CACE            & 0.31 & 31.0 & 11,248 & 0.30 & 31.0 & 35,456 \\
-les         & 0.26 & 23.1 & 11,282 & 0.23 & 22.2 & 35,010 \\
-u       & 0.19 & 17.2 & 11,078 & 0.18 & 16.8 & 34,726 \\
-uQ      & 0.13 & 13.2 & 11,279 & 0.14 & 12.7 & 34,574 \\
-uQiqiu  & 0.19 & 14.9 & 11,360 & 0.10 & 12.3 & 34,825 \\
\hline
\hline
\end{tabular}
\label{tab:ablation}
\end{table}

Table~\ref{tab:cace-param-matched} summarizes the energy and force RMSEs for parameter-count-controlled CACE models on the bulk water benchmark. Two parameter-matched model families with approximately 11k and 35k trainable parameters were considered. In both cases, the force RMSE generally decreases as higher-order electrostatic terms are incorporated, while increasing the number of trainable parameters alone leads to only minor changes in the accuracy. These results indicate that the observed improvements arise primarily from the added electrostatic physics rather than from increased model size.

\begin{figure*}
    \centering
    \includegraphics[width=\linewidth]{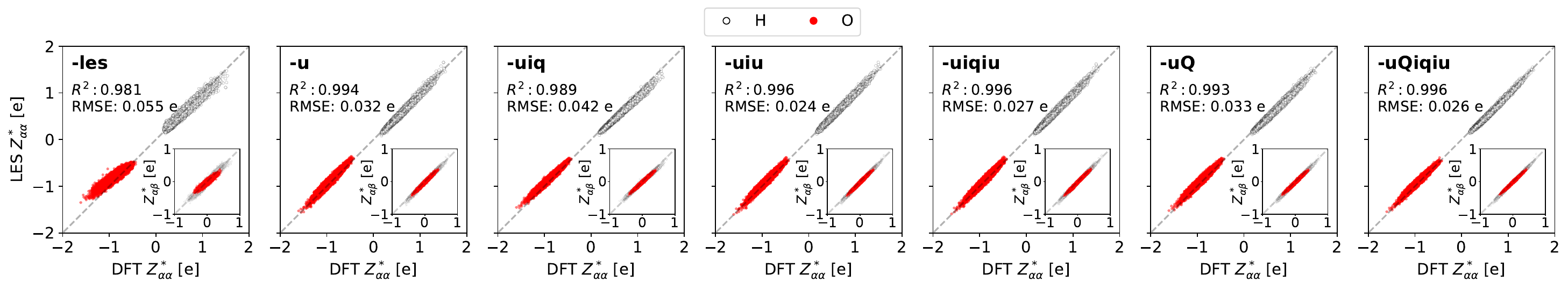}
    \caption{Parity plots comparing the Born effective charge (BEC) tensors predicted from long-range augmented MACE variants ($r = 4.5$~\AA{}, $n_l = 1$) with RPBE-D3 DFT reference data for 100 bulk water configurations~\cite{Schmiedmayer2024}. The main panels compare the diagonal elements ($Z^*_{\alpha\alpha}$), while the insets show the off-diagonal components ($Z^*_{\alpha\beta}$ with $\alpha \neq \beta$). The test RMSE values and $R^2$ coefficients computed over all tensor components are reported in each panel.
    Red and white markers denote oxygen and hydrogen atoms, respectively. 
    }
    \label{fig:water_bec}
\end{figure*}

Fig.~\ref{fig:water_bec} shows the comparison between the reference DFT BEC tensors and those predicted by the long-range augmented MACE variants ($r = 4.5$~\AA{}, $n_l = 1$) via Eqn.~\eqref{eq:z-pbc} for 100 liquid water test configurations~\cite{Schmiedmayer2024}. For variants incorporating polarizability predictions (MACELES-uiu/-uiqiu/-uQiqiu), the effective electronic dielectric constant $\varepsilon_e$
was calculated using Eqn.~\eqref{eq:chi_les} and Eqn.~\eqref{eq:eps_e}, while $\varepsilon_\infty=1.78$ was employed for the remaining models without induced polarization terms. The same procedure was applied to determine $\varepsilon_e$ for the polarizability predictions shown in Figs.\ref{fig:water_panel}c and d. While all augmented models demonstrate high predictive accuracy for both diagonal and off-diagonal components, the inclusion of higher-order terms, specifically the dipole (-u), significantly improves the results. The MACELES-uiu, -uiqiu, and -uQiqiu variants achieve the highest $R^2$ value of 0.996. Furthermore, the addition of the dipole term effectively mitigates the skewness observed at higher magnitudes in the monopole-only models (MACELES). The convergence test of the predicted BEC tensors with respect to both the simulation cell size and the LES parameters, together with a comparison of the IR spectra predicted by the MACELES and MACELES-uiu models, is provided in the SI. These results confirm that the long-range augmented framework provides a reliable and robust route for BEC predictions under periodic boundary conditions.
The acoustic sum rule (ASR) was also evaluated using the MACELES-uiu model on the same BEC test set. The mean per-atom ASR residuals are $1.3\times10^{-3}$~e and $1.9\times10^{-4}$~e for the diagonal and off-diagonal tensor components, respectively, which are small compared to the corresponding DFT BEC ranges of approximately $-1.6$ to $1.6$~e and $-0.8$ to $0.8$~e. The distributions of the ASR residuals are provided in the SI.

\subsection{Details on water simulations}

We trained a MACELES-uiu model ($r=4.5$~\AA{}, $n_l=1$) with anisotropic polarizability tensors $\boldsymbol{\alpha}$ using the same RPBE-D3 bulk water dataset~\cite{Schmiedmayer2024} as in the benchmark. Despite the relatively small receptive field, the model achieves low test RMSEs of 0.2~meV/atom and 10.6~meV/\AA{} for energies and forces, respectively. This level of accuracy is consistent with the benchmark trends discussed above.

We performed equilibrium NVT MD simulations of bulk water employing the MACELES-uiu model within the ASE package~\cite{hjorth_larsen_atomic_2017} at 280, 300, 320, 340, and 360~K, at the corresponding experimental densities ($\approx$ 1000, 997, 989, 980, and 967~kg/m$^3$) using a system of 512 water molecules and a Nos\'e–Hoover thermostat (Fig.\ref{fig:water_panel}). Each simulation used a timestep of 0.25~fs, consisting of 20~ps of equilibration followed by a 100~ps production run used for analysis.

The IR spectrum is obtained from the Fourier transform of the current–current autocorrelation function,
\begin{equation}
I_{\mathrm{IR}}(\omega) \propto \int_0^T dt \left\langle \mathbf{J}(0)\cdot \mathbf{J}(t) \right\rangle e^{-i\omega t},
\label{eq:ir}
\end{equation}
with the time-dependent polarization current of the system given by
$\mathbf{J}(t) = \sum_{i=1}^N \boldsymbol{Z}_i^*(t)\cdot \mathbf{v}_i(t)$,
where $\boldsymbol{Z}_i^*(t)$ is the BEC tensor of atom $i$ and $\mathbf{v}_i(t)$ is its velocity.

\begin{figure}
    \centering
    \includegraphics[width=0.9\linewidth]{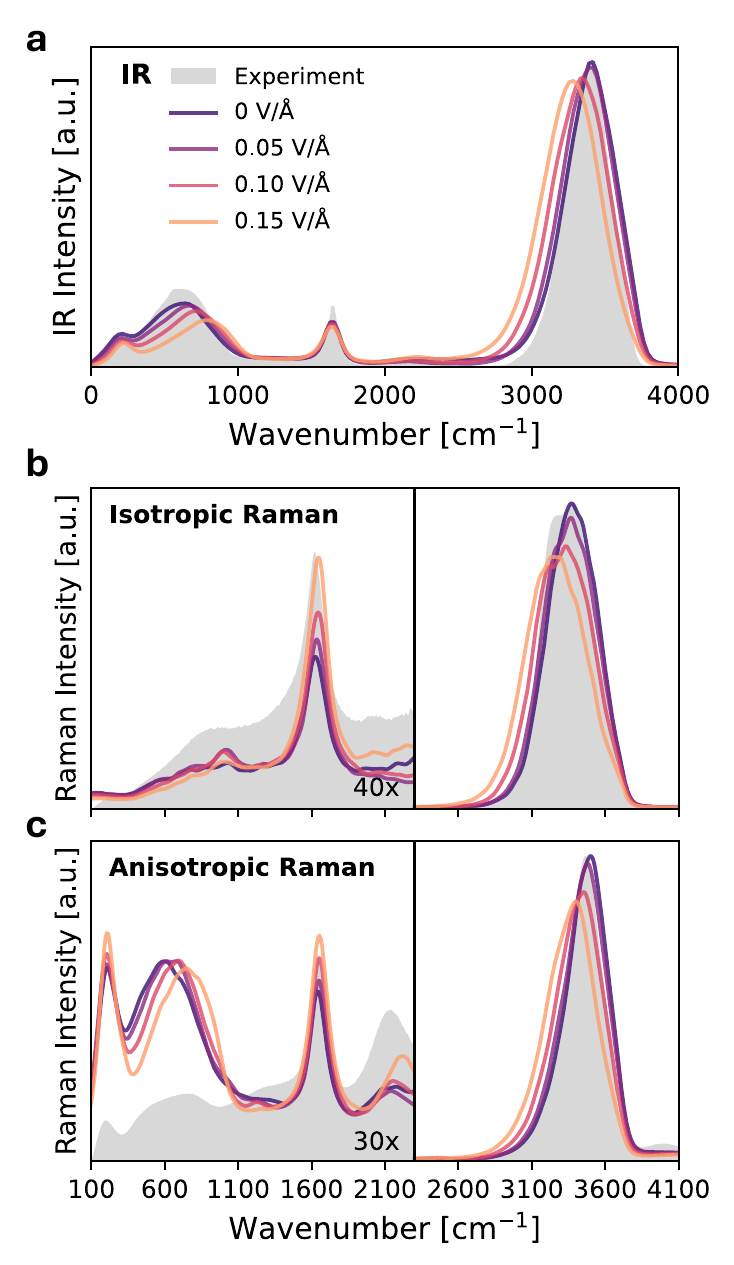}
    \caption{Bulk water infrared (IR) and Raman spectra under external electric fields at 300~K.
    \figLabelCapt{a--c}: Simulated (a) IR, (b) isotropic reduced Raman, and (c) anisotropic reduced Raman spectra in the absence of an external field (black line) and under applied electric fields (colored lines, see legend). The experimental IR spectrum~\cite{Bertie1996Infrared} at 298~K and reduced Raman spectrum~\cite{Pattenaude2018Temperaturea} at 300~K are included for comparison. For clarity, the low-frequency region in the Raman spectrum is scaled by the factor indicated in each panel, while the experimental results are scaled by a factor of 180 and 30 for the isotropic and anisotropic cases, respectively.
    }
    \label{fig:water_e_field}
\end{figure}

The reduced isotropic and anisotropic Raman spectra are computed from the autocorrelation functions of the time-dependent polarizability tensor $\boldsymbol{\alpha}(t)$ with isotropic component $\alpha(t)$ and anisotropic components $\boldsymbol{\beta}(t) = \boldsymbol{\alpha}(t) - \alpha(t)\boldsymbol{I}$.
The reduced isotropic Raman spectrum is given by~\cite{Morawietz2018Interplay}
\begin{equation}
R_{\mathrm{iso}}(\omega) \propto \omega^{2}
\int_0^T dt  \left\langle \alpha(0)\alpha(t) \right\rangle e^{-i\omega t},
\label{eq:raman_iso}
\end{equation}
where $\omega$ is the Raman Stokes-shift frequency.
Similarly, the reduced anisotropic Raman spectrum is given by
\begin{equation}
R_{\mathrm{aniso}}(\omega) \propto \omega^{2}
\int_0^T dt \left\langle \mathrm{Tr}\left[\boldsymbol{\beta}(0)\boldsymbol{\beta}(t)\right] \right\rangle e^{-i\omega t}.
\label{eq:raman_aniso}
\end{equation}
The simulated IR and Raman spectra shown in Figs.~\ref{fig:water_panel} and~\ref{fig:water_e_field} were obtained from the corresponding MD trajectories. The raw spectra were smoothed by convolution with a Gaussian kernel. Experimental reduced Raman spectra shown in Figs.~\ref{fig:water_panel}g and h were reproduced from the results in Ref.~\cite{Pattenaude2018Temperaturea} using the reduced representation defined in Ref.~\cite{Morawietz2018Interplay}. The green dashed line, taken from Ref.~\cite{marsalek2017quantum}, is based on the same underlying measurement~\cite{Pattenaude2018Temperaturea} but employs a different definition of the reduced Raman intensity. The green dotted line is digitized from Ref.~\cite{Brooker1989Ramana}, which also uses an alternative form of the reduced Raman intensity.

Fig.~\ref{fig:water_e_field} shows the IR and reduced Raman spectra of bulk water under static external electric fields ($\mathbf{E}^0 =$ 0.05, 0.10, and 0.15~V/\AA{}) applied along the $z$ direction. The external field was incorporated by adding electrostatic forces derived from the predicted BEC tensors (Eqn.~\eqref{eq:z-pbc}) using $Z^*_{i \alpha\beta}  = \dfrac{\partial \mathcal{F}_{i \beta}}{\partial \mathrm{E}^0_\alpha}$, enabling constant electric field MD simulations under periodic boundary conditions~\cite{zhong2025machine}. To avoid center-of-mass drift due to small residual errors in the sum of BECs over all the atoms in the system, the mean electrostatic force was removed at each step. MD simulations were performed at 300~K and the corresponding experimental density ($\approx$997~kg/m$^3$) using 64 water molecules and a Nos\'e–Hoover thermostat. Each simulation used a timestep of 0.25~fs, consisting of 20~ps of equilibration followed by a 500~ps production run used for spectral analysis. The spectral analysis details follow those described for Fig.~\ref{fig:water_panel}.
Increasing the field strength leads to peak shifts, including a blue shift of the intermolecular librational band (400--750~$\mathrm{cm}^{-1}$) and a red shift of the intramolecular O--H stretching band (3200--3600~$\mathrm{cm}^{-1}$) in both IR and Raman spectra. These trends in both IR and Raman spectra under external electric fields are consistent with a previous study using AIMD by Cassone et al.~\cite{cassone2019ab}, and the IR trends are aligned with another AIMD study~\cite{futera2017Communication} and ML models directly trained on explicit electrostatic properties such as dipoles~\cite{Stocco2025Electricfield} or BEC tensors~\cite{Joll2024}.

\subsection{Details on water interface simulations}

We trained a MACELES-uiu model with anisotropic polarizability tensors
$\boldsymbol{\alpha}$ ($r=5.5$~\AA{}, $n_l=1$) on a dataset of 500 water-air interface configurations (90\% train/10\% test split), each containing 522 water molecules, computed at the revPBE-D3 level of theory~\cite{niblett2021, Cheng2025Latent}. The model achieves test RMSEs of 0.1~meV/atom and 9.8~meV/\AA{} for energies and forces, respectively.

Using the trained model, we performed NVT MD simulations starting from an initial water slab containing 522 water molecules at 300~K using a Nos\'e--Hoover thermostat, with a timestep of 0.25~fs. Each simulation consisted of 20~ps of equilibration followed by a 100~ps production run.
To improve statistical convergence, five independent trajectories were generated from different initial configurations, and the reported spectrum corresponds to the average over these results, with the spread taken as the statistical uncertainty.

The VSFG spectrum was computed using the dielectric-derivative formalism of
Kapil et al.~\cite{Kapil2024Firstprinciples}. For each in-plane axis $a \in \{x, y\}$, we compute the Fourier transform of the cross-correlation between the corresponding diagonal polarizability component $\alpha_{aa}(t)$ and surface-selective time-dependent polarization current along the surface normal $J^M_z(t)$,
\begin{equation}
C_{aa,z}(\omega) \propto
\frac{1}{\omega}\,\mathrm{Re}\!\left[
\int_0^T dt\; e^{-i\omega t}\,
\left\langle \alpha_{aa}(t)\,J^M_z(0) \right\rangle
\right],
\label{eq:vsfg_cross}
\end{equation}
where $J^M_z(t)$ is computed following $\mathbf{J}^M(t) = \sum_{i=1}^N \boldsymbol{Z}_i^*(t)\cdot \mathbf{v}_i(t)\;\sigma(\mathbf{r}_i)$,
where $\boldsymbol{Z}_i^*(t)$ and $\mathbf{v}_i(t)$ denote the BEC tensor and velocity of atom $i$, respectively, and $\sigma(\mathbf{r}_i)$ is the signum function that assigns opposite signs to atoms above and below the slab center. This weighting removes the centrosymmetric bulk contribution while preserving the interfacial response.

The imaginary part of the second-order susceptibility is then obtained by averaging the $xx$ and $yy$ terms,
\begin{equation}
\mathrm{Im}\,\chi^{(2)}(\omega) =
\mathcal{S}\,\frac{1}{2}\left[ C_{xx,z}(\omega) + C_{yy,z}(\omega) \right],
\label{eq:vsfg_chi}
\end{equation}
where averaging over $xx$ and $yy$ components exploits the in-plane isotropy
of the slab to improve statistical convergence, and $\mathcal{S} = \pm 1$ is an overall sign determined by the choice of the surface-normal direction.

\subsection{Details on MAPbI$_3$ simulations}

\begin{figure}
    \centering
    \includegraphics[width=0.9\linewidth]{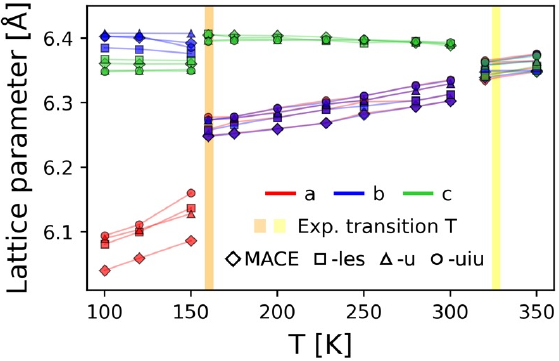}
    \caption{Lattice parameters $a$, $b$, and $c$ of MAPbI$_3$ predicted by MACE models with different levels of long-range augmentation. 
    Solid lines are guides to the eye. The experimentally reported orthorhombic-tetragonal and tetragonal-cubic phase transition temperatures are taken from Ref.~\cite{whitfield2016structures-b43}. 
    }
    \label{fig:mapi_lc}
\end{figure}

\begin{figure}
    \centering
    \includegraphics[width=0.85\linewidth]{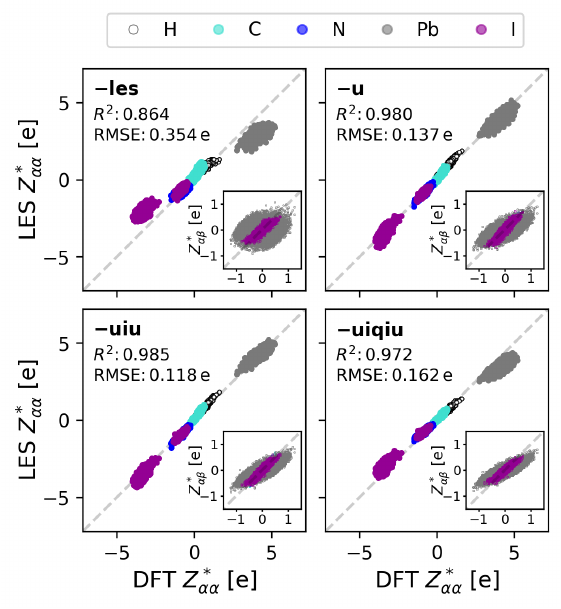}
    \caption{Parity plots of Born effective charge (BEC) tensor $Z^*_{}$ computed from long-range augmented MACE models compared to DFT calculated values. The main panel shows the diagonal components ($Z^*_{\alpha \alpha}$), with the off-diagonal components ($Z^*_{\alpha \beta}$ with $\alpha \ne \beta$) shown in the inset. 
    }
    \label{fig:mapi_bec}
\end{figure}

We employed MACE model ($r = 5.5$~\AA{}, $n_l = 1$) without and with the different levels of long-range augmentation introduced in the benchmark and employed them to perform MD simulations of MAPbI$_3$ perovskite using the ASE package~\cite{hjorth_larsen_atomic_2017}. 
We used $3 \times 3 \times 4$ unit cells (864 atoms) for the orthorhombic/tetragonal phase and $4 \times 4 \times 4$ unit cells (768 atoms) for the cubic phase, and lattice parameters were evaluated from MD simulations in the NPT ensemble over a temperature range of 100 to 350~K and at 1~atm. At 280~K and 300~K, we considered $6 \times 6 \times 6$ unit cells in the lattice constant estimation. Each NPT simulation was carried out for 0.3~ns with a timestep of 1.0~fs. The lattice parameters along each direction were obtained by dividing the simulation cell dimensions by the number of unit cells along the corresponding directions. 
Instabilities were observed in the MD simulations using the MACELES-uiq model, so it was discarded from further analysis.
The resulting lattice parameters for the baseline MACE model and other model variants (-les, -u, -uiu) are summarized in Fig.~\ref{fig:mapi_lc}. 

Fig.~\ref{fig:mapi_bec} shows BEC tensors predicted from the long-range augmented MACE variants compared to the DFT reference values of 300 MAPbI$_3$ test configurations described in the Datasets section above.
We replicate the configurations $2 \times 2 \times 2$ (from 96 atoms to 768 atoms) 
to eliminate the finite-size effects due to finite $k$ when using Eqn.~\ref{eq:z-pbc}.
We note that the finite-size effects are noticeable for the plain LES model with the original system size,
while not significant for the extended variants. The convergence test of the predicted BEC tensors with respect to the simulation cell size is provided in the SI.
For models with induced dipoles (-uiu, -uiqiu), the effective electronic dielectric constant $\varepsilon_e$ for each configuration was computed using Eqn.~\ref{eq:chi_les} and Eqn.~\ref{eq:eps_e} with $\varepsilon_\infty = 4.7$, whereas for other models (-les, -u) without induced terms $\varepsilon_e$ is set to the constant $\varepsilon_\infty$.  
The MACELES-uiu model shows the best accuracy in the BEC predictions,
so it was adopted for the subsequent vibrational spectra computations.
We also show a comparison of the IR spectra predicted by the MACELES and MACELES-uiu models in the SI.
The ASR residual was further assessed for the MACELES-uiu model on the same test configurations. The mean per-atom ASR residuals are $4.6\times10^{-2}$~e and $1.6\times10^{-3}$~e for the diagonal and off-diagonal tensor components, respectively. These values remain small compared to the corresponding DFT BEC ranges of approximately $-4.0$ to $5.2$~e for the diagonal components and $-1.2$ to $1.2$~e for the off-diagonal components. The distributions of the ASR residuals are provided in the SI.

To compute IR and Raman spectra for MAPbI$_3$, we employed the MACELES-uiu model with isotropic latent polarizability. We used $3 \times 3 \times 4$ unit cells for orthorhombic/tetragonal phase and $4 \times 4 \times 4$ unit cells for cubic phase, and MD simulations were carried out at 107~K and 228~K for IR spectra, and 10, 77, 150, 250, 300, 360, and 500~K for Raman spectra at the corresponding experimental densities. Each MD simulation consists of a 20~ps NVT equilibration stage using Langevin thermostat, followed by a 100~ps production run in NVE ensemble with a timestep of 0.5~fs. 
Using Eqn.~\ref{eq:ir}, the IR spectra were computed from the autocorrelation function of the polarization current. The autocorrelation function of the isotropic polarizability $\alpha$ is used to compute the (unreduced) isotropic Raman intensities as
\begin{equation}
I_{\mathrm{iso}}(\omega) \propto \frac{\omega}{1 - e^{-\beta\hbar\omega}}
\int_0^T dt  \left\langle \alpha(0)\alpha(t) \right\rangle e^{-i\omega t},
\label{eq:raman_mapi}
\end{equation}
which is the same as the expression used in Ref.~\cite{zhenbang2022} for direct comparison with the Raman intensities reported in Ref.~\cite{sharma2020elucidating}. 
All spectra obtained from the corresponding MD trajectories were smoothed using a Gaussian convolution kernel.

\subsection{Details on HfO$_2$ simulations}

\begin{table}
\centering
\caption{Energy and force RMSEs and the number of trainable parameters ($N_{\mathrm{params}}$) for MACE models ($r=5.0$~\AA{}, $n_l=2$) with different long-range augmented variants on HfO$_2$ dataset~\cite{chen2026electric-c45}.}
\label{tab:mace-hfo2}
\small
\renewcommand{\arraystretch}{1.2}

\begin{tabular}{l|c|c|c}
\hline
\hline
\ \textbf{Model} \ & \ \textbf{E} (meV/atom) \ & \ \textbf{F} (meV/\AA{}) \ & \ \ \textbf{$N_{\mathrm{params}}$} \ \ \\
\hline
\hline
\ MACE & 0.5 & 34.0 & 655,248 \\
\ -les & 0.5 & 32.1 & 623,136 \\
\ -u & 0.6 & 31.8 & 623,392 \\
\ -uiu & 0.6 & 31.7 & 625,584 \\
\ -uQiu & 0.5 & 25.8 & 1,180,336 \\
\hline
\hline
\end{tabular}
\end{table}

\begin{figure}
    \centering
    \includegraphics[width=0.9\linewidth]{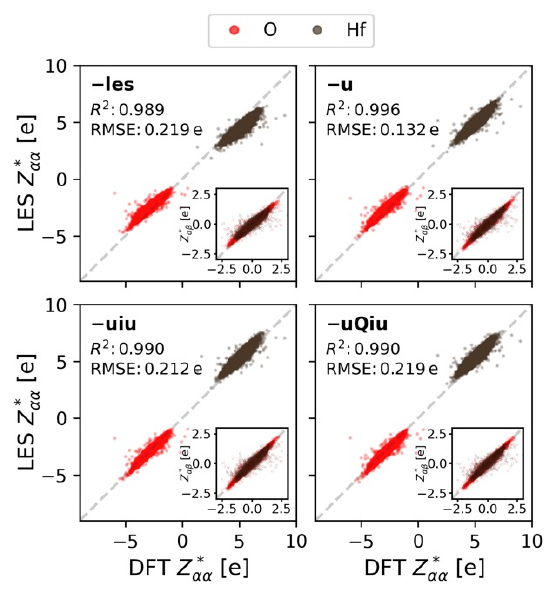}
    \caption{Parity plots of Born effective charge (BEC) tensor $Z^*_{}$ predicted from various long-range augmented MACE models ($r=5.0$~\AA{}, $n_l=2$) compared to DFT calculated values. The main panel compares the diagonal components ($Z^*_{\alpha \alpha}$), while the off-diagonal components ($Z^*_{\alpha \beta}$ with $\alpha \ne \beta$) are shown in the inset. 
    }
    \label{fig:hfo2_bec}
\end{figure}

We randomly split PBEsol DFT HfO$_2$ dataset~\cite{chen2026electric-c45} into 3131 training configurations (10\% validation) and 276 test configurations. 
Each configuration contains either 96 or 108 atoms, and the configurations cover six HfO$_2$ phases--monoclinic ($P2_1/c$), three orthorhombic ($Pca2_1$, $Pbca$, $Pnma$), tetragonal ($P4_2/nmc$), and cubic ($Fm\bar{3}m$)--which are generated from MD trajectories using an on-the-fly learning scheme at temperatures ranging from 1 K to 4000 K. 
We recompute the whole dataset using the same PBEsol functional~\cite{perdew2008pbesol} using the Vienna ab initio simulation package (VASP~\cite{kresse1993,kresse1994liquid,kresse1996efficiency,kresse1996efficient,kresse1999paw} version 6.4.1) with a planewave energy cutoff of 500 eV, a dense k-point mesh in the Brillouin zone with a maximum allowed distance between k-points set to 0.2~$\AA^{-1}$ in reciprocal space including the $\Gamma$ point and a tight electronic convergence threshold for the self-consistent field (SCF) loop ($10^{-8}$~eV). The projection operators are evaluated in reciprocal space (PREC=Accurate, ENCUT=500, GGA=PS, ISIF=2, KSPACING=0.2, KGAMMA=.TRUE., LREAL=.FALSE., EDIFF=1.0e-08, ADDGRID=.TRUE.). The PAW potentials are taken from the VASP library and are: ``PAW\_PBE Hf\_sv 10Jan2008 GW suitable'' for Hafnium and ``PAW\_PBE O 08Apr2002'' for Oxygen.
The BECs and the static dielectric tensors for the test set using DFPT at the PBEsol level are calculated with the same settings (adding LEPSILON=.TRUE.).
The polarizability of each configuration was computed from its dielectric tensor using the Clausius–Mossotti relation. 
We set $\varepsilon_{\infty} = 5.365$, corresponding to the average value from DFPT calculations. 

We trained MACE MLIPs ($r=5.0$~\AA{}, $n_l=2$) without and with different levels of long-range augmented variants. Inclusion of higher-order terms tends to improve the predictive accuracy. The corresponding energy and force RMSEs are listed in Table~\ref{tab:mace-hfo2}.  

Fig.~\ref{fig:hfo2_bec} compares the BEC tensors computed using Eqn.~\eqref{eq:z-pbc} with the corresponding PBEsol DFT values for the 276 test configurations introduced above. Prior to evaluating the BECs, each configuration is replicated by $2\times2\times2$, from 96 or 108 atoms to 768 or 864 atoms, respectively, to eliminate the finite size effects. For models containing induced dipoles, namely MACELES-uiu and MACELES-uQiu, the effective electronic dielectric constant $\varepsilon_e$ was evaluated from Eqn.~\eqref{eq:chi_les} and Eqn.~\eqref{eq:eps_e}. For MACELES and MACELES-u models, which do not include induced polarization terms, $\varepsilon_e$ is set to $\varepsilon_{\infty}$. Although all augmented models exhibit high predictive accuracy, higher-order terms provide further improvement in the results. We further provide a convergence test of the predicted BEC tensors with respect to the simulation cell size in the SI.
The mean per-atom acoustic sum rule (ASR) residuals for the MACELES-uiu model are $1.2\times10^{-1}$~e (diagonal) and $2.7\times10^{-3}$~e (off-diagonal), both of which are small relative to the corresponding DFT BEC ranges ($-6.0$ to $9.0$~e and $-2.5$ to $2.5$~e, respectively). The distributions of the ASR residuals are provided in the SI.

Phonon dispersions of the ferroelectric $Pca2_1$ and tetragonal $P4_2/nmc$ phases of HfO$_2$ were calculated independently using the MACELES-uiu model and PBEsol DFT. 
The DFT calculations for the phonons were performed using the same settings as for the energies, forces, and BEC calculations described above. A 4x4x2 and 2x2x2 supercells are used for the frozen-phonon calculations of the tetragonal and $Pca2_1$ phase, respectively.   
Harmonic force constants were obtained using the finite-displacement method as implemented in the Phonopy package~\cite{togo2015first-f2e}. We employed $2\times2\times2$ and $4\times4\times2$ unit cells for the ferroelectric and tetragonal phases, respectively. For calculations including LO–TO splittings, non-analytical correction (NAC) was employed using the corresponding BECs and high-frequency dielectric constant~\cite{gonze1997dynamical-202, zhong1994giant-6dc}. 

IR spectra of the monoclinic $P2_1/c$, orthorhombic $Pbca$, and ferroelectric $Pca2_1$ phases of HfO$_2$ are calculated from MD trajectories using the MACELES-uiu model with isotropic latent polarizability. All simulations were carried out at 300 K with the ASE package~\cite{hjorth_larsen_atomic_2017} at densities of 10.2, 10.7, and 10.6~g/cm$^3$ for the monoclinic, orthorhombic $Pbca$, and ferroelectric $Pca2_1$ phases, respectively, as obtained from structural relaxation at 1~atm with the model. Each system contains 768 atoms, corresponding to $4\times 4\times 4$, $4\times 4\times 2$, and $4\times 4\times 4$ unit cells for the three phases, respectively. The simulations consist of 20 ps equilibration in the NVT ensemble with a Langevin thermostat, and a subsequent 100 ps production run in the NVE ensemble using a timestep of 0.5 fs. The IR spectra are obtained from the autocorrelation function of the polarization current according to Eqn.~\ref{eq:ir}. 

To evaluate the ferroelectric response of HfO$_2$, we performed NPT MD simulations at $T=300$~K and 1 atm using the MACELES-uiu model and a Nos\'e–Hoover thermostat within the ASE package~\cite{hjorth_larsen_atomic_2017}. We used $4\times 4\times 4$ unit cells of ferroelectric ($Pca2_1$) HfO$_2$, containing 768 atoms, with a timestep of 1 fs. Initial structures with oppositely oriented polarization were prepared by equilibrating the system for 20 ps under electric fields of $E^0=0.12$ and $-0.12\, \mathrm{V/\AA^{}}$, respectively. 
The polarization is calculated as 
\begin{equation}
P_{\alpha} = \frac{e}{V}
\sum_{i\beta} Z^*_{i\alpha\beta} d_{i\beta} 
\label{eq:p_hfo2}
\end{equation}
where $\alpha$ and $\beta$ denote Cartesian directions, $Z^*_{\alpha \beta}$ is the BEC tensor for atom $i$, $V$ is the cell volume, and $d_{i\beta}$ is atomic displacement relative to the non-polar centrosymmetric reference structure.
Eqn.~\ref{eq:p_hfo2} is a commonly used BEC-based expression for estimating the polarization relative to a non-polar reference structure~\cite{rabe2007physics}. To assess the effect of variation of the BECs along a path, we additionally compare this estimate with the polarization obtained by integrating the BECs along a linear interpolation path, as detailed in the SI.

The two equilibrated initial structures were then used as the starting points for the simulations to construct the hysteresis loop. At each value of $E^0$, the system was simulated for an additional 20 ps, and the polarization was obtained by averaging $P_{\alpha}$ over the final 10 ps of each trajectory.

\textbf{Data availability}~
The datasets, training scripts, evaluation scripts, and the trained models are available at \url{https://github.com/ChengUCB/extended_les_fit}.
Source Data are provided with this paper.

\textbf{Code availability}~
The LES library is publicly available at 
\url{https://github.com/ChengUCB/les}.
The CACE package with the LES implementation is available at \url{https://github.com/BingqingCheng/cace}
The MACE package with the LES implementation is available at \url{https://github.com/ChengUCB/mace} and is also included in the official MACE repository at \url{https://github.com/ACEsuit/mace}.
The NequIP and Allegro LES extension package is available at \url{https://github.com/ChengUCB/NequIP-LES}.

\textbf{Acknowledgements}~
The authors thank Giuseppe Fisicaro for help with perovskite datasets, and David Limmer for useful discussions and feedback on the manuscript.
Funding acknowledgement: Research reported in this publication was supported by the National Institute of General Medical Sciences of the National Institutes of Health under Award Number R35GM159986.
The content is solely the responsibility of the authors and does not necessarily represent the official views of the National Institutes of Health.
Part of this work is performed under the auspices of the U.S. Department of Energy by Lawrence Livermore National Laboratory under Contract DE-AC52-07NA27344.

\textbf{Authors contributions statement}
Y.P. performed simulation and analysis for the perovskite and the hafnium dioxide system, and wrote the manuscript.
D.K. implemented code, performed simulation and analysis for the water system, and wrote the manuscript.
D.S.K. performed preliminary exploration of the idea, developed algorithms to readout tensors from equivariant features, implemented code, and wrote the manuscript.
N.D. performed simulations for the interfacial water system.
R.S. and X.W. performed simulations for the perovskite system.
S.H. performed DFT calculations.
B.C designed the research, developed the method, implemented code, and wrote the manuscript.

\textbf{Competing Interests:}
B.C. has an equity stake in AIMATX Inc.
The University of California, Berkeley has filed a non-provisional patent for the Latent Ewald Summation algorithm.


\begin{thebibliography}{125}%
\makeatletter
\providecommand \@ifxundefined [1]{%
 \@ifx{#1\undefined}
}%
\providecommand \@ifnum [1]{%
 \ifnum #1\expandafter \@firstoftwo
 \else \expandafter \@secondoftwo
 \fi
}%
\providecommand \@ifx [1]{%
 \ifx #1\expandafter \@firstoftwo
 \else \expandafter \@secondoftwo
 \fi
}%
\providecommand \natexlab [1]{#1}%
\providecommand \enquote  [1]{``#1''}%
\providecommand \bibnamefont  [1]{#1}%
\providecommand \bibfnamefont [1]{#1}%
\providecommand \citenamefont [1]{#1}%
\providecommand \href@noop [0]{\@secondoftwo}%
\providecommand \href [0]{\begingroup \@sanitize@url \@href}%
\providecommand \@href[1]{\@@startlink{#1}\@@href}%
\providecommand \@@href[1]{\endgroup#1\@@endlink}%
\providecommand \@sanitize@url [0]{\catcode `\\12\catcode `\$12\catcode `\&12\catcode `\#12\catcode `\^12\catcode `\_12\catcode `\%12\relax}%
\providecommand \@@startlink[1]{}%
\providecommand \@@endlink[0]{}%
\providecommand \url  [0]{\begingroup\@sanitize@url \@url }%
\providecommand \@url [1]{\endgroup\@href {#1}{\urlprefix }}%
\providecommand \urlprefix  [0]{URL }%
\providecommand \Eprint [0]{\href }%
\providecommand \doibase [0]{http://dx.doi.org/}%
\providecommand \selectlanguage [0]{\@gobble}%
\providecommand \bibinfo  [0]{\@secondoftwo}%
\providecommand \bibfield  [0]{\@secondoftwo}%
\providecommand \translation [1]{[#1]}%
\providecommand \BibitemOpen [0]{}%
\providecommand \bibitemStop [0]{}%
\providecommand \bibitemNoStop [0]{.\EOS\space}%
\providecommand \EOS [0]{\spacefactor3000\relax}%
\providecommand \BibitemShut  [1]{\csname bibitem#1\endcsname}%
\let\auto@bib@innerbib\@empty
\bibitem [{\citenamefont {Keith}\ \emph {et~al.}(2021)\citenamefont {Keith}, \citenamefont {Vassilev-Galindo}, \citenamefont {Cheng}, \citenamefont {Chmiela}, \citenamefont {Gastegger}, \citenamefont {M{\"u}ller},\ and\ \citenamefont {Tkatchenko}}]{keith2021combining}%
  \BibitemOpen
  \bibfield  {author} {\bibinfo {author} {\bibfnamefont {John~A}\ \bibnamefont {Keith}}, \bibinfo {author} {\bibfnamefont {Valentin}\ \bibnamefont {Vassilev-Galindo}}, \bibinfo {author} {\bibfnamefont {Bingqing}\ \bibnamefont {Cheng}}, \bibinfo {author} {\bibfnamefont {Stefan}\ \bibnamefont {Chmiela}}, \bibinfo {author} {\bibfnamefont {Michael}\ \bibnamefont {Gastegger}}, \bibinfo {author} {\bibfnamefont {Klaus-Robert}\ \bibnamefont {M{\"u}ller}}, \ and\ \bibinfo {author} {\bibfnamefont {Alexandre}\ \bibnamefont {Tkatchenko}},\ }\bibfield  {title} {\enquote {\bibinfo {title} {Combining machine learning and computational chemistry for predictive insights into chemical systems},}\ }\href@noop {} {\bibfield  {journal} {\bibinfo  {journal} {Chemical reviews}\ }\textbf {\bibinfo {volume} {121}},\ \bibinfo {pages} {9816--9872} (\bibinfo {year} {2021})}\BibitemShut {NoStop}%
\bibitem [{\citenamefont {Unke}\ \emph {et~al.}(2021{\natexlab{a}})\citenamefont {Unke}, \citenamefont {Chmiela}, \citenamefont {Sauceda}, \citenamefont {Gastegger}, \citenamefont {Poltavsky}, \citenamefont {Sch{\"u}tt}, \citenamefont {Tkatchenko},\ and\ \citenamefont {M{\"u}ller}}]{unke2021machine}%
  \BibitemOpen
  \bibfield  {author} {\bibinfo {author} {\bibfnamefont {Oliver~T}\ \bibnamefont {Unke}}, \bibinfo {author} {\bibfnamefont {Stefan}\ \bibnamefont {Chmiela}}, \bibinfo {author} {\bibfnamefont {Huziel~E}\ \bibnamefont {Sauceda}}, \bibinfo {author} {\bibfnamefont {Michael}\ \bibnamefont {Gastegger}}, \bibinfo {author} {\bibfnamefont {Igor}\ \bibnamefont {Poltavsky}}, \bibinfo {author} {\bibfnamefont {Kristof~T}\ \bibnamefont {Sch{\"u}tt}}, \bibinfo {author} {\bibfnamefont {Alexandre}\ \bibnamefont {Tkatchenko}}, \ and\ \bibinfo {author} {\bibfnamefont {Klaus-Robert}\ \bibnamefont {M{\"u}ller}},\ }\bibfield  {title} {\enquote {\bibinfo {title} {Machine learning force fields},}\ }\href@noop {} {\bibfield  {journal} {\bibinfo  {journal} {Chemical Reviews}\ }\textbf {\bibinfo {volume} {121}},\ \bibinfo {pages} {10142--10186} (\bibinfo {year} {2021}{\natexlab{a}})}\BibitemShut {NoStop}%
\bibitem [{\citenamefont {Batzner}\ \emph {et~al.}(2022)\citenamefont {Batzner}, \citenamefont {Musaelian}, \citenamefont {Sun}, \citenamefont {Geiger}, \citenamefont {Mailoa}, \citenamefont {Kornbluth}, \citenamefont {Molinari}, \citenamefont {Smidt},\ and\ \citenamefont {Kozinsky}}]{batzner20223}%
  \BibitemOpen
  \bibfield  {author} {\bibinfo {author} {\bibfnamefont {Simon}\ \bibnamefont {Batzner}}, \bibinfo {author} {\bibfnamefont {Albert}\ \bibnamefont {Musaelian}}, \bibinfo {author} {\bibfnamefont {Lixin}\ \bibnamefont {Sun}}, \bibinfo {author} {\bibfnamefont {Mario}\ \bibnamefont {Geiger}}, \bibinfo {author} {\bibfnamefont {Jonathan~P}\ \bibnamefont {Mailoa}}, \bibinfo {author} {\bibfnamefont {Mordechai}\ \bibnamefont {Kornbluth}}, \bibinfo {author} {\bibfnamefont {Nicola}\ \bibnamefont {Molinari}}, \bibinfo {author} {\bibfnamefont {Tess~E}\ \bibnamefont {Smidt}}, \ and\ \bibinfo {author} {\bibfnamefont {Boris}\ \bibnamefont {Kozinsky}},\ }\bibfield  {title} {\enquote {\bibinfo {title} {E (3)-equivariant graph neural networks for data-efficient and accurate interatomic potentials},}\ }\href@noop {} {\bibfield  {journal} {\bibinfo  {journal} {Nature communications}\ }\textbf {\bibinfo {volume} {13}},\ \bibinfo {pages} {2453} (\bibinfo {year} {2022})}\BibitemShut {NoStop}%
\bibitem [{\citenamefont {Batatia}\ \emph {et~al.}(2022)\citenamefont {Batatia}, \citenamefont {Kovacs}, \citenamefont {Simm}, \citenamefont {Ortner},\ and\ \citenamefont {Cs{\'a}nyi}}]{batatia2022mace}%
  \BibitemOpen
  \bibfield  {author} {\bibinfo {author} {\bibfnamefont {Ilyes}\ \bibnamefont {Batatia}}, \bibinfo {author} {\bibfnamefont {David~P}\ \bibnamefont {Kovacs}}, \bibinfo {author} {\bibfnamefont {Gregor}\ \bibnamefont {Simm}}, \bibinfo {author} {\bibfnamefont {Christoph}\ \bibnamefont {Ortner}}, \ and\ \bibinfo {author} {\bibfnamefont {G{\'a}bor}\ \bibnamefont {Cs{\'a}nyi}},\ }\bibfield  {title} {\enquote {\bibinfo {title} {Mace: Higher order equivariant message passing neural networks for fast and accurate force fields},}\ }\href@noop {} {\bibfield  {journal} {\bibinfo  {journal} {Advances in Neural Information Processing Systems}\ }\textbf {\bibinfo {volume} {35}},\ \bibinfo {pages} {11423--11436} (\bibinfo {year} {2022})}\BibitemShut {NoStop}%
\bibitem [{\citenamefont {Baldwin}\ \emph {et~al.}(2026)\citenamefont {Baldwin}, \citenamefont {Batatia}, \citenamefont {Vondrák}, \citenamefont {Margraf},\ and\ \citenamefont {Csányi}}]{Baldwin2026}%
  \BibitemOpen
  \bibfield  {author} {\bibinfo {author} {\bibfnamefont {William~J.}\ \bibnamefont {Baldwin}}, \bibinfo {author} {\bibfnamefont {Ilyes}\ \bibnamefont {Batatia}}, \bibinfo {author} {\bibfnamefont {Martin}\ \bibnamefont {Vondrák}}, \bibinfo {author} {\bibfnamefont {Johannes~T.}\ \bibnamefont {Margraf}}, \ and\ \bibinfo {author} {\bibfnamefont {Gábor}\ \bibnamefont {Csányi}},\ }\bibfield  {title} {\enquote {\bibinfo {title} {Design space of self--consistent electrostatic machine learning interatomic potentials},}\ }\href@noop {} {\bibfield  {journal} {\bibinfo  {journal} {arXiv preprint arXiv.2603.14700}\ } (\bibinfo {year} {2026})}\BibitemShut {NoStop}%
\bibitem [{\citenamefont {Grasselli}\ \emph {et~al.}(2026)\citenamefont {Grasselli}, \citenamefont {Rossi}, \citenamefont {de~Gironcoli},\ and\ \citenamefont {Grisafi}}]{grasselli2026long}%
  \BibitemOpen
  \bibfield  {author} {\bibinfo {author} {\bibfnamefont {Federico}\ \bibnamefont {Grasselli}}, \bibinfo {author} {\bibfnamefont {Kevin}\ \bibnamefont {Rossi}}, \bibinfo {author} {\bibfnamefont {Stefano}\ \bibnamefont {de~Gironcoli}}, \ and\ \bibinfo {author} {\bibfnamefont {Andrea}\ \bibnamefont {Grisafi}},\ }\bibfield  {title} {\enquote {\bibinfo {title} {Long-range electrostatics in atomistic machine learning: a physical perspective},}\ }\href@noop {} {\bibfield  {journal} {\bibinfo  {journal} {arXiv preprint arXiv:2602.11071}\ } (\bibinfo {year} {2026})}\BibitemShut {NoStop}%
\bibitem [{\citenamefont {Kim}\ and\ \citenamefont {Cheng}(2026)}]{Kim2026}%
  \BibitemOpen
  \bibfield  {author} {\bibinfo {author} {\bibfnamefont {Dongjin}\ \bibnamefont {Kim}}\ and\ \bibinfo {author} {\bibfnamefont {Bingqing}\ \bibnamefont {Cheng}},\ }\bibfield  {title} {\enquote {\bibinfo {title} {Long-range electrostatics for machine learning interatomic potentials is easier than we thought},}\ }\href {\doibase 10.1063/5.0316886} {\bibfield  {journal} {\bibinfo  {journal} {The Journal of Chemical Physics}\ }\textbf {\bibinfo {volume} {164}} (\bibinfo {year} {2026}),\ 10.1063/5.0316886}\BibitemShut {NoStop}%
\bibitem [{\citenamefont {Kosmala}\ \emph {et~al.}(2023)\citenamefont {Kosmala}, \citenamefont {Gasteiger}, \citenamefont {Gao},\ and\ \citenamefont {G{\"u}nnemann}}]{kosmala2023ewald}%
  \BibitemOpen
  \bibfield  {author} {\bibinfo {author} {\bibfnamefont {Arthur}\ \bibnamefont {Kosmala}}, \bibinfo {author} {\bibfnamefont {Johannes}\ \bibnamefont {Gasteiger}}, \bibinfo {author} {\bibfnamefont {Nicholas}\ \bibnamefont {Gao}}, \ and\ \bibinfo {author} {\bibfnamefont {Stephan}\ \bibnamefont {G{\"u}nnemann}},\ }\bibfield  {title} {\enquote {\bibinfo {title} {Ewald-based long-range message passing for molecular graphs},}\ }in\ \href@noop {} {\emph {\bibinfo {booktitle} {International Conference on Machine Learning}}}\ (\bibinfo {organization} {PMLR},\ \bibinfo {year} {2023})\ pp.\ \bibinfo {pages} {17544--17563}\BibitemShut {NoStop}%
\bibitem [{\citenamefont {Caruso}\ \emph {et~al.}(2026)\citenamefont {Caruso}, \citenamefont {Venturin}, \citenamefont {Giambagli}, \citenamefont {Rolando}, \citenamefont {{El-Machachi}}, \citenamefont {No{\'e}},\ and\ \citenamefont {Clementi}}]{Caruso2026Extending}%
  \BibitemOpen
  \bibfield  {author} {\bibinfo {author} {\bibfnamefont {Alessandro}\ \bibnamefont {Caruso}}, \bibinfo {author} {\bibfnamefont {Jacopo}\ \bibnamefont {Venturin}}, \bibinfo {author} {\bibfnamefont {Lorenzo}\ \bibnamefont {Giambagli}}, \bibinfo {author} {\bibfnamefont {Edoardo}\ \bibnamefont {Rolando}}, \bibinfo {author} {\bibfnamefont {Zakariya}\ \bibnamefont {{El-Machachi}}}, \bibinfo {author} {\bibfnamefont {Frank}\ \bibnamefont {No{\'e}}}, \ and\ \bibinfo {author} {\bibfnamefont {Cecilia}\ \bibnamefont {Clementi}},\ }\bibfield  {title} {\enquote {\bibinfo {title} {Extending the range of graph neural networks with global encodings},}\ }\href {\doibase 10.1038/s41467-026-69715-3} {\bibfield  {journal} {\bibinfo  {journal} {Nature Communications}\ }\textbf {\bibinfo {volume} {17}},\ \bibinfo {pages} {1855} (\bibinfo {year} {2026})}\BibitemShut {NoStop}%
\bibitem [{\citenamefont {Grisafi}\ and\ \citenamefont {Ceriotti}(2019)}]{grisafi2019incorporating}%
  \BibitemOpen
  \bibfield  {author} {\bibinfo {author} {\bibfnamefont {Andrea}\ \bibnamefont {Grisafi}}\ and\ \bibinfo {author} {\bibfnamefont {Michele}\ \bibnamefont {Ceriotti}},\ }\bibfield  {title} {\enquote {\bibinfo {title} {Incorporating long-range physics in atomic-scale machine learning},}\ }\href@noop {} {\bibfield  {journal} {\bibinfo  {journal} {The Journal of chemical physics}\ }\textbf {\bibinfo {volume} {151}} (\bibinfo {year} {2019})}\BibitemShut {NoStop}%
\bibitem [{\citenamefont {Huguenin-Dumittan}\ \emph {et~al.}(2023)\citenamefont {Huguenin-Dumittan}, \citenamefont {Loche}, \citenamefont {Haoran},\ and\ \citenamefont {Ceriotti}}]{huguenin2023physics}%
  \BibitemOpen
  \bibfield  {author} {\bibinfo {author} {\bibfnamefont {Kevin~K}\ \bibnamefont {Huguenin-Dumittan}}, \bibinfo {author} {\bibfnamefont {Philip}\ \bibnamefont {Loche}}, \bibinfo {author} {\bibfnamefont {Ni}~\bibnamefont {Haoran}}, \ and\ \bibinfo {author} {\bibfnamefont {Michele}\ \bibnamefont {Ceriotti}},\ }\bibfield  {title} {\enquote {\bibinfo {title} {Physics-inspired equivariant descriptors of nonbonded interactions},}\ }\href@noop {} {\bibfield  {journal} {\bibinfo  {journal} {The Journal of Physical Chemistry Letters}\ }\textbf {\bibinfo {volume} {14}},\ \bibinfo {pages} {9612--9618} (\bibinfo {year} {2023})}\BibitemShut {NoStop}%
\bibitem [{\citenamefont {Faller}\ \emph {et~al.}(2024)\citenamefont {Faller}, \citenamefont {Kaltak},\ and\ \citenamefont {Kresse}}]{Faller2024Densitybased}%
  \BibitemOpen
  \bibfield  {author} {\bibinfo {author} {\bibfnamefont {Carolin}\ \bibnamefont {Faller}}, \bibinfo {author} {\bibfnamefont {Merzuk}\ \bibnamefont {Kaltak}}, \ and\ \bibinfo {author} {\bibfnamefont {Georg}\ \bibnamefont {Kresse}},\ }\bibfield  {title} {\enquote {\bibinfo {title} {Density-based long-range electrostatic descriptors for machine learning force fields},}\ }\href {\doibase 10.1063/5.0245615} {\bibfield  {journal} {\bibinfo  {journal} {The Journal of Chemical Physics}\ }\textbf {\bibinfo {volume} {161}},\ \bibinfo {pages} {214701} (\bibinfo {year} {2024})}\BibitemShut {NoStop}%
\bibitem [{\citenamefont {Monacelli}\ and\ \citenamefont {Marzari}(2026)}]{Monacelli2026Electrostatic}%
  \BibitemOpen
  \bibfield  {author} {\bibinfo {author} {\bibfnamefont {Lorenzo}\ \bibnamefont {Monacelli}}\ and\ \bibinfo {author} {\bibfnamefont {Nicola}\ \bibnamefont {Marzari}},\ }\bibfield  {title} {\enquote {\bibinfo {title} {Electrostatic interactions in atomistic and machine-learned potentials for polar materials},}\ }\href {\doibase 10.1103/7ygl-8db2} {\bibfield  {journal} {\bibinfo  {journal} {Physical Review B}\ }\textbf {\bibinfo {volume} {113}},\ \bibinfo {pages} {094101} (\bibinfo {year} {2026})}\BibitemShut {NoStop}%
\bibitem [{\citenamefont {Guo}\ \emph {et~al.}(2026)\citenamefont {Guo}, \citenamefont {Yu}, \citenamefont {Hong}, \citenamefont {Chen}, \citenamefont {Gong},\ and\ \citenamefont {Xiang}}]{Guo2026Capturing}%
  \BibitemOpen
  \bibfield  {author} {\bibinfo {author} {\bibfnamefont {Ruijie}\ \bibnamefont {Guo}}, \bibinfo {author} {\bibfnamefont {Hongyu}\ \bibnamefont {Yu}}, \bibinfo {author} {\bibfnamefont {Liangliang}\ \bibnamefont {Hong}}, \bibinfo {author} {\bibfnamefont {Shiyou}\ \bibnamefont {Chen}}, \bibinfo {author} {\bibfnamefont {Xingao}\ \bibnamefont {Gong}}, \ and\ \bibinfo {author} {\bibfnamefont {Hongjun}\ \bibnamefont {Xiang}},\ }\bibfield  {title} {\enquote {\bibinfo {title} {Capturing long-range interactions with a reciprocal-space neural network},}\ }\href {\doibase 10.1103/7yn2-22h6} {\bibfield  {journal} {\bibinfo  {journal} {Physical Review B}\ }\textbf {\bibinfo {volume} {113}},\ \bibinfo {pages} {174101} (\bibinfo {year} {2026})}\BibitemShut {NoStop}%
\bibitem [{\citenamefont {Rumiantsev}\ \emph {et~al.}()\citenamefont {Rumiantsev}, \citenamefont {Langer}, \citenamefont {Sodjargal}, \citenamefont {Ceriotti},\ and\ \citenamefont {Loche}}]{Rumiantsev2025Learning}%
  \BibitemOpen
  \bibfield  {author} {\bibinfo {author} {\bibfnamefont {Egor}\ \bibnamefont {Rumiantsev}}, \bibinfo {author} {\bibfnamefont {Marcel~F.}\ \bibnamefont {Langer}}, \bibinfo {author} {\bibfnamefont {Tulga-Erdene}\ \bibnamefont {Sodjargal}}, \bibinfo {author} {\bibfnamefont {Michele}\ \bibnamefont {Ceriotti}}, \ and\ \bibinfo {author} {\bibfnamefont {Philip}\ \bibnamefont {Loche}},\ }\href {\doibase 10.48550/arXiv.2507.19382} {\enquote {\bibinfo {title} {Learning long-range representations with equivariant messages},}\ }\Eprint {http://arxiv.org/abs/2507.19382} {2507.19382} \BibitemShut {NoStop}%
\bibitem [{\citenamefont {Frank}\ \emph {et~al.}(2026)\citenamefont {Frank}, \citenamefont {Chmiela}, \citenamefont {M{\"u}ller},\ and\ \citenamefont {Unke}}]{frank2026machine}%
  \BibitemOpen
  \bibfield  {author} {\bibinfo {author} {\bibfnamefont {J~Thorben}\ \bibnamefont {Frank}}, \bibinfo {author} {\bibfnamefont {Stefan}\ \bibnamefont {Chmiela}}, \bibinfo {author} {\bibfnamefont {Klaus-Robert}\ \bibnamefont {M{\"u}ller}}, \ and\ \bibinfo {author} {\bibfnamefont {Oliver~T}\ \bibnamefont {Unke}},\ }\bibfield  {title} {\enquote {\bibinfo {title} {Machine learning global atomic representations with euclidean fast attention},}\ }\href@noop {} {\bibfield  {journal} {\bibinfo  {journal} {Nature Machine Intelligence}\ }\textbf {\bibinfo {volume} {8}},\ \bibinfo {pages} {388--402} (\bibinfo {year} {2026})}\BibitemShut {NoStop}%
\bibitem [{\citenamefont {Zhang}\ \emph {et~al.}(2022)\citenamefont {Zhang}, \citenamefont {Wang}, \citenamefont {Muniz}, \citenamefont {Panagiotopoulos}, \citenamefont {Car},\ and\ \citenamefont {E}}]{zhang2022deep}%
  \BibitemOpen
  \bibfield  {author} {\bibinfo {author} {\bibfnamefont {Linfeng}\ \bibnamefont {Zhang}}, \bibinfo {author} {\bibfnamefont {Han}\ \bibnamefont {Wang}}, \bibinfo {author} {\bibfnamefont {Maria~Carolina}\ \bibnamefont {Muniz}}, \bibinfo {author} {\bibfnamefont {Athanassios~Z.}\ \bibnamefont {Panagiotopoulos}}, \bibinfo {author} {\bibfnamefont {Roberto}\ \bibnamefont {Car}}, \ and\ \bibinfo {author} {\bibfnamefont {Weinan}\ \bibnamefont {E}},\ }\bibfield  {title} {\enquote {\bibinfo {title} {A deep potential model with long-range electrostatic interactions},}\ }\href@noop {} {\bibfield  {journal} {\bibinfo  {journal} {The Journal of Chemical Physics}\ }\textbf {\bibinfo {volume} {156}} (\bibinfo {year} {2022})}\BibitemShut {NoStop}%
\bibitem [{\citenamefont {Morawietz}\ \emph {et~al.}(2012)\citenamefont {Morawietz}, \citenamefont {Sharma},\ and\ \citenamefont {Behler}}]{Morawietz2012neural}%
  \BibitemOpen
  \bibfield  {author} {\bibinfo {author} {\bibfnamefont {Tobias}\ \bibnamefont {Morawietz}}, \bibinfo {author} {\bibfnamefont {Vikas}\ \bibnamefont {Sharma}}, \ and\ \bibinfo {author} {\bibfnamefont {J{\"o}rg}\ \bibnamefont {Behler}},\ }\bibfield  {title} {\enquote {\bibinfo {title} {A neural network potential-energy surface for the water dimer based on environment-dependent atomic energies and charges},}\ }\href {\doibase 10.1063/1.3682557} {\bibfield  {journal} {\bibinfo  {journal} {The Journal of Chemical Physics}\ }\textbf {\bibinfo {volume} {136}},\ \bibinfo {pages} {064103} (\bibinfo {year} {2012})}\BibitemShut {NoStop}%
\bibitem [{\citenamefont {Ko}\ \emph {et~al.}(2021)\citenamefont {Ko}, \citenamefont {Finkler}, \citenamefont {Goedecker},\ and\ \citenamefont {Behler}}]{ko2021fourth}%
  \BibitemOpen
  \bibfield  {author} {\bibinfo {author} {\bibfnamefont {Tsz~Wai}\ \bibnamefont {Ko}}, \bibinfo {author} {\bibfnamefont {Jonas~A}\ \bibnamefont {Finkler}}, \bibinfo {author} {\bibfnamefont {Stefan}\ \bibnamefont {Goedecker}}, \ and\ \bibinfo {author} {\bibfnamefont {J{\"o}rg}\ \bibnamefont {Behler}},\ }\bibfield  {title} {\enquote {\bibinfo {title} {A fourth-generation high-dimensional neural network potential with accurate electrostatics including non-local charge transfer},}\ }\href@noop {} {\bibfield  {journal} {\bibinfo  {journal} {Nature communications}\ }\textbf {\bibinfo {volume} {12}},\ \bibinfo {pages} {398} (\bibinfo {year} {2021})}\BibitemShut {NoStop}%
\bibitem [{\citenamefont {Unke}\ and\ \citenamefont {Meuwly}(2019)}]{unke2019physnet}%
  \BibitemOpen
  \bibfield  {author} {\bibinfo {author} {\bibfnamefont {Oliver~T}\ \bibnamefont {Unke}}\ and\ \bibinfo {author} {\bibfnamefont {Markus}\ \bibnamefont {Meuwly}},\ }\bibfield  {title} {\enquote {\bibinfo {title} {Physnet: A neural network for predicting energies, forces, dipole moments, and partial charges},}\ }\href@noop {} {\bibfield  {journal} {\bibinfo  {journal} {Journal of chemical theory and computation}\ }\textbf {\bibinfo {volume} {15}},\ \bibinfo {pages} {3678--3693} (\bibinfo {year} {2019})}\BibitemShut {NoStop}%
\bibitem [{\citenamefont {Unke}\ \emph {et~al.}(2021{\natexlab{b}})\citenamefont {Unke}, \citenamefont {Chmiela}, \citenamefont {Gastegger}, \citenamefont {Sch{\"u}tt}, \citenamefont {Sauceda},\ and\ \citenamefont {M{\"u}ller}}]{unke2021spookynet}%
  \BibitemOpen
  \bibfield  {author} {\bibinfo {author} {\bibfnamefont {Oliver~T}\ \bibnamefont {Unke}}, \bibinfo {author} {\bibfnamefont {Stefan}\ \bibnamefont {Chmiela}}, \bibinfo {author} {\bibfnamefont {Michael}\ \bibnamefont {Gastegger}}, \bibinfo {author} {\bibfnamefont {Kristof~T}\ \bibnamefont {Sch{\"u}tt}}, \bibinfo {author} {\bibfnamefont {Huziel~E}\ \bibnamefont {Sauceda}}, \ and\ \bibinfo {author} {\bibfnamefont {Klaus-Robert}\ \bibnamefont {M{\"u}ller}},\ }\bibfield  {title} {\enquote {\bibinfo {title} {Spookynet: Learning force fields with electronic degrees of freedom and nonlocal effects},}\ }\href@noop {} {\bibfield  {journal} {\bibinfo  {journal} {Nature communications}\ }\textbf {\bibinfo {volume} {12}},\ \bibinfo {pages} {7273} (\bibinfo {year} {2021}{\natexlab{b}})}\BibitemShut {NoStop}%
\bibitem [{\citenamefont {Cheng}(2025)}]{Cheng2025Latent}%
  \BibitemOpen
  \bibfield  {author} {\bibinfo {author} {\bibfnamefont {Bingqing}\ \bibnamefont {Cheng}},\ }\bibfield  {title} {\enquote {\bibinfo {title} {Latent ewald summation for machine learning of long-range interactions},}\ }\href {\doibase 10.1038/s41524-025-01577-7} {\bibfield  {journal} {\bibinfo  {journal} {npj Computational Materials}\ }\textbf {\bibinfo {volume} {11}},\ \bibinfo {pages} {80} (\bibinfo {year} {2025})}\BibitemShut {NoStop}%
\bibitem [{\citenamefont {King}\ \emph {et~al.}(2025)\citenamefont {King}, \citenamefont {Kim}, \citenamefont {Zhong},\ and\ \citenamefont {Cheng}}]{King2025Machine}%
  \BibitemOpen
  \bibfield  {author} {\bibinfo {author} {\bibfnamefont {Daniel~S.}\ \bibnamefont {King}}, \bibinfo {author} {\bibfnamefont {Dongjin}\ \bibnamefont {Kim}}, \bibinfo {author} {\bibfnamefont {Peichen}\ \bibnamefont {Zhong}}, \ and\ \bibinfo {author} {\bibfnamefont {Bingqing}\ \bibnamefont {Cheng}},\ }\bibfield  {title} {\enquote {\bibinfo {title} {Machine learning of charges and long-range interactions from energies and forces},}\ }\href {\doibase 10.1038/s41467-025-63852-x} {\bibfield  {journal} {\bibinfo  {journal} {Nature Communications}\ }\textbf {\bibinfo {volume} {16}},\ \bibinfo {pages} {8763} (\bibinfo {year} {2025})}\BibitemShut {NoStop}%
\bibitem [{\citenamefont {Zhong}\ \emph {et~al.}(2025)\citenamefont {Zhong}, \citenamefont {Kim}, \citenamefont {King},\ and\ \citenamefont {Cheng}}]{zhong2025machine}%
  \BibitemOpen
  \bibfield  {author} {\bibinfo {author} {\bibfnamefont {Peichen}\ \bibnamefont {Zhong}}, \bibinfo {author} {\bibfnamefont {Dongjin}\ \bibnamefont {Kim}}, \bibinfo {author} {\bibfnamefont {Daniel~S.}\ \bibnamefont {King}}, \ and\ \bibinfo {author} {\bibfnamefont {Bingqing}\ \bibnamefont {Cheng}},\ }\bibfield  {title} {\enquote {\bibinfo {title} {Machine learning interatomic potential can infer electrical response},}\ }\href {\doibase 10.1038/s41524-025-01911-z} {\bibfield  {journal} {\bibinfo  {journal} {npj Computational Materials}\ }\textbf {\bibinfo {volume} {11}} (\bibinfo {year} {2025}),\ 10.1038/s41524-025-01911-z}\BibitemShut {NoStop}%
\bibitem [{\citenamefont {Rappe}\ and\ \citenamefont {Goddard~III}(1991)}]{rappe1991charge}%
  \BibitemOpen
  \bibfield  {author} {\bibinfo {author} {\bibfnamefont {Anthony~K}\ \bibnamefont {Rappe}}\ and\ \bibinfo {author} {\bibfnamefont {William~A}\ \bibnamefont {Goddard~III}},\ }\bibfield  {title} {\enquote {\bibinfo {title} {Charge equilibration for molecular dynamics simulations},}\ }\href@noop {} {\bibfield  {journal} {\bibinfo  {journal} {The Journal of Physical Chemistry}\ }\textbf {\bibinfo {volume} {95}},\ \bibinfo {pages} {3358--3363} (\bibinfo {year} {1991})}\BibitemShut {NoStop}%
\bibitem [{\citenamefont {Ghasemi}\ \emph {et~al.}(2015)\citenamefont {Ghasemi}, \citenamefont {Hofstetter}, \citenamefont {Saha},\ and\ \citenamefont {Goedecker}}]{Ghasemi2015Interatomic}%
  \BibitemOpen
  \bibfield  {author} {\bibinfo {author} {\bibfnamefont {S.~Alireza}\ \bibnamefont {Ghasemi}}, \bibinfo {author} {\bibfnamefont {Albert}\ \bibnamefont {Hofstetter}}, \bibinfo {author} {\bibfnamefont {Santanu}\ \bibnamefont {Saha}}, \ and\ \bibinfo {author} {\bibfnamefont {Stefan}\ \bibnamefont {Goedecker}},\ }\bibfield  {title} {\enquote {\bibinfo {title} {Interatomic potentials for ionic systems with density functional accuracy based on charge densities obtained by a neural network},}\ }\href {\doibase 10.1103/PhysRevB.92.045131} {\bibfield  {journal} {\bibinfo  {journal} {Physical Review B}\ }\textbf {\bibinfo {volume} {92}},\ \bibinfo {pages} {045131} (\bibinfo {year} {2015})}\BibitemShut {NoStop}%
\bibitem [{\citenamefont {Fuchs}\ \emph {et~al.}(2025)\citenamefont {Fuchs}, \citenamefont {Sanocki},\ and\ \citenamefont {Zavadlav}}]{fuchs2025Learninga}%
  \BibitemOpen
  \bibfield  {author} {\bibinfo {author} {\bibfnamefont {Paul}\ \bibnamefont {Fuchs}}, \bibinfo {author} {\bibfnamefont {Micha{\l}}\ \bibnamefont {Sanocki}}, \ and\ \bibinfo {author} {\bibfnamefont {Julija}\ \bibnamefont {Zavadlav}},\ }\bibfield  {title} {\enquote {\bibinfo {title} {Learning non-local molecular interactions via equivariant local representations and charge equilibration},}\ }\href {\doibase 10.1038/s41524-025-01790-4} {\bibfield  {journal} {\bibinfo  {journal} {npj Computational Materials}\ }\textbf {\bibinfo {volume} {11}},\ \bibinfo {pages} {287} (\bibinfo {year} {2025})}\BibitemShut {NoStop}%
\bibitem [{\citenamefont {Gong}\ \emph {et~al.}(2025)\citenamefont {Gong}, \citenamefont {Zhang}, \citenamefont {Mu}, \citenamefont {Pu}, \citenamefont {Wang}, \citenamefont {Han}, \citenamefont {Yu}, \citenamefont {Chen}, \citenamefont {Zheng}, \citenamefont {Wang}, \citenamefont {Chen}, \citenamefont {Yang}, \citenamefont {Wu}, \citenamefont {Shi}, \citenamefont {Gao}, \citenamefont {Yan},\ and\ \citenamefont {Xiang}}]{gong_predictive_2025}%
  \BibitemOpen
  \bibfield  {author} {\bibinfo {author} {\bibfnamefont {Sheng}\ \bibnamefont {Gong}}, \bibinfo {author} {\bibfnamefont {Yumin}\ \bibnamefont {Zhang}}, \bibinfo {author} {\bibfnamefont {Zhenliang}\ \bibnamefont {Mu}}, \bibinfo {author} {\bibfnamefont {Zhichen}\ \bibnamefont {Pu}}, \bibinfo {author} {\bibfnamefont {Hongyi}\ \bibnamefont {Wang}}, \bibinfo {author} {\bibfnamefont {Xu}~\bibnamefont {Han}}, \bibinfo {author} {\bibfnamefont {Zhiao}\ \bibnamefont {Yu}}, \bibinfo {author} {\bibfnamefont {Mengyi}\ \bibnamefont {Chen}}, \bibinfo {author} {\bibfnamefont {Tianze}\ \bibnamefont {Zheng}}, \bibinfo {author} {\bibfnamefont {Zhi}\ \bibnamefont {Wang}}, \bibinfo {author} {\bibfnamefont {Lifei}\ \bibnamefont {Chen}}, \bibinfo {author} {\bibfnamefont {Zhenze}\ \bibnamefont {Yang}}, \bibinfo {author} {\bibfnamefont {Xiaojie}\ \bibnamefont {Wu}}, \bibinfo {author} {\bibfnamefont {Shaochen}\ \bibnamefont {Shi}}, \bibinfo {author} {\bibfnamefont {Weihao}\ \bibnamefont {Gao}}, \bibinfo {author} {\bibfnamefont {Wen}\
  \bibnamefont {Yan}}, \ and\ \bibinfo {author} {\bibfnamefont {Liang}\ \bibnamefont {Xiang}},\ }\bibfield  {title} {\enquote {\bibinfo {title} {A predictive machine learning force-field framework for liquid electrolyte development},}\ }\href {\doibase 10.1038/s42256-025-01009-7} {\bibfield  {journal} {\bibinfo  {journal} {Nature Machine Intelligence}\ }\textbf {\bibinfo {volume} {7}},\ \bibinfo {pages} {543--552} (\bibinfo {year} {2025})}\BibitemShut {NoStop}%
\bibitem [{\citenamefont {Anstine}\ \emph {et~al.}(2025)\citenamefont {Anstine}, \citenamefont {Zubatyuk},\ and\ \citenamefont {Isayev}}]{Anstine2025}%
  \BibitemOpen
  \bibfield  {author} {\bibinfo {author} {\bibfnamefont {Dylan~M.}\ \bibnamefont {Anstine}}, \bibinfo {author} {\bibfnamefont {Roman}\ \bibnamefont {Zubatyuk}}, \ and\ \bibinfo {author} {\bibfnamefont {Olexandr}\ \bibnamefont {Isayev}},\ }\bibfield  {title} {\enquote {\bibinfo {title} {Aimnet2: a neural network potential to meet your neutral, charged, organic, and elemental-organic needs},}\ }\href {\doibase 10.1039/d4sc08572h} {\bibfield  {journal} {\bibinfo  {journal} {Chemical Science}\ }\textbf {\bibinfo {volume} {16}},\ \bibinfo {pages} {10228–10244} (\bibinfo {year} {2025})}\BibitemShut {NoStop}%
\bibitem [{\citenamefont {Batatia}\ \emph {et~al.}(2026)\citenamefont {Batatia}, \citenamefont {Baldwin}, \citenamefont {Kuryla}, \citenamefont {Hart}, \citenamefont {Kasoar}, \citenamefont {Elena}, \citenamefont {Moore}, \citenamefont {Gawkowski}, \citenamefont {Shi}, \citenamefont {Kapil} \emph {et~al.}}]{batatia2026mace}%
  \BibitemOpen
  \bibfield  {author} {\bibinfo {author} {\bibfnamefont {Ilyes}\ \bibnamefont {Batatia}}, \bibinfo {author} {\bibfnamefont {William~J}\ \bibnamefont {Baldwin}}, \bibinfo {author} {\bibfnamefont {Domantas}\ \bibnamefont {Kuryla}}, \bibinfo {author} {\bibfnamefont {Joseph}\ \bibnamefont {Hart}}, \bibinfo {author} {\bibfnamefont {Elliott}\ \bibnamefont {Kasoar}}, \bibinfo {author} {\bibfnamefont {Alin~M}\ \bibnamefont {Elena}}, \bibinfo {author} {\bibfnamefont {Harry}\ \bibnamefont {Moore}}, \bibinfo {author} {\bibfnamefont {Miko{\l}aj~J}\ \bibnamefont {Gawkowski}}, \bibinfo {author} {\bibfnamefont {Benjamin~X}\ \bibnamefont {Shi}}, \bibinfo {author} {\bibfnamefont {Venkat}\ \bibnamefont {Kapil}},  \emph {et~al.},\ }\bibfield  {title} {\enquote {\bibinfo {title} {Mace-polar-1: A polarisable electrostatic foundation model for molecular chemistry},}\ }\href@noop {} {\bibfield  {journal} {\bibinfo  {journal} {arXiv preprint arXiv:2602.19411}\ } (\bibinfo {year} {2026})}\BibitemShut {NoStop}%
\bibitem [{\citenamefont {Jensen}(2023)}]{jensen2023unifying}%
  \BibitemOpen
  \bibfield  {author} {\bibinfo {author} {\bibfnamefont {Frank}\ \bibnamefont {Jensen}},\ }\bibfield  {title} {\enquote {\bibinfo {title} {Unifying charge-flow polarization models},}\ }\href@noop {} {\bibfield  {journal} {\bibinfo  {journal} {Journal of Chemical Theory and Computation}\ }\textbf {\bibinfo {volume} {19}},\ \bibinfo {pages} {4047--4073} (\bibinfo {year} {2023})}\BibitemShut {NoStop}%
\bibitem [{\citenamefont {Lee~Warren}\ \emph {et~al.}(2008)\citenamefont {Lee~Warren}, \citenamefont {Davis},\ and\ \citenamefont {Patel}}]{lee2008origin}%
  \BibitemOpen
  \bibfield  {author} {\bibinfo {author} {\bibfnamefont {G}~\bibnamefont {Lee~Warren}}, \bibinfo {author} {\bibfnamefont {Joseph~E}\ \bibnamefont {Davis}}, \ and\ \bibinfo {author} {\bibfnamefont {Sandeep}\ \bibnamefont {Patel}},\ }\bibfield  {title} {\enquote {\bibinfo {title} {Origin and control of superlinear polarizability scaling in chemical potential equalization methods},}\ }\href@noop {} {\bibfield  {journal} {\bibinfo  {journal} {The Journal of chemical physics}\ }\textbf {\bibinfo {volume} {128}} (\bibinfo {year} {2008})}\BibitemShut {NoStop}%
\bibitem [{\citenamefont {Gao}\ and\ \citenamefont {Remsing}(2022)}]{gao2022self}%
  \BibitemOpen
  \bibfield  {author} {\bibinfo {author} {\bibfnamefont {Ang}\ \bibnamefont {Gao}}\ and\ \bibinfo {author} {\bibfnamefont {Richard~C}\ \bibnamefont {Remsing}},\ }\bibfield  {title} {\enquote {\bibinfo {title} {Self-consistent determination of long-range electrostatics in neural network potentials},}\ }\href@noop {} {\bibfield  {journal} {\bibinfo  {journal} {Nature communications}\ }\textbf {\bibinfo {volume} {13}},\ \bibinfo {pages} {1572} (\bibinfo {year} {2022})}\BibitemShut {NoStop}%
\bibitem [{\citenamefont {Leontyev}\ \emph {et~al.}(2003)\citenamefont {Leontyev}, \citenamefont {Vener}, \citenamefont {Rostov}, \citenamefont {Basilevsky},\ and\ \citenamefont {Newton}}]{leontyev2003continuum}%
  \BibitemOpen
  \bibfield  {author} {\bibinfo {author} {\bibfnamefont {IV}~\bibnamefont {Leontyev}}, \bibinfo {author} {\bibfnamefont {MV}~\bibnamefont {Vener}}, \bibinfo {author} {\bibfnamefont {IV}~\bibnamefont {Rostov}}, \bibinfo {author} {\bibfnamefont {MV}~\bibnamefont {Basilevsky}}, \ and\ \bibinfo {author} {\bibfnamefont {Marshall~D}\ \bibnamefont {Newton}},\ }\bibfield  {title} {\enquote {\bibinfo {title} {Continuum level treatment of electronic polarization in the framework of molecular simulations of solvation effects},}\ }\href@noop {} {\bibfield  {journal} {\bibinfo  {journal} {The Journal of chemical physics}\ }\textbf {\bibinfo {volume} {119}},\ \bibinfo {pages} {8024--8037} (\bibinfo {year} {2003})}\BibitemShut {NoStop}%
\bibitem [{\citenamefont {Leontyev}\ and\ \citenamefont {Stuchebrukhov}(2014)}]{leontyev2014polarizable}%
  \BibitemOpen
  \bibfield  {author} {\bibinfo {author} {\bibfnamefont {Igor~V}\ \bibnamefont {Leontyev}}\ and\ \bibinfo {author} {\bibfnamefont {Alexei~A}\ \bibnamefont {Stuchebrukhov}},\ }\bibfield  {title} {\enquote {\bibinfo {title} {Polarizable molecular interactions in condensed phase and their equivalent nonpolarizable models},}\ }\href@noop {} {\bibfield  {journal} {\bibinfo  {journal} {The Journal of chemical physics}\ }\textbf {\bibinfo {volume} {141}} (\bibinfo {year} {2014})}\BibitemShut {NoStop}%
\bibitem [{\citenamefont {Jorge}(2024)}]{jorge2024theoretically}%
  \BibitemOpen
  \bibfield  {author} {\bibinfo {author} {\bibfnamefont {Miguel}\ \bibnamefont {Jorge}},\ }\bibfield  {title} {\enquote {\bibinfo {title} {Theoretically grounded approaches to account for polarization effects in fixed-charge force fields},}\ }\href@noop {} {\bibfield  {journal} {\bibinfo  {journal} {The Journal of Chemical Physics}\ }\textbf {\bibinfo {volume} {161}} (\bibinfo {year} {2024})}\BibitemShut {NoStop}%
\bibitem [{\citenamefont {Loche}\ \emph {et~al.}(2025)\citenamefont {Loche}, \citenamefont {Huguenin-Dumittan}, \citenamefont {Honarmand}, \citenamefont {Xu}, \citenamefont {Rumiantsev}, \citenamefont {How}, \citenamefont {Langer},\ and\ \citenamefont {Ceriotti}}]{loche2025fast}%
  \BibitemOpen
  \bibfield  {author} {\bibinfo {author} {\bibfnamefont {Philip}\ \bibnamefont {Loche}}, \bibinfo {author} {\bibfnamefont {Kevin~K}\ \bibnamefont {Huguenin-Dumittan}}, \bibinfo {author} {\bibfnamefont {Melika}\ \bibnamefont {Honarmand}}, \bibinfo {author} {\bibfnamefont {Qianjun}\ \bibnamefont {Xu}}, \bibinfo {author} {\bibfnamefont {Egor}\ \bibnamefont {Rumiantsev}}, \bibinfo {author} {\bibfnamefont {Wei~Bin}\ \bibnamefont {How}}, \bibinfo {author} {\bibfnamefont {Marcel~F}\ \bibnamefont {Langer}}, \ and\ \bibinfo {author} {\bibfnamefont {Michele}\ \bibnamefont {Ceriotti}},\ }\bibfield  {title} {\enquote {\bibinfo {title} {Fast and flexible long-range models for atomistic machine learning},}\ }\href@noop {} {\bibfield  {journal} {\bibinfo  {journal} {The Journal of Chemical Physics}\ }\textbf {\bibinfo {volume} {162}} (\bibinfo {year} {2025})}\BibitemShut {NoStop}%
\bibitem [{\citenamefont {Kim}\ \emph {et~al.}(2025)\citenamefont {Kim}, \citenamefont {Wang}, \citenamefont {Vargas}, \citenamefont {Zhong}, \citenamefont {King}, \citenamefont {Inizan},\ and\ \citenamefont {Cheng}}]{Kim2025Universalb}%
  \BibitemOpen
  \bibfield  {author} {\bibinfo {author} {\bibfnamefont {Dongjin}\ \bibnamefont {Kim}}, \bibinfo {author} {\bibfnamefont {Xiaoyu}\ \bibnamefont {Wang}}, \bibinfo {author} {\bibfnamefont {Santiago}\ \bibnamefont {Vargas}}, \bibinfo {author} {\bibfnamefont {Peichen}\ \bibnamefont {Zhong}}, \bibinfo {author} {\bibfnamefont {Daniel~S.}\ \bibnamefont {King}}, \bibinfo {author} {\bibfnamefont {Theo~Jaffrelot}\ \bibnamefont {Inizan}}, \ and\ \bibinfo {author} {\bibfnamefont {Bingqing}\ \bibnamefont {Cheng}},\ }\bibfield  {title} {\enquote {\bibinfo {title} {A universal augmentation framework for long-range electrostatics in machine learning interatomic potentials},}\ }\href {\doibase 10.1021/acs.jctc.5c01400} {\bibfield  {journal} {\bibinfo  {journal} {Journal of Chemical Theory and Computation}\ } (\bibinfo {year} {2025}),\ 10.1021/acs.jctc.5c01400}\BibitemShut {NoStop}%
\bibitem [{\citenamefont {Sala}\ \emph {et~al.}(2010)\citenamefont {Sala}, \citenamefont {Guàrdia},\ and\ \citenamefont {Masia}}]{Sala2010}%
  \BibitemOpen
  \bibfield  {author} {\bibinfo {author} {\bibfnamefont {Jonàs}\ \bibnamefont {Sala}}, \bibinfo {author} {\bibfnamefont {Elvira}\ \bibnamefont {Guàrdia}}, \ and\ \bibinfo {author} {\bibfnamefont {Marco}\ \bibnamefont {Masia}},\ }\bibfield  {title} {\enquote {\bibinfo {title} {The polarizable point dipoles method with electrostatic damping: Implementation on a model system},}\ }\href {\doibase 10.1063/1.3511713} {\bibfield  {journal} {\bibinfo  {journal} {The Journal of Chemical Physics}\ }\textbf {\bibinfo {volume} {133}} (\bibinfo {year} {2010}),\ 10.1063/1.3511713}\BibitemShut {NoStop}%
\bibitem [{\citenamefont {Riera}\ \emph {et~al.}(2023)\citenamefont {Riera}, \citenamefont {Knight}, \citenamefont {Bull-Vulpe}, \citenamefont {Zhu}, \citenamefont {Agnew}, \citenamefont {Smith}, \citenamefont {Simmonett},\ and\ \citenamefont {Paesani}}]{Riera2023}%
  \BibitemOpen
  \bibfield  {author} {\bibinfo {author} {\bibfnamefont {Marc}\ \bibnamefont {Riera}}, \bibinfo {author} {\bibfnamefont {Christopher}\ \bibnamefont {Knight}}, \bibinfo {author} {\bibfnamefont {Ethan~F.}\ \bibnamefont {Bull-Vulpe}}, \bibinfo {author} {\bibfnamefont {Xuanyu}\ \bibnamefont {Zhu}}, \bibinfo {author} {\bibfnamefont {Henry}\ \bibnamefont {Agnew}}, \bibinfo {author} {\bibfnamefont {Daniel G.~A.}\ \bibnamefont {Smith}}, \bibinfo {author} {\bibfnamefont {Andrew~C.}\ \bibnamefont {Simmonett}}, \ and\ \bibinfo {author} {\bibfnamefont {Francesco}\ \bibnamefont {Paesani}},\ }\bibfield  {title} {\enquote {\bibinfo {title} {Mbx: A many-body energy and force calculator for data-driven many-body simulations},}\ }\href {\doibase 10.1063/5.0156036} {\bibfield  {journal} {\bibinfo  {journal} {The Journal of Chemical Physics}\ }\textbf {\bibinfo {volume} {159}} (\bibinfo {year} {2023}),\ 10.1063/5.0156036}\BibitemShut {NoStop}%
\bibitem [{\citenamefont {C{\'a}rdenas}(2011)}]{cardenas2011fukui}%
  \BibitemOpen
  \bibfield  {author} {\bibinfo {author} {\bibfnamefont {Carlos}\ \bibnamefont {C{\'a}rdenas}},\ }\bibfield  {title} {\enquote {\bibinfo {title} {The fukui potential is a measure of the chemical hardness},}\ }\href@noop {} {\bibfield  {journal} {\bibinfo  {journal} {Chemical Physics Letters}\ }\textbf {\bibinfo {volume} {513}},\ \bibinfo {pages} {127--129} (\bibinfo {year} {2011})}\BibitemShut {NoStop}%
\bibitem [{\citenamefont {Farahvash}\ \emph {et~al.}(2018)\citenamefont {Farahvash}, \citenamefont {Leontyev},\ and\ \citenamefont {Stuchebrukhov}}]{farahvash2018dynamic}%
  \BibitemOpen
  \bibfield  {author} {\bibinfo {author} {\bibfnamefont {Ardavan}\ \bibnamefont {Farahvash}}, \bibinfo {author} {\bibfnamefont {Igor}\ \bibnamefont {Leontyev}}, \ and\ \bibinfo {author} {\bibfnamefont {Alexei}\ \bibnamefont {Stuchebrukhov}},\ }\bibfield  {title} {\enquote {\bibinfo {title} {Dynamic and electronic polarization corrections to the dielectric constant of water},}\ }\href@noop {} {\bibfield  {journal} {\bibinfo  {journal} {The Journal of Physical Chemistry A}\ }\textbf {\bibinfo {volume} {122}},\ \bibinfo {pages} {9243--9250} (\bibinfo {year} {2018})}\BibitemShut {NoStop}%
\bibitem [{\citenamefont {Resta}\ and\ \citenamefont {Vanderbilt}(2007)}]{resta2007theory}%
  \BibitemOpen
  \bibfield  {author} {\bibinfo {author} {\bibfnamefont {Raffaele}\ \bibnamefont {Resta}}\ and\ \bibinfo {author} {\bibfnamefont {David}\ \bibnamefont {Vanderbilt}},\ }\bibfield  {title} {\enquote {\bibinfo {title} {Theory of polarization: a modern approach},}\ }in\ \href@noop {} {\emph {\bibinfo {booktitle} {Physics of ferroelectrics: a modern perspective}}}\ (\bibinfo  {publisher} {Springer},\ \bibinfo {year} {2007})\ pp.\ \bibinfo {pages} {31--68}\BibitemShut {NoStop}%
\bibitem [{\citenamefont {Siepmann}\ and\ \citenamefont {Sprik}(1995)}]{siepmann1995influence}%
  \BibitemOpen
  \bibfield  {author} {\bibinfo {author} {\bibfnamefont {J~Ilja}\ \bibnamefont {Siepmann}}\ and\ \bibinfo {author} {\bibfnamefont {Michiel}\ \bibnamefont {Sprik}},\ }\bibfield  {title} {\enquote {\bibinfo {title} {Influence of surface topology and electrostatic potential on water/electrode systems},}\ }\href@noop {} {\bibfield  {journal} {\bibinfo  {journal} {The Journal of chemical physics}\ }\textbf {\bibinfo {volume} {102}},\ \bibinfo {pages} {511--524} (\bibinfo {year} {1995})}\BibitemShut {NoStop}%
\bibitem [{\citenamefont {Wang}\ \emph {et~al.}(2025)\citenamefont {Wang}, \citenamefont {Chen}, \citenamefont {Zeng}, \citenamefont {Stein}, \citenamefont {Lim},\ and\ \citenamefont {Cheng}}]{Wang2025Ionmodulateda}%
  \BibitemOpen
  \bibfield  {author} {\bibinfo {author} {\bibfnamefont {Xiaoyu}\ \bibnamefont {Wang}}, \bibinfo {author} {\bibfnamefont {Junmin}\ \bibnamefont {Chen}}, \bibinfo {author} {\bibfnamefont {Zezhu}\ \bibnamefont {Zeng}}, \bibinfo {author} {\bibfnamefont {Frederick}\ \bibnamefont {Stein}}, \bibinfo {author} {\bibfnamefont {Junho}\ \bibnamefont {Lim}}, \ and\ \bibinfo {author} {\bibfnamefont {Bingqing}\ \bibnamefont {Cheng}},\ }\bibfield  {title} {\enquote {\bibinfo {title} {Ion-modulated structure, proton transfer, and capacitance in the pt(111)/water electric double layer},}\ }\href@noop {} {\bibfield  {journal} {\bibinfo  {journal} {arXiv preprint arXiv:2509.13727}\ } (\bibinfo {year} {2025})}\BibitemShut {NoStop}%
\bibitem [{\citenamefont {Semelak}\ \emph {et~al.}(2025)\citenamefont {Semelak}, \citenamefont {Pickering}, \citenamefont {Huddleston}, \citenamefont {Olmos}, \citenamefont {Grassano}, \citenamefont {Clemente}, \citenamefont {Drusin}, \citenamefont {Marti}, \citenamefont {Gonzalez~Lebrero}, \citenamefont {Roitberg},\ and\ \citenamefont {Estrin}}]{Semelak2025Advancing}%
  \BibitemOpen
  \bibfield  {author} {\bibinfo {author} {\bibfnamefont {Jonathan~A.}\ \bibnamefont {Semelak}}, \bibinfo {author} {\bibfnamefont {Ignacio}\ \bibnamefont {Pickering}}, \bibinfo {author} {\bibfnamefont {Kate}\ \bibnamefont {Huddleston}}, \bibinfo {author} {\bibfnamefont {Justo}\ \bibnamefont {Olmos}}, \bibinfo {author} {\bibfnamefont {Juan~Santiago}\ \bibnamefont {Grassano}}, \bibinfo {author} {\bibfnamefont {Camila~Mara}\ \bibnamefont {Clemente}}, \bibinfo {author} {\bibfnamefont {Salvador~I.}\ \bibnamefont {Drusin}}, \bibinfo {author} {\bibfnamefont {Marcelo}\ \bibnamefont {Marti}}, \bibinfo {author} {\bibfnamefont {Mariano~Camilo}\ \bibnamefont {Gonzalez~Lebrero}}, \bibinfo {author} {\bibfnamefont {Adrian~E.}\ \bibnamefont {Roitberg}}, \ and\ \bibinfo {author} {\bibfnamefont {Dario~A.}\ \bibnamefont {Estrin}},\ }\bibfield  {title} {\enquote {\bibinfo {title} {Advancing multiscale molecular modeling with machine learning-derived electrostatics},}\ }\href {\doibase 10.1021/acs.jctc.4c01792} {\bibfield
  {journal} {\bibinfo  {journal} {Journal of Chemical Theory and Computation}\ }\textbf {\bibinfo {volume} {21}},\ \bibinfo {pages} {5194--5207} (\bibinfo {year} {2025})}\BibitemShut {NoStop}%
\bibitem [{\citenamefont {Morado}\ \emph {et~al.}(2025)\citenamefont {Morado}, \citenamefont {Zinovjev}, \citenamefont {Hedges}, \citenamefont {Cole},\ and\ \citenamefont {Michel}}]{morado2025enhancing}%
  \BibitemOpen
  \bibfield  {author} {\bibinfo {author} {\bibfnamefont {Jo{\~a}o}\ \bibnamefont {Morado}}, \bibinfo {author} {\bibfnamefont {Kirill}\ \bibnamefont {Zinovjev}}, \bibinfo {author} {\bibfnamefont {Lester~O}\ \bibnamefont {Hedges}}, \bibinfo {author} {\bibfnamefont {Daniel~J}\ \bibnamefont {Cole}}, \ and\ \bibinfo {author} {\bibfnamefont {Julien}\ \bibnamefont {Michel}},\ }\bibfield  {title} {\enquote {\bibinfo {title} {Enhancing electrostatic embedding for ml/mm free energy calculations},}\ }\href@noop {} {\bibfield  {journal} {\bibinfo  {journal} {Journal of Chemical Theory and Computation}\ } (\bibinfo {year} {2025})}\BibitemShut {NoStop}%
\bibitem [{\citenamefont {Cheng}(2024)}]{cheng2024cartesian}%
  \BibitemOpen
  \bibfield  {author} {\bibinfo {author} {\bibfnamefont {Bingqing}\ \bibnamefont {Cheng}},\ }\bibfield  {title} {\enquote {\bibinfo {title} {Cartesian atomic cluster expansion for machine learning interatomic potentials},}\ }\href@noop {} {\bibfield  {journal} {\bibinfo  {journal} {npj Computational Materials}\ }\textbf {\bibinfo {volume} {10}},\ \bibinfo {pages} {157} (\bibinfo {year} {2024})}\BibitemShut {NoStop}%
\bibitem [{\citenamefont {Musaelian}\ \emph {et~al.}(2023)\citenamefont {Musaelian}, \citenamefont {Batzner}, \citenamefont {Johansson}, \citenamefont {Sun}, \citenamefont {Owen}, \citenamefont {Kornbluth},\ and\ \citenamefont {Kozinsky}}]{musaelian2023learning}%
  \BibitemOpen
  \bibfield  {author} {\bibinfo {author} {\bibfnamefont {Albert}\ \bibnamefont {Musaelian}}, \bibinfo {author} {\bibfnamefont {Simon}\ \bibnamefont {Batzner}}, \bibinfo {author} {\bibfnamefont {Anders}\ \bibnamefont {Johansson}}, \bibinfo {author} {\bibfnamefont {Lixin}\ \bibnamefont {Sun}}, \bibinfo {author} {\bibfnamefont {Cameron~J}\ \bibnamefont {Owen}}, \bibinfo {author} {\bibfnamefont {Mordechai}\ \bibnamefont {Kornbluth}}, \ and\ \bibinfo {author} {\bibfnamefont {Boris}\ \bibnamefont {Kozinsky}},\ }\bibfield  {title} {\enquote {\bibinfo {title} {Learning local equivariant representations for large-scale atomistic dynamics},}\ }\href@noop {} {\bibfield  {journal} {\bibinfo  {journal} {Nature Communications}\ }\textbf {\bibinfo {volume} {14}},\ \bibinfo {pages} {579} (\bibinfo {year} {2023})}\BibitemShut {NoStop}%
\bibitem [{\citenamefont {Tan}\ \emph {et~al.}(2026)\citenamefont {Tan}, \citenamefont {Descoteaux}, \citenamefont {Kotak}, \citenamefont {De~Miranda~Nascimento}, \citenamefont {Kavanagh}, \citenamefont {Zichi}, \citenamefont {Wang}, \citenamefont {Saluja}, \citenamefont {Hu}, \citenamefont {Smidt}, \citenamefont {Johansson}, \citenamefont {Witt}, \citenamefont {Kozinsky},\ and\ \citenamefont {Musaelian}}]{Tan2026Highperformance}%
  \BibitemOpen
  \bibfield  {author} {\bibinfo {author} {\bibfnamefont {Chuin~Wei}\ \bibnamefont {Tan}}, \bibinfo {author} {\bibfnamefont {Marc~L.}\ \bibnamefont {Descoteaux}}, \bibinfo {author} {\bibfnamefont {Mit}\ \bibnamefont {Kotak}}, \bibinfo {author} {\bibfnamefont {Gabriel}\ \bibnamefont {De~Miranda~Nascimento}}, \bibinfo {author} {\bibfnamefont {Se{\'a}n~R.}\ \bibnamefont {Kavanagh}}, \bibinfo {author} {\bibfnamefont {Laura}\ \bibnamefont {Zichi}}, \bibinfo {author} {\bibfnamefont {Menghang}\ \bibnamefont {Wang}}, \bibinfo {author} {\bibfnamefont {Aadit}\ \bibnamefont {Saluja}}, \bibinfo {author} {\bibfnamefont {Yizhong~R.}\ \bibnamefont {Hu}}, \bibinfo {author} {\bibfnamefont {Tess}\ \bibnamefont {Smidt}}, \bibinfo {author} {\bibfnamefont {Anders}\ \bibnamefont {Johansson}}, \bibinfo {author} {\bibfnamefont {William~C.}\ \bibnamefont {Witt}}, \bibinfo {author} {\bibfnamefont {Boris}\ \bibnamefont {Kozinsky}}, \ and\ \bibinfo {author} {\bibfnamefont {Albert}\ \bibnamefont {Musaelian}},\ }\bibfield  {title}
  {\enquote {\bibinfo {title} {High-performance training and inference for deep equivariant interatomic potentials},}\ }\href {\doibase 10.1039/D5DD00423C} {\bibfield  {journal} {\bibinfo  {journal} {Digital Discovery}\ }\textbf {\bibinfo {volume} {5}},\ \bibinfo {pages} {1558--1567} (\bibinfo {year} {2026})}\BibitemShut {NoStop}%
\bibitem [{\citenamefont {Schmiedmayer}\ and\ \citenamefont {Kresse}(2024)}]{Schmiedmayer2024}%
  \BibitemOpen
  \bibfield  {author} {\bibinfo {author} {\bibfnamefont {Bernhard}\ \bibnamefont {Schmiedmayer}}\ and\ \bibinfo {author} {\bibfnamefont {Georg}\ \bibnamefont {Kresse}},\ }\bibfield  {title} {\enquote {\bibinfo {title} {Derivative learning of tensorial quantities—predicting finite temperature infrared spectra from first principles},}\ }\href {http://dx.doi.org/10.1063/5.0217243} {\bibfield  {journal} {\bibinfo  {journal} {The Journal of Chemical Physics}\ }\textbf {\bibinfo {volume} {161}} (\bibinfo {year} {2024})}\BibitemShut {NoStop}%
\bibitem [{\citenamefont {Bertie}\ and\ \citenamefont {Lan}(1996)}]{Bertie1996Infrared}%
  \BibitemOpen
  \bibfield  {author} {\bibinfo {author} {\bibfnamefont {John~E.}\ \bibnamefont {Bertie}}\ and\ \bibinfo {author} {\bibfnamefont {Zhida}\ \bibnamefont {Lan}},\ }\bibfield  {title} {\enquote {\bibinfo {title} {Infrared intensities of liquids xx: The intensity of the oh stretching band of liquid water revisited, and the best current values of the optical constants of h2o(l) at 25{$^\circ$}c between 15,000 and 1 cm-1},}\ }\href {\doibase 10.1366/0003702963905385} {\bibfield  {journal} {\bibinfo  {journal} {Applied Spectroscopy}\ }\textbf {\bibinfo {volume} {50}},\ \bibinfo {pages} {1047--1057} (\bibinfo {year} {1996})}\BibitemShut {NoStop}%
\bibitem [{\citenamefont {Grisafi}\ \emph {et~al.}(2018)\citenamefont {Grisafi}, \citenamefont {Wilkins}, \citenamefont {Cs{\'a}nyi},\ and\ \citenamefont {Ceriotti}}]{grisafi2018symmetry}%
  \BibitemOpen
  \bibfield  {author} {\bibinfo {author} {\bibfnamefont {Andrea}\ \bibnamefont {Grisafi}}, \bibinfo {author} {\bibfnamefont {David~M}\ \bibnamefont {Wilkins}}, \bibinfo {author} {\bibfnamefont {G{\'a}bor}\ \bibnamefont {Cs{\'a}nyi}}, \ and\ \bibinfo {author} {\bibfnamefont {Michele}\ \bibnamefont {Ceriotti}},\ }\bibfield  {title} {\enquote {\bibinfo {title} {Symmetry-adapted machine learning for tensorial properties of atomistic systems},}\ }\href@noop {} {\bibfield  {journal} {\bibinfo  {journal} {Physical review letters}\ }\textbf {\bibinfo {volume} {120}},\ \bibinfo {pages} {036002} (\bibinfo {year} {2018})}\BibitemShut {NoStop}%
\bibitem [{\citenamefont {Pattenaude}\ \emph {et~al.}(2018)\citenamefont {Pattenaude}, \citenamefont {Streacker},\ and\ \citenamefont {Ben-Amotz}}]{Pattenaude2018Temperaturea}%
  \BibitemOpen
  \bibfield  {author} {\bibinfo {author} {\bibfnamefont {Shannon~R.}\ \bibnamefont {Pattenaude}}, \bibinfo {author} {\bibfnamefont {Louis~M.}\ \bibnamefont {Streacker}}, \ and\ \bibinfo {author} {\bibfnamefont {Dor}\ \bibnamefont {Ben-Amotz}},\ }\bibfield  {title} {\enquote {\bibinfo {title} {Temperature and polarization dependent raman spectra of liquid {\textsc{ h{\textsubscript{2}} o }} and {\textsc{ d{\textsubscript{2}} o }}},}\ }\href {\doibase 10.1002/jrs.5465} {\bibfield  {journal} {\bibinfo  {journal} {Journal of Raman Spectroscopy}\ }\textbf {\bibinfo {volume} {49}},\ \bibinfo {pages} {1860--1866} (\bibinfo {year} {2018})}\BibitemShut {NoStop}%
\bibitem [{\citenamefont {Morawietz}\ \emph {et~al.}(2018)\citenamefont {Morawietz}, \citenamefont {Marsalek}, \citenamefont {Pattenaude}, \citenamefont {Streacker}, \citenamefont {{Ben-Amotz}},\ and\ \citenamefont {Markland}}]{Morawietz2018Interplay}%
  \BibitemOpen
  \bibfield  {author} {\bibinfo {author} {\bibfnamefont {Tobias}\ \bibnamefont {Morawietz}}, \bibinfo {author} {\bibfnamefont {Ondrej}\ \bibnamefont {Marsalek}}, \bibinfo {author} {\bibfnamefont {Shannon~R.}\ \bibnamefont {Pattenaude}}, \bibinfo {author} {\bibfnamefont {Louis~M.}\ \bibnamefont {Streacker}}, \bibinfo {author} {\bibfnamefont {Dor}\ \bibnamefont {{Ben-Amotz}}}, \ and\ \bibinfo {author} {\bibfnamefont {Thomas~E.}\ \bibnamefont {Markland}},\ }\bibfield  {title} {\enquote {\bibinfo {title} {The interplay of structure and dynamics in the raman spectrum of liquid water over the full frequency and temperature range},}\ }\href {\doibase 10.1021/acs.jpclett.8b00133} {\bibfield  {journal} {\bibinfo  {journal} {The Journal of Physical Chemistry Letters}\ }\textbf {\bibinfo {volume} {9}},\ \bibinfo {pages} {851--857} (\bibinfo {year} {2018})}\BibitemShut {NoStop}%
\bibitem [{\citenamefont {Marsalek}\ and\ \citenamefont {Markland}(2017)}]{marsalek2017quantum}%
  \BibitemOpen
  \bibfield  {author} {\bibinfo {author} {\bibfnamefont {Ondrej}\ \bibnamefont {Marsalek}}\ and\ \bibinfo {author} {\bibfnamefont {Thomas~E}\ \bibnamefont {Markland}},\ }\bibfield  {title} {\enquote {\bibinfo {title} {Quantum dynamics and spectroscopy of ab initio liquid water: The interplay of nuclear and electronic quantum effects},}\ }\href@noop {} {\bibfield  {journal} {\bibinfo  {journal} {The journal of physical chemistry letters}\ }\textbf {\bibinfo {volume} {8}},\ \bibinfo {pages} {1545--1551} (\bibinfo {year} {2017})}\BibitemShut {NoStop}%
\bibitem [{\citenamefont {Brooker}\ \emph {et~al.}(1989)\citenamefont {Brooker}, \citenamefont {Hancock}, \citenamefont {Rice},\ and\ \citenamefont {Shapter}}]{Brooker1989Ramana}%
  \BibitemOpen
  \bibfield  {author} {\bibinfo {author} {\bibfnamefont {M.~H.}\ \bibnamefont {Brooker}}, \bibinfo {author} {\bibfnamefont {G.}~\bibnamefont {Hancock}}, \bibinfo {author} {\bibfnamefont {B.~C.}\ \bibnamefont {Rice}}, \ and\ \bibinfo {author} {\bibfnamefont {J.}~\bibnamefont {Shapter}},\ }\bibfield  {title} {\enquote {\bibinfo {title} {Raman frequency and intensity studies of liquid h{\textsubscript{2}} o, h{\textsubscript{2}}{\textsuperscript{18}} o and d{\textsubscript{2}} o},}\ }\href {\doibase 10.1002/jrs.1250201009} {\bibfield  {journal} {\bibinfo  {journal} {Journal of Raman Spectroscopy}\ }\textbf {\bibinfo {volume} {20}},\ \bibinfo {pages} {683--694} (\bibinfo {year} {1989})}\BibitemShut {NoStop}%
\bibitem [{\citenamefont {Tang}\ \emph {et~al.}(2020)\citenamefont {Tang}, \citenamefont {Ohto}, \citenamefont {Sun}, \citenamefont {Rouxel}, \citenamefont {Imoto}, \citenamefont {Backus}, \citenamefont {Mukamel}, \citenamefont {Bonn},\ and\ \citenamefont {Nagata}}]{tang2020Molecular}%
  \BibitemOpen
  \bibfield  {author} {\bibinfo {author} {\bibfnamefont {Fujie}\ \bibnamefont {Tang}}, \bibinfo {author} {\bibfnamefont {Tatsuhiko}\ \bibnamefont {Ohto}}, \bibinfo {author} {\bibfnamefont {Shumei}\ \bibnamefont {Sun}}, \bibinfo {author} {\bibfnamefont {J{\'e}r{\'e}my~R.}\ \bibnamefont {Rouxel}}, \bibinfo {author} {\bibfnamefont {Sho}\ \bibnamefont {Imoto}}, \bibinfo {author} {\bibfnamefont {Ellen H.~G.}\ \bibnamefont {Backus}}, \bibinfo {author} {\bibfnamefont {Shaul}\ \bibnamefont {Mukamel}}, \bibinfo {author} {\bibfnamefont {Mischa}\ \bibnamefont {Bonn}}, \ and\ \bibinfo {author} {\bibfnamefont {Yuki}\ \bibnamefont {Nagata}},\ }\bibfield  {title} {\enquote {\bibinfo {title} {Molecular structure and modeling of water--air and ice--air interfaces monitored by sum-frequency generation},}\ }\href {\doibase 10.1021/acs.chemrev.9b00512} {\bibfield  {journal} {\bibinfo  {journal} {Chemical Reviews}\ }\textbf {\bibinfo {volume} {120}},\ \bibinfo {pages} {3633--3667} (\bibinfo {year} {2020})}\BibitemShut {NoStop}%
\bibitem [{\citenamefont {Sun}\ \emph {et~al.}(2016)\citenamefont {Sun}, \citenamefont {Liang}, \citenamefont {Xu}, \citenamefont {Zhu}, \citenamefont {Shen},\ and\ \citenamefont {Tian}}]{sun2016Phase}%
  \BibitemOpen
  \bibfield  {author} {\bibinfo {author} {\bibfnamefont {Shumei}\ \bibnamefont {Sun}}, \bibinfo {author} {\bibfnamefont {Rongda}\ \bibnamefont {Liang}}, \bibinfo {author} {\bibfnamefont {Xiaofan}\ \bibnamefont {Xu}}, \bibinfo {author} {\bibfnamefont {Heyuan}\ \bibnamefont {Zhu}}, \bibinfo {author} {\bibfnamefont {Y.~Ron}\ \bibnamefont {Shen}}, \ and\ \bibinfo {author} {\bibfnamefont {Chuanshan}\ \bibnamefont {Tian}},\ }\bibfield  {title} {\enquote {\bibinfo {title} {Phase reference in phase-sensitive sum-frequency vibrational spectroscopy},}\ }\href {\doibase 10.1063/1.4954824} {\bibfield  {journal} {\bibinfo  {journal} {The Journal of Chemical Physics}\ }\textbf {\bibinfo {volume} {144}},\ \bibinfo {pages} {244711} (\bibinfo {year} {2016})}\BibitemShut {NoStop}%
\bibitem [{\citenamefont {Litman}\ \emph {et~al.}(2023)\citenamefont {Litman}, \citenamefont {Lan}, \citenamefont {Nagata},\ and\ \citenamefont {Wilkins}}]{Litman2023Fully}%
  \BibitemOpen
  \bibfield  {author} {\bibinfo {author} {\bibfnamefont {Yair}\ \bibnamefont {Litman}}, \bibinfo {author} {\bibfnamefont {Jinggang}\ \bibnamefont {Lan}}, \bibinfo {author} {\bibfnamefont {Yuki}\ \bibnamefont {Nagata}}, \ and\ \bibinfo {author} {\bibfnamefont {David~M.}\ \bibnamefont {Wilkins}},\ }\bibfield  {title} {\enquote {\bibinfo {title} {Fully first-principles surface spectroscopy with machine learning},}\ }\href {\doibase 10.1021/acs.jpclett.3c01989} {\bibfield  {journal} {\bibinfo  {journal} {The Journal of Physical Chemistry Letters}\ }\textbf {\bibinfo {volume} {14}},\ \bibinfo {pages} {8175--8182} (\bibinfo {year} {2023})}\BibitemShut {NoStop}%
\bibitem [{\citenamefont {Sommers}\ \emph {et~al.}(2020)\citenamefont {Sommers}, \citenamefont {Andrade}, \citenamefont {Zhang}, \citenamefont {Wang},\ and\ \citenamefont {Car}}]{Sommers2020Raman}%
  \BibitemOpen
  \bibfield  {author} {\bibinfo {author} {\bibfnamefont {Grace~M.}\ \bibnamefont {Sommers}}, \bibinfo {author} {\bibfnamefont {Marcos F.~Calegari}\ \bibnamefont {Andrade}}, \bibinfo {author} {\bibfnamefont {Linfeng}\ \bibnamefont {Zhang}}, \bibinfo {author} {\bibfnamefont {Han}\ \bibnamefont {Wang}}, \ and\ \bibinfo {author} {\bibfnamefont {Roberto}\ \bibnamefont {Car}},\ }\bibfield  {title} {\enquote {\bibinfo {title} {Raman spectrum and polarizability of liquid water from deep neural networks},}\ }\href {\doibase 10.1039/D0CP01893G} {\bibfield  {journal} {\bibinfo  {journal} {Physical Chemistry Chemical Physics}\ }\textbf {\bibinfo {volume} {22}},\ \bibinfo {pages} {10592--10602} (\bibinfo {year} {2020})}\BibitemShut {NoStop}%
\bibitem [{\citenamefont {Wang}\ \emph {et~al.}(2024)\citenamefont {Wang}, \citenamefont {Wang}, \citenamefont {Zhang}, \citenamefont {Yang}, \citenamefont {Liang}, \citenamefont {Shi}, \citenamefont {Wang}, \citenamefont {Xing},\ and\ \citenamefont {Sun}}]{wang2024n}%
  \BibitemOpen
  \bibfield  {author} {\bibinfo {author} {\bibfnamefont {Junjie}\ \bibnamefont {Wang}}, \bibinfo {author} {\bibfnamefont {Yong}\ \bibnamefont {Wang}}, \bibinfo {author} {\bibfnamefont {Haoting}\ \bibnamefont {Zhang}}, \bibinfo {author} {\bibfnamefont {Ziyang}\ \bibnamefont {Yang}}, \bibinfo {author} {\bibfnamefont {Zhixin}\ \bibnamefont {Liang}}, \bibinfo {author} {\bibfnamefont {Jiuyang}\ \bibnamefont {Shi}}, \bibinfo {author} {\bibfnamefont {Hui-Tian}\ \bibnamefont {Wang}}, \bibinfo {author} {\bibfnamefont {Dingyu}\ \bibnamefont {Xing}}, \ and\ \bibinfo {author} {\bibfnamefont {Jian}\ \bibnamefont {Sun}},\ }\bibfield  {title} {\enquote {\bibinfo {title} {E (n)-equivariant cartesian tensor message passing interatomic potential},}\ }\href@noop {} {\bibfield  {journal} {\bibinfo  {journal} {Nature communications}\ }\textbf {\bibinfo {volume} {15}},\ \bibinfo {pages} {7607} (\bibinfo {year} {2024})}\BibitemShut {NoStop}%
\bibitem [{\citenamefont {Medders}\ and\ \citenamefont {Paesani}(2015)}]{medders2015Infrareda}%
  \BibitemOpen
  \bibfield  {author} {\bibinfo {author} {\bibfnamefont {Gregory~R.}\ \bibnamefont {Medders}}\ and\ \bibinfo {author} {\bibfnamefont {Francesco}\ \bibnamefont {Paesani}},\ }\bibfield  {title} {\enquote {\bibinfo {title} {Infrared and raman spectroscopy of liquid water through ``first-principles'' many-body molecular dynamics},}\ }\href {\doibase 10.1021/ct501131j} {\bibfield  {journal} {\bibinfo  {journal} {Journal of Chemical Theory and Computation}\ }\textbf {\bibinfo {volume} {11}},\ \bibinfo {pages} {1145--1154} (\bibinfo {year} {2015})}\BibitemShut {NoStop}%
\bibitem [{\citenamefont {Cassone}\ \emph {et~al.}(2019)\citenamefont {Cassone}, \citenamefont {Sponer}, \citenamefont {Trusso},\ and\ \citenamefont {Saija}}]{cassone2019ab}%
  \BibitemOpen
  \bibfield  {author} {\bibinfo {author} {\bibfnamefont {Giuseppe}\ \bibnamefont {Cassone}}, \bibinfo {author} {\bibfnamefont {Jiri}\ \bibnamefont {Sponer}}, \bibinfo {author} {\bibfnamefont {Sebastiano}\ \bibnamefont {Trusso}}, \ and\ \bibinfo {author} {\bibfnamefont {Franz}\ \bibnamefont {Saija}},\ }\bibfield  {title} {\enquote {\bibinfo {title} {Ab initio spectroscopy of water under electric fields},}\ }\href@noop {} {\bibfield  {journal} {\bibinfo  {journal} {Physical Chemistry Chemical Physics}\ }\textbf {\bibinfo {volume} {21}},\ \bibinfo {pages} {21205--21212} (\bibinfo {year} {2019})}\BibitemShut {NoStop}%
\bibitem [{\citenamefont {Wan}\ \emph {et~al.}(2013)\citenamefont {Wan}, \citenamefont {Spanu}, \citenamefont {Galli},\ and\ \citenamefont {Gygi}}]{wan2013Raman}%
  \BibitemOpen
  \bibfield  {author} {\bibinfo {author} {\bibfnamefont {Quan}\ \bibnamefont {Wan}}, \bibinfo {author} {\bibfnamefont {Leonardo}\ \bibnamefont {Spanu}}, \bibinfo {author} {\bibfnamefont {Giulia~A.}\ \bibnamefont {Galli}}, \ and\ \bibinfo {author} {\bibfnamefont {Fran{\c c}ois}\ \bibnamefont {Gygi}},\ }\bibfield  {title} {\enquote {\bibinfo {title} {Raman spectra of liquid water from ab initio molecular dynamics: Vibrational signatures of charge fluctuations in the hydrogen bond network},}\ }\href {\doibase 10.1021/ct4005307} {\bibfield  {journal} {\bibinfo  {journal} {Journal of Chemical Theory and Computation}\ }\textbf {\bibinfo {volume} {9}},\ \bibinfo {pages} {4124--4130} (\bibinfo {year} {2013})}\BibitemShut {NoStop}%
\bibitem [{\citenamefont {LaCour}\ \emph {et~al.}(2023)\citenamefont {LaCour}, \citenamefont {Heindel},\ and\ \citenamefont {{Head-Gordon}}}]{LaCour2023Predicting}%
  \BibitemOpen
  \bibfield  {author} {\bibinfo {author} {\bibfnamefont {R.~Allen}\ \bibnamefont {LaCour}}, \bibinfo {author} {\bibfnamefont {Joseph~P.}\ \bibnamefont {Heindel}}, \ and\ \bibinfo {author} {\bibfnamefont {Teresa}\ \bibnamefont {{Head-Gordon}}},\ }\bibfield  {title} {\enquote {\bibinfo {title} {Predicting the raman spectra of liquid water with a monomer-field model},}\ }\href {\doibase 10.1021/acs.jpclett.3c02873} {\bibfield  {journal} {\bibinfo  {journal} {The Journal of Physical Chemistry Letters}\ }\textbf {\bibinfo {volume} {14}},\ \bibinfo {pages} {11742--11749} (\bibinfo {year} {2023})}\BibitemShut {NoStop}%
\bibitem [{\citenamefont {Kapil}\ \emph {et~al.}(2024)\citenamefont {Kapil}, \citenamefont {P{\'e}ter~Kov{\'a}cs}, \citenamefont {Cs{\'a}nyi},\ and\ \citenamefont {Michaelides}}]{Kapil2024Firstprinciples}%
  \BibitemOpen
  \bibfield  {author} {\bibinfo {author} {\bibfnamefont {Venkat}\ \bibnamefont {Kapil}}, \bibinfo {author} {\bibfnamefont {D{\'a}vid}\ \bibnamefont {P{\'e}ter~Kov{\'a}cs}}, \bibinfo {author} {\bibfnamefont {G{\'a}bor}\ \bibnamefont {Cs{\'a}nyi}}, \ and\ \bibinfo {author} {\bibfnamefont {Angelos}\ \bibnamefont {Michaelides}},\ }\bibfield  {title} {\enquote {\bibinfo {title} {First-principles spectroscopy of aqueous interfaces using machine-learned electronic and quantum nuclear effects},}\ }\href {\doibase 10.1039/D3FD00113J} {\bibfield  {journal} {\bibinfo  {journal} {Faraday Discussions}\ }\textbf {\bibinfo {volume} {249}},\ \bibinfo {pages} {50--68} (\bibinfo {year} {2024})}\BibitemShut {NoStop}%
\bibitem [{\citenamefont {Moberg}\ \emph {et~al.}(2018)\citenamefont {Moberg}, \citenamefont {Straight},\ and\ \citenamefont {Paesani}}]{moberg2018Temperature}%
  \BibitemOpen
  \bibfield  {author} {\bibinfo {author} {\bibfnamefont {Daniel~R.}\ \bibnamefont {Moberg}}, \bibinfo {author} {\bibfnamefont {Shelby~C.}\ \bibnamefont {Straight}}, \ and\ \bibinfo {author} {\bibfnamefont {Francesco}\ \bibnamefont {Paesani}},\ }\bibfield  {title} {\enquote {\bibinfo {title} {Temperature dependence of the air/water interface revealed by polarization sensitive sum-frequency generation spectroscopy},}\ }\href {\doibase 10.1021/acs.jpcb.8b01726} {\bibfield  {journal} {\bibinfo  {journal} {The Journal of Physical Chemistry B}\ }\textbf {\bibinfo {volume} {122}},\ \bibinfo {pages} {4356--4365} (\bibinfo {year} {2018})}\BibitemShut {NoStop}%
\bibitem [{\citenamefont {Murphy}\ \emph {et~al.}(1989)\citenamefont {Murphy}, \citenamefont {Brooker}, \citenamefont {Nielsen}, \citenamefont {Praestgaard},\ and\ \citenamefont {Bertie}}]{Murphy1989Further}%
  \BibitemOpen
  \bibfield  {author} {\bibinfo {author} {\bibfnamefont {W.~F.}\ \bibnamefont {Murphy}}, \bibinfo {author} {\bibfnamefont {M.~H.}\ \bibnamefont {Brooker}}, \bibinfo {author} {\bibfnamefont {O.~Faurskov}\ \bibnamefont {Nielsen}}, \bibinfo {author} {\bibfnamefont {E.}~\bibnamefont {Praestgaard}}, \ and\ \bibinfo {author} {\bibfnamefont {John~E.}\ \bibnamefont {Bertie}},\ }\bibfield  {title} {\enquote {\bibinfo {title} {Further assessment of reduction procedures for raman spectra},}\ }\href {\doibase 10.1002/jrs.1250201010} {\bibfield  {journal} {\bibinfo  {journal} {Journal of Raman Spectroscopy}\ }\textbf {\bibinfo {volume} {20}},\ \bibinfo {pages} {695--699} (\bibinfo {year} {1989})}\BibitemShut {NoStop}%
\bibitem [{\citenamefont {McCoy}(2014)}]{McCoy2014Role}%
  \BibitemOpen
  \bibfield  {author} {\bibinfo {author} {\bibfnamefont {Anne~B.}\ \bibnamefont {McCoy}},\ }\bibfield  {title} {\enquote {\bibinfo {title} {The role of electrical anharmonicity in the association band in the water spectrum},}\ }\href {\doibase 10.1021/jp501647e} {\bibfield  {journal} {\bibinfo  {journal} {The Journal of Physical Chemistry B}\ }\textbf {\bibinfo {volume} {118}},\ \bibinfo {pages} {8286--8294} (\bibinfo {year} {2014})}\BibitemShut {NoStop}%
\bibitem [{\citenamefont {Mar{\'e}chal}(2011)}]{marechal2011molecular}%
  \BibitemOpen
  \bibfield  {author} {\bibinfo {author} {\bibfnamefont {Yves}\ \bibnamefont {Mar{\'e}chal}},\ }\bibfield  {title} {\enquote {\bibinfo {title} {The molecular structure of liquid water delivered by absorption spectroscopy in the whole ir region completed with thermodynamics data},}\ }\href {\doibase 10.1016/j.molstruc.2011.07.054} {\bibfield  {journal} {\bibinfo  {journal} {Journal of Molecular Structure}\ }\textbf {\bibinfo {volume} {1004}},\ \bibinfo {pages} {146--155} (\bibinfo {year} {2011})}\BibitemShut {NoStop}%
\bibitem [{\citenamefont {Walrafen}\ and\ \citenamefont {Pugh}(2004)}]{Walrafen2004Raman}%
  \BibitemOpen
  \bibfield  {author} {\bibinfo {author} {\bibfnamefont {G.~E.}\ \bibnamefont {Walrafen}}\ and\ \bibinfo {author} {\bibfnamefont {Elijah}\ \bibnamefont {Pugh}},\ }\bibfield  {title} {\enquote {\bibinfo {title} {Raman combinations and stretching overtones from water, heavy water, and nacl in water at shifts to ca. 7000 cm-1},}\ }\href {\doibase 10.1023/B:JOSL.0000026646.33891.a8} {\bibfield  {journal} {\bibinfo  {journal} {Journal of Solution Chemistry}\ }\textbf {\bibinfo {volume} {33}},\ \bibinfo {pages} {81--97} (\bibinfo {year} {2004})}\BibitemShut {NoStop}%
\bibitem [{\citenamefont {Ojha}\ \emph {et~al.}(2018)\citenamefont {Ojha}, \citenamefont {Karhan},\ and\ \citenamefont {K{\"u}hne}}]{ojha2018Hydrogen}%
  \BibitemOpen
  \bibfield  {author} {\bibinfo {author} {\bibfnamefont {Deepak}\ \bibnamefont {Ojha}}, \bibinfo {author} {\bibfnamefont {Kristof}\ \bibnamefont {Karhan}}, \ and\ \bibinfo {author} {\bibfnamefont {Thomas~D.}\ \bibnamefont {K{\"u}hne}},\ }\bibfield  {title} {\enquote {\bibinfo {title} {On the hydrogen bond strength and vibrational spectroscopy of liquid water},}\ }\href {\doibase 10.1038/s41598-018-35357-9} {\bibfield  {journal} {\bibinfo  {journal} {Scientific Reports}\ }\textbf {\bibinfo {volume} {8}},\ \bibinfo {pages} {16888} (\bibinfo {year} {2018})}\BibitemShut {NoStop}%
\bibitem [{\citenamefont {Wang}\ \emph {et~al.}(2004)\citenamefont {Wang}, \citenamefont {Pakoulev}, \citenamefont {Pang},\ and\ \citenamefont {Dlott}}]{wang2004Vibrational}%
  \BibitemOpen
  \bibfield  {author} {\bibinfo {author} {\bibfnamefont {Zhaohui}\ \bibnamefont {Wang}}, \bibinfo {author} {\bibfnamefont {Andrei}\ \bibnamefont {Pakoulev}}, \bibinfo {author} {\bibfnamefont {Yoonsoo}\ \bibnamefont {Pang}}, \ and\ \bibinfo {author} {\bibfnamefont {Dana~D.}\ \bibnamefont {Dlott}},\ }\bibfield  {title} {\enquote {\bibinfo {title} {Vibrational substructure in the oh stretching transition of water and hod},}\ }\href {\doibase 10.1021/jp048545t} {\bibfield  {journal} {\bibinfo  {journal} {The Journal of Physical Chemistry A}\ }\textbf {\bibinfo {volume} {108}},\ \bibinfo {pages} {9054--9063} (\bibinfo {year} {2004})}\BibitemShut {NoStop}%
\bibitem [{\citenamefont {Scherer}\ \emph {et~al.}(1974)\citenamefont {Scherer}, \citenamefont {Go},\ and\ \citenamefont {Kint}}]{Scherer1974Raman}%
  \BibitemOpen
  \bibfield  {author} {\bibinfo {author} {\bibfnamefont {James~R.}\ \bibnamefont {Scherer}}, \bibinfo {author} {\bibfnamefont {Man~K.}\ \bibnamefont {Go}}, \ and\ \bibinfo {author} {\bibfnamefont {Saima}\ \bibnamefont {Kint}},\ }\bibfield  {title} {\enquote {\bibinfo {title} {Raman spectra and structure of water from -10 to 90.deg.}}\ }\href {\doibase 10.1021/j100606a013} {\bibfield  {journal} {\bibinfo  {journal} {The Journal of Physical Chemistry}\ }\textbf {\bibinfo {volume} {78}},\ \bibinfo {pages} {1304--1313} (\bibinfo {year} {1974})}\BibitemShut {NoStop}%
\bibitem [{\citenamefont {Reddy}\ \emph {et~al.}(2017)\citenamefont {Reddy}, \citenamefont {Moberg}, \citenamefont {Straight},\ and\ \citenamefont {Paesani}}]{Reddy2017Temperaturedependent}%
  \BibitemOpen
  \bibfield  {author} {\bibinfo {author} {\bibfnamefont {Sandeep~K.}\ \bibnamefont {Reddy}}, \bibinfo {author} {\bibfnamefont {Daniel~R.}\ \bibnamefont {Moberg}}, \bibinfo {author} {\bibfnamefont {Shelby~C.}\ \bibnamefont {Straight}}, \ and\ \bibinfo {author} {\bibfnamefont {Francesco}\ \bibnamefont {Paesani}},\ }\bibfield  {title} {\enquote {\bibinfo {title} {Temperature-dependent vibrational spectra and structure of liquid water from classical and quantum simulations with the mb-pol potential energy function},}\ }\href {\doibase 10.1063/1.5006480} {\bibfield  {journal} {\bibinfo  {journal} {The Journal of Chemical Physics}\ }\textbf {\bibinfo {volume} {147}},\ \bibinfo {pages} {244504} (\bibinfo {year} {2017})}\BibitemShut {NoStop}%
\bibitem [{\citenamefont {Walrafen}\ \emph {et~al.}(1986)\citenamefont {Walrafen}, \citenamefont {Hokmabadi},\ and\ \citenamefont {Yang}}]{Walrafen1986Raman}%
  \BibitemOpen
  \bibfield  {author} {\bibinfo {author} {\bibfnamefont {G.~E.}\ \bibnamefont {Walrafen}}, \bibinfo {author} {\bibfnamefont {M.~S.}\ \bibnamefont {Hokmabadi}}, \ and\ \bibinfo {author} {\bibfnamefont {W.-H.}\ \bibnamefont {Yang}},\ }\bibfield  {title} {\enquote {\bibinfo {title} {Raman isosbestic points from liquid water},}\ }\href {\doibase 10.1063/1.451383} {\bibfield  {journal} {\bibinfo  {journal} {The Journal of Chemical Physics}\ }\textbf {\bibinfo {volume} {85}},\ \bibinfo {pages} {6964--6969} (\bibinfo {year} {1986})}\BibitemShut {NoStop}%
\bibitem [{\citenamefont {Geissler}(2013)}]{Geissler2013Water}%
  \BibitemOpen
  \bibfield  {author} {\bibinfo {author} {\bibfnamefont {Phillip~L.}\ \bibnamefont {Geissler}},\ }\bibfield  {title} {\enquote {\bibinfo {title} {Water interfaces, solvation, and spectroscopy},}\ }\href {\doibase 10.1146/annurev-physchem-040412-110153} {\bibfield  {journal} {\bibinfo  {journal} {Annual Review of Physical Chemistry}\ }\textbf {\bibinfo {volume} {64}},\ \bibinfo {pages} {317--337} (\bibinfo {year} {2013})}\BibitemShut {NoStop}%
\bibitem [{\citenamefont {Geissler}(2005)}]{Geissler2005Temperature}%
  \BibitemOpen
  \bibfield  {author} {\bibinfo {author} {\bibfnamefont {Phillip~L.}\ \bibnamefont {Geissler}},\ }\bibfield  {title} {\enquote {\bibinfo {title} {Temperature dependence of inhomogeneous broadening: On the meaning of isosbestic points},}\ }\href {\doibase 10.1021/ja0545214} {\bibfield  {journal} {\bibinfo  {journal} {Journal of the American Chemical Society}\ }\textbf {\bibinfo {volume} {127}},\ \bibinfo {pages} {14930--14935} (\bibinfo {year} {2005})}\BibitemShut {NoStop}%
\bibitem [{\citenamefont {Niblett}\ \emph {et~al.}(2021)\citenamefont {Niblett}, \citenamefont {Galib},\ and\ \citenamefont {Limmer}}]{niblett2021}%
  \BibitemOpen
  \bibfield  {author} {\bibinfo {author} {\bibfnamefont {Samuel~P.}\ \bibnamefont {Niblett}}, \bibinfo {author} {\bibfnamefont {Mirza}\ \bibnamefont {Galib}}, \ and\ \bibinfo {author} {\bibfnamefont {David~T.}\ \bibnamefont {Limmer}},\ }\bibfield  {title} {\enquote {\bibinfo {title} {Learning intermolecular forces at liquid–vapor interfaces},}\ }\href@noop {} {\bibfield  {journal} {\bibinfo  {journal} {The Journal of Chemical Physics}\ }\textbf {\bibinfo {volume} {155}},\ \bibinfo {pages} {164101} (\bibinfo {year} {2021})}\BibitemShut {NoStop}%
\bibitem [{\citenamefont {Ji}\ \emph {et~al.}(2008)\citenamefont {Ji}, \citenamefont {Ostroverkhov}, \citenamefont {Tian},\ and\ \citenamefont {Shen}}]{Ji2008Characterization}%
  \BibitemOpen
  \bibfield  {author} {\bibinfo {author} {\bibfnamefont {N.}~\bibnamefont {Ji}}, \bibinfo {author} {\bibfnamefont {V.}~\bibnamefont {Ostroverkhov}}, \bibinfo {author} {\bibfnamefont {C.~S.}\ \bibnamefont {Tian}}, \ and\ \bibinfo {author} {\bibfnamefont {Y.~R.}\ \bibnamefont {Shen}},\ }\bibfield  {title} {\enquote {\bibinfo {title} {Characterization of vibrational resonances of water-vapor interfaces by phase-sensitive sum-frequency spectroscopy},}\ }\href {\doibase 10.1103/PhysRevLett.100.096102} {\bibfield  {journal} {\bibinfo  {journal} {Physical Review Letters}\ }\textbf {\bibinfo {volume} {100}},\ \bibinfo {pages} {096102} (\bibinfo {year} {2008})}\BibitemShut {NoStop}%
\bibitem [{\citenamefont {Bonn}\ \emph {et~al.}(2015)\citenamefont {Bonn}, \citenamefont {Nagata},\ and\ \citenamefont {Backus}}]{Bonn2015Molecular}%
  \BibitemOpen
  \bibfield  {author} {\bibinfo {author} {\bibfnamefont {Mischa}\ \bibnamefont {Bonn}}, \bibinfo {author} {\bibfnamefont {Yuki}\ \bibnamefont {Nagata}}, \ and\ \bibinfo {author} {\bibfnamefont {Ellen H.~G.}\ \bibnamefont {Backus}},\ }\bibfield  {title} {\enquote {\bibinfo {title} {Molecular structure and dynamics of water at the water--air interface studied with surface-specific vibrational spectroscopy},}\ }\href {\doibase 10.1002/anie.201411188} {\bibfield  {journal} {\bibinfo  {journal} {Angewandte Chemie International Edition}\ }\textbf {\bibinfo {volume} {54}},\ \bibinfo {pages} {5560--5576} (\bibinfo {year} {2015})}\BibitemShut {NoStop}%
\bibitem [{\citenamefont {Du}\ \emph {et~al.}(1993)\citenamefont {Du}, \citenamefont {Superfine}, \citenamefont {Freysz},\ and\ \citenamefont {Shen}}]{Du1993Vibrational}%
  \BibitemOpen
  \bibfield  {author} {\bibinfo {author} {\bibfnamefont {Q.}~\bibnamefont {Du}}, \bibinfo {author} {\bibfnamefont {R.}~\bibnamefont {Superfine}}, \bibinfo {author} {\bibfnamefont {E.}~\bibnamefont {Freysz}}, \ and\ \bibinfo {author} {\bibfnamefont {Y.~R.}\ \bibnamefont {Shen}},\ }\bibfield  {title} {\enquote {\bibinfo {title} {Vibrational spectroscopy of water at the vapor/water interface},}\ }\href {\doibase 10.1103/PhysRevLett.70.2313} {\bibfield  {journal} {\bibinfo  {journal} {Physical Review Letters}\ }\textbf {\bibinfo {volume} {70}},\ \bibinfo {pages} {2313--2316} (\bibinfo {year} {1993})}\BibitemShut {NoStop}%
\bibitem [{\citenamefont {Whitfield}\ \emph {et~al.}(2016)\citenamefont {Whitfield}, \citenamefont {Herron}, \citenamefont {Guise}, \citenamefont {Page}, \citenamefont {Cheng}, \citenamefont {Milas},\ and\ \citenamefont {Crawford}}]{whitfield2016structures-b43}%
  \BibitemOpen
  \bibfield  {author} {\bibinfo {author} {\bibfnamefont {P.~S.}\ \bibnamefont {Whitfield}}, \bibinfo {author} {\bibfnamefont {N.}~\bibnamefont {Herron}}, \bibinfo {author} {\bibfnamefont {W.~E.}\ \bibnamefont {Guise}}, \bibinfo {author} {\bibfnamefont {K.}~\bibnamefont {Page}}, \bibinfo {author} {\bibfnamefont {Y.~Q.}\ \bibnamefont {Cheng}}, \bibinfo {author} {\bibfnamefont {I.}~\bibnamefont {Milas}}, \ and\ \bibinfo {author} {\bibfnamefont {M.~K.}\ \bibnamefont {Crawford}},\ }\bibfield  {title} {\enquote {\bibinfo {title} {Structures, phase transitions and tricritical behavior of the hybrid perovskite methyl ammonium lead iodide},}\ }\href {\doibase 10.1038/srep35685} {\bibfield  {journal} {\bibinfo  {journal} {Scientific Reports}\ }\textbf {\bibinfo {volume} {6}},\ \bibinfo {pages} {35685} (\bibinfo {year} {2016})}\BibitemShut {NoStop}%
\bibitem [{\citenamefont {Schuck}\ \emph {et~al.}(2018)\citenamefont {Schuck}, \citenamefont {Többens}, \citenamefont {Koch-Müller}, \citenamefont {Efthimiopoulos},\ and\ \citenamefont {Schorr}}]{schuck2018infrared-342}%
  \BibitemOpen
  \bibfield  {author} {\bibinfo {author} {\bibfnamefont {Götz}\ \bibnamefont {Schuck}}, \bibinfo {author} {\bibfnamefont {Daniel~M.}\ \bibnamefont {Többens}}, \bibinfo {author} {\bibfnamefont {Monika}\ \bibnamefont {Koch-Müller}}, \bibinfo {author} {\bibfnamefont {Ilias}\ \bibnamefont {Efthimiopoulos}}, \ and\ \bibinfo {author} {\bibfnamefont {Susan}\ \bibnamefont {Schorr}},\ }\bibfield  {title} {\enquote {\bibinfo {title} {Infrared spectroscopic study of vibrational modes across the orthorhombic–tetragonal phase transition in methylammonium lead halide single crystals},}\ }\href {\doibase 10.1021/acs.jpcc.7b11499} {\bibfield  {journal} {\bibinfo  {journal} {The Journal of Physical Chemistry C}\ }\textbf {\bibinfo {volume} {122}},\ \bibinfo {pages} {5227--5237} (\bibinfo {year} {2018})},\ \Eprint {http://arxiv.org/abs/1803.10721} {1803.10721} \BibitemShut {NoStop}%
\bibitem [{\citenamefont {Sharma}\ \emph {et~al.}(2020)\citenamefont {Sharma}, \citenamefont {Dai}, \citenamefont {Gao}, \citenamefont {Brenner}, \citenamefont {Yadgarov}, \citenamefont {Zhang}, \citenamefont {Rakita}, \citenamefont {Korobko}, \citenamefont {Rappe},\ and\ \citenamefont {Yaffe}}]{sharma2020elucidating}%
  \BibitemOpen
  \bibfield  {author} {\bibinfo {author} {\bibfnamefont {Rituraj}\ \bibnamefont {Sharma}}, \bibinfo {author} {\bibfnamefont {Zhenbang}\ \bibnamefont {Dai}}, \bibinfo {author} {\bibfnamefont {Lingyuan}\ \bibnamefont {Gao}}, \bibinfo {author} {\bibfnamefont {Thomas~M}\ \bibnamefont {Brenner}}, \bibinfo {author} {\bibfnamefont {Lena}\ \bibnamefont {Yadgarov}}, \bibinfo {author} {\bibfnamefont {Jiahao}\ \bibnamefont {Zhang}}, \bibinfo {author} {\bibfnamefont {Yevgeny}\ \bibnamefont {Rakita}}, \bibinfo {author} {\bibfnamefont {Roman}\ \bibnamefont {Korobko}}, \bibinfo {author} {\bibfnamefont {Andrew~M}\ \bibnamefont {Rappe}}, \ and\ \bibinfo {author} {\bibfnamefont {Omer}\ \bibnamefont {Yaffe}},\ }\bibfield  {title} {\enquote {\bibinfo {title} {Elucidating the atomistic origin of anharmonicity in tetragonal ch 3 nh 3 pbi 3 with raman scattering},}\ }\href@noop {} {\bibfield  {journal} {\bibinfo  {journal} {Physical Review Materials}\ }\textbf {\bibinfo {volume} {4}},\ \bibinfo {pages} {092401} (\bibinfo {year}
  {2020})}\BibitemShut {NoStop}%
\bibitem [{\citenamefont {P{\'e}rez-Osorio}\ \emph {et~al.}(2015)\citenamefont {P{\'e}rez-Osorio}, \citenamefont {Milot}, \citenamefont {Filip}, \citenamefont {Patel}, \citenamefont {Herz}, \citenamefont {Johnston},\ and\ \citenamefont {Giustino}}]{perez2015vibrational}%
  \BibitemOpen
  \bibfield  {author} {\bibinfo {author} {\bibfnamefont {Miguel~A}\ \bibnamefont {P{\'e}rez-Osorio}}, \bibinfo {author} {\bibfnamefont {Rebecca~L}\ \bibnamefont {Milot}}, \bibinfo {author} {\bibfnamefont {Marina~R}\ \bibnamefont {Filip}}, \bibinfo {author} {\bibfnamefont {Jay~B}\ \bibnamefont {Patel}}, \bibinfo {author} {\bibfnamefont {Laura~M}\ \bibnamefont {Herz}}, \bibinfo {author} {\bibfnamefont {Michael~B}\ \bibnamefont {Johnston}}, \ and\ \bibinfo {author} {\bibfnamefont {Feliciano}\ \bibnamefont {Giustino}},\ }\bibfield  {title} {\enquote {\bibinfo {title} {Vibrational properties of the organic--inorganic halide perovskite ch3nh3pbi3 from theory and experiment: factor group analysis, first-principles calculations, and low-temperature infrared spectra},}\ }\href@noop {} {\bibfield  {journal} {\bibinfo  {journal} {The Journal of Physical Chemistry C}\ }\textbf {\bibinfo {volume} {119}},\ \bibinfo {pages} {25703--25718} (\bibinfo {year} {2015})}\BibitemShut {NoStop}%
\bibitem [{\citenamefont {Park}\ \emph {et~al.}(2015)\citenamefont {Park}, \citenamefont {Jain}, \citenamefont {Zhang}, \citenamefont {Hagfeldt}, \citenamefont {Boschloo},\ and\ \citenamefont {Edvinsson}}]{park2015resonance}%
  \BibitemOpen
  \bibfield  {author} {\bibinfo {author} {\bibfnamefont {Byung-wook}\ \bibnamefont {Park}}, \bibinfo {author} {\bibfnamefont {Sagar~M}\ \bibnamefont {Jain}}, \bibinfo {author} {\bibfnamefont {Xiaoliang}\ \bibnamefont {Zhang}}, \bibinfo {author} {\bibfnamefont {Anders}\ \bibnamefont {Hagfeldt}}, \bibinfo {author} {\bibfnamefont {Gerrit}\ \bibnamefont {Boschloo}}, \ and\ \bibinfo {author} {\bibfnamefont {Tomas}\ \bibnamefont {Edvinsson}},\ }\bibfield  {title} {\enquote {\bibinfo {title} {Resonance raman and excitation energy dependent charge transfer mechanism in halide-substituted hybrid perovskite solar cells},}\ }\href@noop {} {\bibfield  {journal} {\bibinfo  {journal} {ACS nano}\ }\textbf {\bibinfo {volume} {9}},\ \bibinfo {pages} {2088--2101} (\bibinfo {year} {2015})}\BibitemShut {NoStop}%
\bibitem [{\citenamefont {Dai}(2022)}]{zhenbang2022}%
  \BibitemOpen
  \bibfield  {author} {\bibinfo {author} {\bibfnamefont {Zhenbang}\ \bibnamefont {Dai}},\ }\emph {\bibinfo {title} {First-principles investigations on raman spectroscopy and bulk photovoltaic effect}},\ \href@noop {} {Ph.D. thesis},\ \bibinfo  {school} {University of Pennsylvania} (\bibinfo {year} {2022})\BibitemShut {NoStop}%
\bibitem [{\citenamefont {Quarti}\ \emph {et~al.}(2013)\citenamefont {Quarti}, \citenamefont {Grancini}, \citenamefont {Mosconi}, \citenamefont {Bruno}, \citenamefont {Ball}, \citenamefont {Lee}, \citenamefont {Snaith}, \citenamefont {Petrozza},\ and\ \citenamefont {De~Angelis}}]{quarti2013raman}%
  \BibitemOpen
  \bibfield  {author} {\bibinfo {author} {\bibfnamefont {Claudio}\ \bibnamefont {Quarti}}, \bibinfo {author} {\bibfnamefont {Giulia}\ \bibnamefont {Grancini}}, \bibinfo {author} {\bibfnamefont {Edoardo}\ \bibnamefont {Mosconi}}, \bibinfo {author} {\bibfnamefont {Paola}\ \bibnamefont {Bruno}}, \bibinfo {author} {\bibfnamefont {James~M}\ \bibnamefont {Ball}}, \bibinfo {author} {\bibfnamefont {Michael~M}\ \bibnamefont {Lee}}, \bibinfo {author} {\bibfnamefont {Henry~J}\ \bibnamefont {Snaith}}, \bibinfo {author} {\bibfnamefont {Annamaria}\ \bibnamefont {Petrozza}}, \ and\ \bibinfo {author} {\bibfnamefont {Filippo}\ \bibnamefont {De~Angelis}},\ }\bibfield  {title} {\enquote {\bibinfo {title} {The raman spectrum of the ch3nh3pbi3 hybrid perovskite: interplay of theory and experiment},}\ }\href@noop {} {\bibfield  {journal} {\bibinfo  {journal} {The journal of physical chemistry letters}\ }\textbf {\bibinfo {volume} {5}},\ \bibinfo {pages} {279--284} (\bibinfo {year} {2013})}\BibitemShut {NoStop}%
\bibitem [{\citenamefont {Brivio}\ \emph {et~al.}(2015)\citenamefont {Brivio}, \citenamefont {Frost}, \citenamefont {Skelton}, \citenamefont {Jackson}, \citenamefont {Weber}, \citenamefont {Weller}, \citenamefont {Goni}, \citenamefont {Leguy}, \citenamefont {Barnes},\ and\ \citenamefont {Walsh}}]{brivio2015lattice}%
  \BibitemOpen
  \bibfield  {author} {\bibinfo {author} {\bibfnamefont {Federico}\ \bibnamefont {Brivio}}, \bibinfo {author} {\bibfnamefont {Jarvist~M}\ \bibnamefont {Frost}}, \bibinfo {author} {\bibfnamefont {Jonathan~M}\ \bibnamefont {Skelton}}, \bibinfo {author} {\bibfnamefont {Adam~J}\ \bibnamefont {Jackson}}, \bibinfo {author} {\bibfnamefont {Oliver~J}\ \bibnamefont {Weber}}, \bibinfo {author} {\bibfnamefont {Mark~T}\ \bibnamefont {Weller}}, \bibinfo {author} {\bibfnamefont {Alejandro~R}\ \bibnamefont {Goni}}, \bibinfo {author} {\bibfnamefont {Aur{\'e}lien~MA}\ \bibnamefont {Leguy}}, \bibinfo {author} {\bibfnamefont {Piers~RF}\ \bibnamefont {Barnes}}, \ and\ \bibinfo {author} {\bibfnamefont {Aron}\ \bibnamefont {Walsh}},\ }\bibfield  {title} {\enquote {\bibinfo {title} {Lattice dynamics and vibrational spectra of the orthorhombic, tetragonal, and cubic phases of methylammonium lead iodide},}\ }\href@noop {} {\bibfield  {journal} {\bibinfo  {journal} {Physical Review B}\ }\textbf {\bibinfo {volume} {92}},\ \bibinfo
  {pages} {144308} (\bibinfo {year} {2015})}\BibitemShut {NoStop}%
\bibitem [{\citenamefont {P{\'e}rez-Osorio}\ \emph {et~al.}(2018)\citenamefont {P{\'e}rez-Osorio}, \citenamefont {Lin}, \citenamefont {Phillips}, \citenamefont {Milot}, \citenamefont {Herz}, \citenamefont {Johnston},\ and\ \citenamefont {Giustino}}]{perez2018raman}%
  \BibitemOpen
  \bibfield  {author} {\bibinfo {author} {\bibfnamefont {Miguel~A}\ \bibnamefont {P{\'e}rez-Osorio}}, \bibinfo {author} {\bibfnamefont {Qianqian}\ \bibnamefont {Lin}}, \bibinfo {author} {\bibfnamefont {Richard~T}\ \bibnamefont {Phillips}}, \bibinfo {author} {\bibfnamefont {Rebecca~L}\ \bibnamefont {Milot}}, \bibinfo {author} {\bibfnamefont {Laura~M}\ \bibnamefont {Herz}}, \bibinfo {author} {\bibfnamefont {Michael~B}\ \bibnamefont {Johnston}}, \ and\ \bibinfo {author} {\bibfnamefont {Feliciano}\ \bibnamefont {Giustino}},\ }\bibfield  {title} {\enquote {\bibinfo {title} {Raman spectrum of the organic--inorganic halide perovskite ch3nh3pbi3 from first principles and high-resolution low-temperature raman measurements},}\ }\href@noop {} {\bibfield  {journal} {\bibinfo  {journal} {The Journal of Physical Chemistry C}\ }\textbf {\bibinfo {volume} {122}},\ \bibinfo {pages} {21703--21717} (\bibinfo {year} {2018})}\BibitemShut {NoStop}%
\bibitem [{\citenamefont {Yue}\ \emph {et~al.}(2021)\citenamefont {Yue}, \citenamefont {Muniz}, \citenamefont {Calegari~Andrade}, \citenamefont {Zhang}, \citenamefont {Car},\ and\ \citenamefont {Panagiotopoulos}}]{yue2021short}%
  \BibitemOpen
  \bibfield  {author} {\bibinfo {author} {\bibfnamefont {Shuwen}\ \bibnamefont {Yue}}, \bibinfo {author} {\bibfnamefont {Maria~Carolina}\ \bibnamefont {Muniz}}, \bibinfo {author} {\bibfnamefont {Marcos~F}\ \bibnamefont {Calegari~Andrade}}, \bibinfo {author} {\bibfnamefont {Linfeng}\ \bibnamefont {Zhang}}, \bibinfo {author} {\bibfnamefont {Roberto}\ \bibnamefont {Car}}, \ and\ \bibinfo {author} {\bibfnamefont {Athanassios~Z}\ \bibnamefont {Panagiotopoulos}},\ }\bibfield  {title} {\enquote {\bibinfo {title} {When do short-range atomistic machine-learning models fall short?}}\ }\href {\doibase 10.1063/5.0031215} {\bibfield  {journal} {\bibinfo  {journal} {J. Chem. Phys.}\ }\textbf {\bibinfo {volume} {154}},\ \bibinfo {pages} {034111} (\bibinfo {year} {2021})}\BibitemShut {NoStop}%
\bibitem [{\citenamefont {Fransson}\ \emph {et~al.}(2023)\citenamefont {Fransson}, \citenamefont {Rahm}, \citenamefont {Wiktor},\ and\ \citenamefont {Erhart}}]{fransson2023revealing-ad5}%
  \BibitemOpen
  \bibfield  {author} {\bibinfo {author} {\bibfnamefont {Erik}\ \bibnamefont {Fransson}}, \bibinfo {author} {\bibfnamefont {J.~Magnus}\ \bibnamefont {Rahm}}, \bibinfo {author} {\bibfnamefont {Julia}\ \bibnamefont {Wiktor}}, \ and\ \bibinfo {author} {\bibfnamefont {Paul}\ \bibnamefont {Erhart}},\ }\bibfield  {title} {\enquote {\bibinfo {title} {Revealing the free energy landscape of halide perovskites: Metastability and transition characters in {CsPbBr}3 and {MAPbI}3},}\ }\href {\doibase 10.1021/acs.chemmater.3c01740} {\bibfield  {journal} {\bibinfo  {journal} {Chemistry of Materials}\ }\textbf {\bibinfo {volume} {35}},\ \bibinfo {pages} {8229--8238} (\bibinfo {year} {2023})},\ \Eprint {http://arxiv.org/abs/2307.12100} {2307.12100} \BibitemShut {NoStop}%
\bibitem [{\citenamefont {Jinnouchi}\ \emph {et~al.}(2019)\citenamefont {Jinnouchi}, \citenamefont {Lahnsteiner}, \citenamefont {Karsai}, \citenamefont {Kresse},\ and\ \citenamefont {Bokdam}}]{jinnouchi2019phase-507}%
  \BibitemOpen
  \bibfield  {author} {\bibinfo {author} {\bibfnamefont {Ryosuke}\ \bibnamefont {Jinnouchi}}, \bibinfo {author} {\bibfnamefont {Jonathan}\ \bibnamefont {Lahnsteiner}}, \bibinfo {author} {\bibfnamefont {Ferenc}\ \bibnamefont {Karsai}}, \bibinfo {author} {\bibfnamefont {Georg}\ \bibnamefont {Kresse}}, \ and\ \bibinfo {author} {\bibfnamefont {Menno}\ \bibnamefont {Bokdam}},\ }\bibfield  {title} {\enquote {\bibinfo {title} {Phase transitions of hybrid perovskites simulated by machine-learning force fields trained on the fly with bayesian inference},}\ }\href {\doibase 10.1103/physrevlett.122.225701} {\bibfield  {journal} {\bibinfo  {journal} {Physical Review Letters}\ }\textbf {\bibinfo {volume} {122}},\ \bibinfo {pages} {225701} (\bibinfo {year} {2019})},\ \Eprint {http://arxiv.org/abs/1903.09613} {1903.09613} \BibitemShut {NoStop}%
\bibitem [{\citenamefont {Toriumi}\ \emph {et~al.}(2006)\citenamefont {Toriumi}, \citenamefont {Kita}, \citenamefont {Tomida},\ and\ \citenamefont {Yamamoto}}]{toriumi2006doped-e52}%
  \BibitemOpen
  \bibfield  {author} {\bibinfo {author} {\bibfnamefont {Akira}\ \bibnamefont {Toriumi}}, \bibinfo {author} {\bibfnamefont {Koji}\ \bibnamefont {Kita}}, \bibinfo {author} {\bibfnamefont {Kazuyuki}\ \bibnamefont {Tomida}}, \ and\ \bibinfo {author} {\bibfnamefont {Yoshiki}\ \bibnamefont {Yamamoto}},\ }\bibfield  {title} {\enquote {\bibinfo {title} {Doped {HfO}2 for higher-k dielectrics},}\ }\href {\doibase 10.1149/1.2209268} {\bibfield  {journal} {\bibinfo  {journal} {Electrochemical Society Transactions}\ }\textbf {\bibinfo {volume} {1}},\ \bibinfo {pages} {185--197} (\bibinfo {year} {2006})}\BibitemShut {NoStop}%
\bibitem [{\citenamefont {Zhou}\ \emph {et~al.}(2014)\citenamefont {Zhou}, \citenamefont {Shi}, \citenamefont {Zhang}, \citenamefont {Su},\ and\ \citenamefont {Jiang}}]{zhou2014simulated-42d}%
  \BibitemOpen
  \bibfield  {author} {\bibinfo {author} {\bibfnamefont {B}~\bibnamefont {Zhou}}, \bibinfo {author} {\bibfnamefont {H}~\bibnamefont {Shi}}, \bibinfo {author} {\bibfnamefont {X~D}\ \bibnamefont {Zhang}}, \bibinfo {author} {\bibfnamefont {Q}~\bibnamefont {Su}}, \ and\ \bibinfo {author} {\bibfnamefont {Z~Y}\ \bibnamefont {Jiang}},\ }\bibfield  {title} {\enquote {\bibinfo {title} {The simulated vibrational spectra of {HfO}2 polymorphs},}\ }\href {\doibase 10.1088/0022-3727/47/11/115502} {\bibfield  {journal} {\bibinfo  {journal} {Journal of Physics D: Applied Physics}\ }\textbf {\bibinfo {volume} {47}},\ \bibinfo {pages} {115502} (\bibinfo {year} {2014})}\BibitemShut {NoStop}%
\bibitem [{\citenamefont {Lee}\ \emph {et~al.}(2023)\citenamefont {Lee}, \citenamefont {Lee}, \citenamefont {Cho}, \citenamefont {Choi}, \citenamefont {Kim},\ and\ \citenamefont {Park}}]{lee2023unveiled-8a5}%
  \BibitemOpen
  \bibfield  {author} {\bibinfo {author} {\bibfnamefont {Dong~Hyun}\ \bibnamefont {Lee}}, \bibinfo {author} {\bibfnamefont {Younghwan}\ \bibnamefont {Lee}}, \bibinfo {author} {\bibfnamefont {Yong~Hyeon}\ \bibnamefont {Cho}}, \bibinfo {author} {\bibfnamefont {Hyojun}\ \bibnamefont {Choi}}, \bibinfo {author} {\bibfnamefont {Se~Hyun}\ \bibnamefont {Kim}}, \ and\ \bibinfo {author} {\bibfnamefont {Min~Hyuk}\ \bibnamefont {Park}},\ }\bibfield  {title} {\enquote {\bibinfo {title} {Unveiled ferroelectricity in well‐known non‐ferroelectric materials and their semiconductor applications},}\ }\href {\doibase 10.1002/adfm.202303956} {\bibfield  {journal} {\bibinfo  {journal} {Advanced Functional Materials}\ }\textbf {\bibinfo {volume} {33}} (\bibinfo {year} {2023}),\ 10.1002/adfm.202303956}\BibitemShut {NoStop}%
\bibitem [{\citenamefont {Schroeder}\ \emph {et~al.}(2022)\citenamefont {Schroeder}, \citenamefont {Park}, \citenamefont {Mikolajick},\ and\ \citenamefont {Hwang}}]{schroeder2022fundamentals-ce3}%
  \BibitemOpen
  \bibfield  {author} {\bibinfo {author} {\bibfnamefont {Uwe}\ \bibnamefont {Schroeder}}, \bibinfo {author} {\bibfnamefont {Min~Hyuk}\ \bibnamefont {Park}}, \bibinfo {author} {\bibfnamefont {Thomas}\ \bibnamefont {Mikolajick}}, \ and\ \bibinfo {author} {\bibfnamefont {Cheol~Seong}\ \bibnamefont {Hwang}},\ }\bibfield  {title} {\enquote {\bibinfo {title} {The fundamentals and applications of ferroelectric {HfO}2},}\ }\href {\doibase 10.1038/s41578-022-00431-2} {\bibfield  {journal} {\bibinfo  {journal} {Nature Reviews Materials}\ }\textbf {\bibinfo {volume} {7}},\ \bibinfo {pages} {653--669} (\bibinfo {year} {2022})}\BibitemShut {NoStop}%
\bibitem [{\citenamefont {Zhu}\ \emph {et~al.}(2024)\citenamefont {Zhu}, \citenamefont {Ma}, \citenamefont {Deng},\ and\ \citenamefont {Liu}}]{zhu2024progress-cd2}%
  \BibitemOpen
  \bibfield  {author} {\bibinfo {author} {\bibfnamefont {Tianyuan}\ \bibnamefont {Zhu}}, \bibinfo {author} {\bibfnamefont {Liyang}\ \bibnamefont {Ma}}, \bibinfo {author} {\bibfnamefont {Shiqing}\ \bibnamefont {Deng}}, \ and\ \bibinfo {author} {\bibfnamefont {Shi}\ \bibnamefont {Liu}},\ }\bibfield  {title} {\enquote {\bibinfo {title} {Progress in computational understanding of ferroelectric mechanisms in {HfO}2},}\ }\href {\doibase 10.1038/s41524-024-01352-0} {\bibfield  {journal} {\bibinfo  {journal} {npj Computational Materials}\ }\textbf {\bibinfo {volume} {10}},\ \bibinfo {pages} {188} (\bibinfo {year} {2024})},\ \Eprint {http://arxiv.org/abs/2405.03558} {2405.03558} \BibitemShut {NoStop}%
\bibitem [{\citenamefont {Zhao}\ \emph {et~al.}(2022)\citenamefont {Zhao}, \citenamefont {Chen},\ and\ \citenamefont {Liao}}]{zhao2022microstructural-5dc}%
  \BibitemOpen
  \bibfield  {author} {\bibinfo {author} {\bibfnamefont {Dou}\ \bibnamefont {Zhao}}, \bibinfo {author} {\bibfnamefont {Zibin}\ \bibnamefont {Chen}}, \ and\ \bibinfo {author} {\bibfnamefont {Xiaozhou}\ \bibnamefont {Liao}},\ }\bibfield  {title} {\enquote {\bibinfo {title} {Microstructural evolution and ferroelectricity in {HfO} 2 films},}\ }\href {\doibase 10.20517/microstructures.2021.11} {\bibfield  {journal} {\bibinfo  {journal} {Microstructures}\ }\textbf {\bibinfo {volume} {2}},\ \bibinfo {pages} {2022007} (\bibinfo {year} {2022})}\BibitemShut {NoStop}%
\bibitem [{\citenamefont {Park}\ \emph {et~al.}(2024)\citenamefont {Park}, \citenamefont {Kim}, \citenamefont {Lee}, \citenamefont {Lee}, \citenamefont {Kim}, \citenamefont {Kang}, \citenamefont {Lin}, \citenamefont {Pattison}, \citenamefont {Theis}, \citenamefont {Kim}, \citenamefont {Choi}, \citenamefont {Cho}, \citenamefont {Ercius}, \citenamefont {Lee}, \citenamefont {Chae},\ and\ \citenamefont {Park}}]{park2024atomic-scale-b11}%
  \BibitemOpen
  \bibfield  {author} {\bibinfo {author} {\bibfnamefont {Kunwoo}\ \bibnamefont {Park}}, \bibinfo {author} {\bibfnamefont {Dongmin}\ \bibnamefont {Kim}}, \bibinfo {author} {\bibfnamefont {Kyoungjun}\ \bibnamefont {Lee}}, \bibinfo {author} {\bibfnamefont {Hyun-Jae}\ \bibnamefont {Lee}}, \bibinfo {author} {\bibfnamefont {Jihoon}\ \bibnamefont {Kim}}, \bibinfo {author} {\bibfnamefont {Sungsu}\ \bibnamefont {Kang}}, \bibinfo {author} {\bibfnamefont {Alex}\ \bibnamefont {Lin}}, \bibinfo {author} {\bibfnamefont {Alexander~J.}\ \bibnamefont {Pattison}}, \bibinfo {author} {\bibfnamefont {Wolfgang}\ \bibnamefont {Theis}}, \bibinfo {author} {\bibfnamefont {Chang~Hoon}\ \bibnamefont {Kim}}, \bibinfo {author} {\bibfnamefont {Hyesung}\ \bibnamefont {Choi}}, \bibinfo {author} {\bibfnamefont {Jung~Woo}\ \bibnamefont {Cho}}, \bibinfo {author} {\bibfnamefont {Peter}\ \bibnamefont {Ercius}}, \bibinfo {author} {\bibfnamefont {Jun~Hee}\ \bibnamefont {Lee}}, \bibinfo {author} {\bibfnamefont {Seung~Chul}\ \bibnamefont {Chae}}, \
  and\ \bibinfo {author} {\bibfnamefont {Jungwon}\ \bibnamefont {Park}},\ }\bibfield  {title} {\enquote {\bibinfo {title} {Atomic-scale scanning of domain network in the ferroelectric {HfO}2 thin film},}\ }\href {\doibase 10.1021/acsnano.4c08721} {\bibfield  {journal} {\bibinfo  {journal} {{ACS} Nano}\ }\textbf {\bibinfo {volume} {18}},\ \bibinfo {pages} {26315--26326} (\bibinfo {year} {2024})}\BibitemShut {NoStop}%
\bibitem [{\citenamefont {Chen}\ and\ \citenamefont {Mizoguchi}(2026)}]{chen2026electric-c45}%
  \BibitemOpen
  \bibfield  {author} {\bibinfo {author} {\bibfnamefont {Po-Yen}\ \bibnamefont {Chen}}\ and\ \bibinfo {author} {\bibfnamefont {Teruyasu}\ \bibnamefont {Mizoguchi}},\ }\bibfield  {title} {\enquote {\bibinfo {title} {Electric field-induced phase transitions and hysteresis in ferroelectric {HfO$_2$} captured with machine learning potential},}\ }\href {\doibase 10.1016/j.mtelec.2026.100228} {\bibfield  {journal} {\bibinfo  {journal} {Materials Today Electronics}\ }\textbf {\bibinfo {volume} {17}},\ \bibinfo {pages} {100228} (\bibinfo {year} {2026})}\BibitemShut {NoStop}%
\bibitem [{\citenamefont {Batatia}\ \emph {et~al.}(2023)\citenamefont {Batatia}, \citenamefont {Benner}, \citenamefont {Chiang}, \citenamefont {Elena}, \citenamefont {Kov{\'a}cs}, \citenamefont {Riebesell}, \citenamefont {Advincula}, \citenamefont {Asta}, \citenamefont {Baldwin}, \citenamefont {Bernstein} \emph {et~al.}}]{batatia2023foundation}%
  \BibitemOpen
  \bibfield  {author} {\bibinfo {author} {\bibfnamefont {Ilyes}\ \bibnamefont {Batatia}}, \bibinfo {author} {\bibfnamefont {Philipp}\ \bibnamefont {Benner}}, \bibinfo {author} {\bibfnamefont {Yuan}\ \bibnamefont {Chiang}}, \bibinfo {author} {\bibfnamefont {Alin~M}\ \bibnamefont {Elena}}, \bibinfo {author} {\bibfnamefont {D{\'a}vid~P}\ \bibnamefont {Kov{\'a}cs}}, \bibinfo {author} {\bibfnamefont {Janosh}\ \bibnamefont {Riebesell}}, \bibinfo {author} {\bibfnamefont {Xavier~R}\ \bibnamefont {Advincula}}, \bibinfo {author} {\bibfnamefont {Mark}\ \bibnamefont {Asta}}, \bibinfo {author} {\bibfnamefont {William~J}\ \bibnamefont {Baldwin}}, \bibinfo {author} {\bibfnamefont {Noam}\ \bibnamefont {Bernstein}},  \emph {et~al.},\ }\bibfield  {title} {\enquote {\bibinfo {title} {A foundation model for atomistic materials chemistry},}\ }\href@noop {} {\bibfield  {journal} {\bibinfo  {journal} {arXiv preprint arXiv:2401.00096}\ } (\bibinfo {year} {2023})}\BibitemShut {NoStop}%
\bibitem [{\citenamefont {Gonze}\ and\ \citenamefont {Lee}(1997)}]{gonze1997dynamical-202}%
  \BibitemOpen
  \bibfield  {author} {\bibinfo {author} {\bibfnamefont {Xavier}\ \bibnamefont {Gonze}}\ and\ \bibinfo {author} {\bibfnamefont {Changyol}\ \bibnamefont {Lee}},\ }\bibfield  {title} {\enquote {\bibinfo {title} {Dynamical matrices, born effective charges, dielectric permittivity tensors, and interatomic force constants from density-functional perturbation theory},}\ }\href {\doibase 10.1103/physrevb.55.10355} {\bibfield  {journal} {\bibinfo  {journal} {Physical Review B}\ }\textbf {\bibinfo {volume} {55}},\ \bibinfo {pages} {10355--10368} (\bibinfo {year} {1997})}\BibitemShut {NoStop}%
\bibitem [{\citenamefont {Zhong}\ \emph {et~al.}(1994)\citenamefont {Zhong}, \citenamefont {King-Smith},\ and\ \citenamefont {Vanderbilt}}]{zhong1994giant-6dc}%
  \BibitemOpen
  \bibfield  {author} {\bibinfo {author} {\bibfnamefont {W.}~\bibnamefont {Zhong}}, \bibinfo {author} {\bibfnamefont {R.~D.}\ \bibnamefont {King-Smith}}, \ and\ \bibinfo {author} {\bibfnamefont {David}\ \bibnamefont {Vanderbilt}},\ }\bibfield  {title} {\enquote {\bibinfo {title} {Giant {LO}-{TO} splittings in perovskite ferroelectrics},}\ }\href {\doibase 10.1103/physrevlett.72.3618} {\bibfield  {journal} {\bibinfo  {journal} {Physical Review Letters}\ }\textbf {\bibinfo {volume} {72}},\ \bibinfo {pages} {3618--3621} (\bibinfo {year} {1994})}\BibitemShut {NoStop}%
\bibitem [{\citenamefont {Fan}\ \emph {et~al.}(2022)\citenamefont {Fan}, \citenamefont {Singh}, \citenamefont {Xu}, \citenamefont {Park}, \citenamefont {Qi}, \citenamefont {Cheong}, \citenamefont {Vanderbilt}, \citenamefont {Rabe},\ and\ \citenamefont {Musfeldt}}]{fan2022vibrational-811}%
  \BibitemOpen
  \bibfield  {author} {\bibinfo {author} {\bibfnamefont {Shiyu}\ \bibnamefont {Fan}}, \bibinfo {author} {\bibfnamefont {Sobhit}\ \bibnamefont {Singh}}, \bibinfo {author} {\bibfnamefont {Xianghan}\ \bibnamefont {Xu}}, \bibinfo {author} {\bibfnamefont {Kiman}\ \bibnamefont {Park}}, \bibinfo {author} {\bibfnamefont {Yubo}\ \bibnamefont {Qi}}, \bibinfo {author} {\bibfnamefont {S.~W.}\ \bibnamefont {Cheong}}, \bibinfo {author} {\bibfnamefont {David}\ \bibnamefont {Vanderbilt}}, \bibinfo {author} {\bibfnamefont {Karin~M.}\ \bibnamefont {Rabe}}, \ and\ \bibinfo {author} {\bibfnamefont {J.~L.}\ \bibnamefont {Musfeldt}},\ }\bibfield  {title} {\enquote {\bibinfo {title} {Vibrational fingerprints of ferroelectric {HfO}2},}\ }\href {\doibase 10.1038/s41535-022-00436-8} {\bibfield  {journal} {\bibinfo  {journal} {npj Quantum Materials}\ }\textbf {\bibinfo {volume} {7}},\ \bibinfo {pages} {32} (\bibinfo {year} {2022})}\BibitemShut {NoStop}%
\bibitem [{\citenamefont {Shimizu}\ \emph {et~al.}(2018)\citenamefont {Shimizu}, \citenamefont {Mimura}, \citenamefont {Kiguchi}, \citenamefont {Shiraishi}, \citenamefont {Konno}, \citenamefont {Katsuya}, \citenamefont {Sakata},\ and\ \citenamefont {Funakubo}}]{shimizu2018ferroelectricity-7a5}%
  \BibitemOpen
  \bibfield  {author} {\bibinfo {author} {\bibfnamefont {Takao}\ \bibnamefont {Shimizu}}, \bibinfo {author} {\bibfnamefont {Takanori}\ \bibnamefont {Mimura}}, \bibinfo {author} {\bibfnamefont {Takanori}\ \bibnamefont {Kiguchi}}, \bibinfo {author} {\bibfnamefont {Takahisa}\ \bibnamefont {Shiraishi}}, \bibinfo {author} {\bibfnamefont {Toyohiko}\ \bibnamefont {Konno}}, \bibinfo {author} {\bibfnamefont {Yoshio}\ \bibnamefont {Katsuya}}, \bibinfo {author} {\bibfnamefont {Osami}\ \bibnamefont {Sakata}}, \ and\ \bibinfo {author} {\bibfnamefont {Hiroshi}\ \bibnamefont {Funakubo}},\ }\bibfield  {title} {\enquote {\bibinfo {title} {Ferroelectricity mediated by ferroelastic domain switching in {HfO}2-based epitaxial thin films},}\ }\href {\doibase 10.1063/1.5055258} {\bibfield  {journal} {\bibinfo  {journal} {Applied Physics Letters}\ }\textbf {\bibinfo {volume} {113}},\ \bibinfo {pages} {212901} (\bibinfo {year} {2018})}\BibitemShut {NoStop}%
\bibitem [{\citenamefont {Hu}\ \emph {et~al.}(2025)\citenamefont {Hu}, \citenamefont {Lv}, \citenamefont {Tsai}, \citenamefont {Xue}, \citenamefont {Jing}, \citenamefont {Lin}, \citenamefont {Tong}, \citenamefont {Cao}, \citenamefont {Teobaldi},\ and\ \citenamefont {Liu}}]{hu2025mapping-961}%
  \BibitemOpen
  \bibfield  {author} {\bibinfo {author} {\bibfnamefont {Qi}~\bibnamefont {Hu}}, \bibinfo {author} {\bibfnamefont {Shuning}\ \bibnamefont {Lv}}, \bibinfo {author} {\bibfnamefont {Hsiaoyi}\ \bibnamefont {Tsai}}, \bibinfo {author} {\bibfnamefont {Yufeng}\ \bibnamefont {Xue}}, \bibinfo {author} {\bibfnamefont {Xixiang}\ \bibnamefont {Jing}}, \bibinfo {author} {\bibfnamefont {Fanrong}\ \bibnamefont {Lin}}, \bibinfo {author} {\bibfnamefont {Chuanjia}\ \bibnamefont {Tong}}, \bibinfo {author} {\bibfnamefont {Tengfei}\ \bibnamefont {Cao}}, \bibinfo {author} {\bibfnamefont {Gilberto}\ \bibnamefont {Teobaldi}}, \ and\ \bibinfo {author} {\bibfnamefont {Li-Min}\ \bibnamefont {Liu}},\ }\bibfield  {title} {\enquote {\bibinfo {title} {Mapping of the full polarization switching pathways for {HfO}2 and its implications},}\ }\href {\doibase 10.1073/pnas.2419685122} {\bibfield  {journal} {\bibinfo  {journal} {Proceedings of the National Academy of Sciences}\ }\textbf {\bibinfo {volume} {122}},\ \bibinfo {pages} {e2419685122}
  (\bibinfo {year} {2025})}\BibitemShut {NoStop}%
\bibitem [{\citenamefont {Ma}\ \emph {et~al.}(2023)\citenamefont {Ma}, \citenamefont {Wu}, \citenamefont {Zhu}, \citenamefont {Huang}, \citenamefont {Lu},\ and\ \citenamefont {Liu}}]{ma2023ultrahigh-8db}%
  \BibitemOpen
  \bibfield  {author} {\bibinfo {author} {\bibfnamefont {Liyang}\ \bibnamefont {Ma}}, \bibinfo {author} {\bibfnamefont {Jing}\ \bibnamefont {Wu}}, \bibinfo {author} {\bibfnamefont {Tianyuan}\ \bibnamefont {Zhu}}, \bibinfo {author} {\bibfnamefont {Yiwei}\ \bibnamefont {Huang}}, \bibinfo {author} {\bibfnamefont {Qiyang}\ \bibnamefont {Lu}}, \ and\ \bibinfo {author} {\bibfnamefont {Shi}\ \bibnamefont {Liu}},\ }\bibfield  {title} {\enquote {\bibinfo {title} {Ultrahigh oxygen ion mobility in ferroelectric hafnia},}\ }\href {\doibase 10.1103/physrevlett.131.256801} {\bibfield  {journal} {\bibinfo  {journal} {Physical Review Letters}\ }\textbf {\bibinfo {volume} {131}},\ \bibinfo {pages} {256801} (\bibinfo {year} {2023})},\ \Eprint {http://arxiv.org/abs/2305.02952} {2305.02952} \BibitemShut {NoStop}%
\bibitem [{\citenamefont {Cuevas-Zuvir{\'\i}a}\ and\ \citenamefont {Pacios}(2020)}]{cuevas2020analytical}%
  \BibitemOpen
  \bibfield  {author} {\bibinfo {author} {\bibfnamefont {Bruno}\ \bibnamefont {Cuevas-Zuvir{\'\i}a}}\ and\ \bibinfo {author} {\bibfnamefont {Luis~F}\ \bibnamefont {Pacios}},\ }\bibfield  {title} {\enquote {\bibinfo {title} {Analytical model of electron density and its machine learning inference},}\ }\href@noop {} {\bibfield  {journal} {\bibinfo  {journal} {Journal of Chemical Information and Modeling}\ }\textbf {\bibinfo {volume} {60}},\ \bibinfo {pages} {3831--3842} (\bibinfo {year} {2020})}\BibitemShut {NoStop}%
\bibitem [{\citenamefont {Yuan}\ \emph {et~al.}(2026)\citenamefont {Yuan}, \citenamefont {Liu}, \citenamefont {Chen}, \citenamefont {Zhong}, \citenamefont {Raja}, \citenamefont {Kreiman}, \citenamefont {Vargas}, \citenamefont {Xu}, \citenamefont {Head-Gordon}, \citenamefont {Yang} \emph {et~al.}}]{yuan2026foundation}%
  \BibitemOpen
  \bibfield  {author} {\bibinfo {author} {\bibfnamefont {Eric C-Y}\ \bibnamefont {Yuan}}, \bibinfo {author} {\bibfnamefont {Yunsheng}\ \bibnamefont {Liu}}, \bibinfo {author} {\bibfnamefont {Junmin}\ \bibnamefont {Chen}}, \bibinfo {author} {\bibfnamefont {Peichen}\ \bibnamefont {Zhong}}, \bibinfo {author} {\bibfnamefont {Sanjeev}\ \bibnamefont {Raja}}, \bibinfo {author} {\bibfnamefont {Tobias}\ \bibnamefont {Kreiman}}, \bibinfo {author} {\bibfnamefont {Santiago}\ \bibnamefont {Vargas}}, \bibinfo {author} {\bibfnamefont {Wenbin}\ \bibnamefont {Xu}}, \bibinfo {author} {\bibfnamefont {Martin}\ \bibnamefont {Head-Gordon}}, \bibinfo {author} {\bibfnamefont {Chao}\ \bibnamefont {Yang}},  \emph {et~al.},\ }\bibfield  {title} {\enquote {\bibinfo {title} {Foundation models for atomistic simulation of chemistry and materials},}\ }\href@noop {} {\bibfield  {journal} {\bibinfo  {journal} {Nature Reviews Chemistry}\ ,\ \bibinfo {pages} {1--19}} (\bibinfo {year} {2026})}\BibitemShut {NoStop}%
\bibitem [{\citenamefont {Paszke}\ \emph {et~al.}(2019)\citenamefont {Paszke}, \citenamefont {Gross}, \citenamefont {Massa}, \citenamefont {Lerer}, \citenamefont {Bradbury}, \citenamefont {Chanan}, \citenamefont {Killeen}, \citenamefont {Lin}, \citenamefont {Gimelshein}, \citenamefont {Antiga}, \citenamefont {Desmaison}, \citenamefont {Kopf}, \citenamefont {Yang}, \citenamefont {DeVito}, \citenamefont {Raison}, \citenamefont {Tejani}, \citenamefont {Chilamkurthy}, \citenamefont {Steiner}, \citenamefont {Fang}, \citenamefont {Bai},\ and\ \citenamefont {Chintala}}]{Paszke2019PyTorch}%
  \BibitemOpen
  \bibfield  {author} {\bibinfo {author} {\bibfnamefont {Adam}\ \bibnamefont {Paszke}}, \bibinfo {author} {\bibfnamefont {Sam}\ \bibnamefont {Gross}}, \bibinfo {author} {\bibfnamefont {Francisco}\ \bibnamefont {Massa}}, \bibinfo {author} {\bibfnamefont {Adam}\ \bibnamefont {Lerer}}, \bibinfo {author} {\bibfnamefont {James}\ \bibnamefont {Bradbury}}, \bibinfo {author} {\bibfnamefont {Gregory}\ \bibnamefont {Chanan}}, \bibinfo {author} {\bibfnamefont {Trevor}\ \bibnamefont {Killeen}}, \bibinfo {author} {\bibfnamefont {Zeming}\ \bibnamefont {Lin}}, \bibinfo {author} {\bibfnamefont {Natalia}\ \bibnamefont {Gimelshein}}, \bibinfo {author} {\bibfnamefont {Luca}\ \bibnamefont {Antiga}}, \bibinfo {author} {\bibfnamefont {Alban}\ \bibnamefont {Desmaison}}, \bibinfo {author} {\bibfnamefont {Andreas}\ \bibnamefont {Kopf}}, \bibinfo {author} {\bibfnamefont {Edward}\ \bibnamefont {Yang}}, \bibinfo {author} {\bibfnamefont {Zachary}\ \bibnamefont {DeVito}}, \bibinfo {author} {\bibfnamefont {Martin}\ \bibnamefont {Raison}},
  \bibinfo {author} {\bibfnamefont {Alykhan}\ \bibnamefont {Tejani}}, \bibinfo {author} {\bibfnamefont {Sasank}\ \bibnamefont {Chilamkurthy}}, \bibinfo {author} {\bibfnamefont {Benoit}\ \bibnamefont {Steiner}}, \bibinfo {author} {\bibfnamefont {Lu}~\bibnamefont {Fang}}, \bibinfo {author} {\bibfnamefont {Junjie}\ \bibnamefont {Bai}}, \ and\ \bibinfo {author} {\bibfnamefont {Soumith}\ \bibnamefont {Chintala}},\ }\bibfield  {title} {\enquote {\bibinfo {title} {Pytorch: An imperative style, high-performance deep learning library},}\ }in\ \href@noop {} {\emph {\bibinfo {booktitle} {Advances in Neural Information Processing Systems}}},\ Vol.~\bibinfo {volume} {32}\ (\bibinfo  {publisher} {Curran Associates, Inc.},\ \bibinfo {year} {2019})\BibitemShut {NoStop}%
\bibitem [{\citenamefont {Hjorth~Larsen}\ \emph {et~al.}(2017)\citenamefont {Hjorth~Larsen}, \citenamefont {Jørgen~Mortensen}, \citenamefont {Blomqvist}, \citenamefont {Castelli}, \citenamefont {Christensen}, \citenamefont {Dułak}, \citenamefont {Friis}, \citenamefont {Groves}, \citenamefont {Hammer}, \citenamefont {Hargus}, \citenamefont {Hermes}, \citenamefont {Jennings}, \citenamefont {Bjerre~Jensen}, \citenamefont {Kermode}, \citenamefont {Kitchin}, \citenamefont {Leonhard~Kolsbjerg}, \citenamefont {Kubal}, \citenamefont {Kaasbjerg}, \citenamefont {Lysgaard}, \citenamefont {Bergmann~Maronsson}, \citenamefont {Maxson}, \citenamefont {Olsen}, \citenamefont {Pastewka}, \citenamefont {Peterson}, \citenamefont {Rostgaard}, \citenamefont {Schiøtz}, \citenamefont {Schütt}, \citenamefont {Strange}, \citenamefont {Thygesen}, \citenamefont {Vegge}, \citenamefont {Vilhelmsen}, \citenamefont {Walter}, \citenamefont {Zeng},\ and\ \citenamefont {Jacobsen}}]{hjorth_larsen_atomic_2017}%
  \BibitemOpen
  \bibfield  {author} {\bibinfo {author} {\bibfnamefont {Ask}\ \bibnamefont {Hjorth~Larsen}}, \bibinfo {author} {\bibfnamefont {Jens}\ \bibnamefont {Jørgen~Mortensen}}, \bibinfo {author} {\bibfnamefont {Jakob}\ \bibnamefont {Blomqvist}}, \bibinfo {author} {\bibfnamefont {Ivano~E}\ \bibnamefont {Castelli}}, \bibinfo {author} {\bibfnamefont {Rune}\ \bibnamefont {Christensen}}, \bibinfo {author} {\bibfnamefont {Marcin}\ \bibnamefont {Dułak}}, \bibinfo {author} {\bibfnamefont {Jesper}\ \bibnamefont {Friis}}, \bibinfo {author} {\bibfnamefont {Michael~N}\ \bibnamefont {Groves}}, \bibinfo {author} {\bibfnamefont {Bjørk}\ \bibnamefont {Hammer}}, \bibinfo {author} {\bibfnamefont {Cory}\ \bibnamefont {Hargus}}, \bibinfo {author} {\bibfnamefont {Eric~D}\ \bibnamefont {Hermes}}, \bibinfo {author} {\bibfnamefont {Paul~C}\ \bibnamefont {Jennings}}, \bibinfo {author} {\bibfnamefont {Peter}\ \bibnamefont {Bjerre~Jensen}}, \bibinfo {author} {\bibfnamefont {James}\ \bibnamefont {Kermode}}, \bibinfo {author} {\bibfnamefont
  {John~R}\ \bibnamefont {Kitchin}}, \bibinfo {author} {\bibfnamefont {Esben}\ \bibnamefont {Leonhard~Kolsbjerg}}, \bibinfo {author} {\bibfnamefont {Joseph}\ \bibnamefont {Kubal}}, \bibinfo {author} {\bibfnamefont {Kristen}\ \bibnamefont {Kaasbjerg}}, \bibinfo {author} {\bibfnamefont {Steen}\ \bibnamefont {Lysgaard}}, \bibinfo {author} {\bibfnamefont {Jón}\ \bibnamefont {Bergmann~Maronsson}}, \bibinfo {author} {\bibfnamefont {Tristan}\ \bibnamefont {Maxson}}, \bibinfo {author} {\bibfnamefont {Thomas}\ \bibnamefont {Olsen}}, \bibinfo {author} {\bibfnamefont {Lars}\ \bibnamefont {Pastewka}}, \bibinfo {author} {\bibfnamefont {Andrew}\ \bibnamefont {Peterson}}, \bibinfo {author} {\bibfnamefont {Carsten}\ \bibnamefont {Rostgaard}}, \bibinfo {author} {\bibfnamefont {Jakob}\ \bibnamefont {Schiøtz}}, \bibinfo {author} {\bibfnamefont {Ole}\ \bibnamefont {Schütt}}, \bibinfo {author} {\bibfnamefont {Mikkel}\ \bibnamefont {Strange}}, \bibinfo {author} {\bibfnamefont {Kristian~S}\ \bibnamefont {Thygesen}}, \bibinfo
  {author} {\bibfnamefont {Tejs}\ \bibnamefont {Vegge}}, \bibinfo {author} {\bibfnamefont {Lasse}\ \bibnamefont {Vilhelmsen}}, \bibinfo {author} {\bibfnamefont {Michael}\ \bibnamefont {Walter}}, \bibinfo {author} {\bibfnamefont {Zhenhua}\ \bibnamefont {Zeng}}, \ and\ \bibinfo {author} {\bibfnamefont {Karsten~W}\ \bibnamefont {Jacobsen}},\ }\bibfield  {title} {\enquote {\bibinfo {title} {The atomic simulation environment—a {Python} library for working with atoms},}\ }\href {\doibase 10.1088/1361-648X/aa680e} {\bibfield  {journal} {\bibinfo  {journal} {Journal of Physics: Condensed Matter}\ }\textbf {\bibinfo {volume} {29}},\ \bibinfo {pages} {273002} (\bibinfo {year} {2017})}\BibitemShut {NoStop}%
\bibitem [{\citenamefont {Futera}\ and\ \citenamefont {English}(2017)}]{futera2017Communication}%
  \BibitemOpen
  \bibfield  {author} {\bibinfo {author} {\bibfnamefont {Zdenek}\ \bibnamefont {Futera}}\ and\ \bibinfo {author} {\bibfnamefont {Niall~J.}\ \bibnamefont {English}},\ }\bibfield  {title} {\enquote {\bibinfo {title} {Communication: Influence of external static and alternating electric fields on water from long-time non-equilibrium ab initio molecular dynamics},}\ }\href {\doibase 10.1063/1.4994694} {\bibfield  {journal} {\bibinfo  {journal} {The Journal of Chemical Physics}\ }\textbf {\bibinfo {volume} {147}},\ \bibinfo {pages} {031102} (\bibinfo {year} {2017})}\BibitemShut {NoStop}%
\bibitem [{\citenamefont {Stocco}\ \emph {et~al.}(2025)\citenamefont {Stocco}, \citenamefont {Carbogno},\ and\ \citenamefont {Rossi}}]{Stocco2025Electricfield}%
  \BibitemOpen
  \bibfield  {author} {\bibinfo {author} {\bibfnamefont {Elia}\ \bibnamefont {Stocco}}, \bibinfo {author} {\bibfnamefont {Christian}\ \bibnamefont {Carbogno}}, \ and\ \bibinfo {author} {\bibfnamefont {Mariana}\ \bibnamefont {Rossi}},\ }\bibfield  {title} {\enquote {\bibinfo {title} {Electric-field driven nuclear dynamics of liquids and solids from a multi-valued machine-learned dipolar model},}\ }\href {\doibase 10.1038/s41524-025-01751-x} {\bibfield  {journal} {\bibinfo  {journal} {npj Computational Materials}\ }\textbf {\bibinfo {volume} {11}},\ \bibinfo {pages} {304} (\bibinfo {year} {2025})}\BibitemShut {NoStop}%
\bibitem [{\citenamefont {Joll}\ \emph {et~al.}(2024)\citenamefont {Joll}, \citenamefont {Schienbein}, \citenamefont {Rosso},\ and\ \citenamefont {Blumberger}}]{Joll2024}%
  \BibitemOpen
  \bibfield  {author} {\bibinfo {author} {\bibfnamefont {Kit}\ \bibnamefont {Joll}}, \bibinfo {author} {\bibfnamefont {Philipp}\ \bibnamefont {Schienbein}}, \bibinfo {author} {\bibfnamefont {Kevin~M.}\ \bibnamefont {Rosso}}, \ and\ \bibinfo {author} {\bibfnamefont {Jochen}\ \bibnamefont {Blumberger}},\ }\bibfield  {title} {\enquote {\bibinfo {title} {Machine learning the electric field response of condensed phase systems using perturbed neural network potentials},}\ }\href {\doibase 10.1038/s41467-024-52491-3} {\bibfield  {journal} {\bibinfo  {journal} {Nature Communications}\ }\textbf {\bibinfo {volume} {15}} (\bibinfo {year} {2024}),\ 10.1038/s41467-024-52491-3}\BibitemShut {NoStop}%
\bibitem [{\citenamefont {Perdew}\ \emph {et~al.}(2008)\citenamefont {Perdew}, \citenamefont {Ruzsinszky}, \citenamefont {Csonka}, \citenamefont {Vydrov}, \citenamefont {Scuseria}, \citenamefont {Constantin}, \citenamefont {Zhou},\ and\ \citenamefont {Burke}}]{perdew2008pbesol}%
  \BibitemOpen
  \bibfield  {author} {\bibinfo {author} {\bibfnamefont {John~P.}\ \bibnamefont {Perdew}}, \bibinfo {author} {\bibfnamefont {Adrienn}\ \bibnamefont {Ruzsinszky}}, \bibinfo {author} {\bibfnamefont {Gabor~I.}\ \bibnamefont {Csonka}}, \bibinfo {author} {\bibfnamefont {Oleg~A.}\ \bibnamefont {Vydrov}}, \bibinfo {author} {\bibfnamefont {Gustavo~E.}\ \bibnamefont {Scuseria}}, \bibinfo {author} {\bibfnamefont {Lucian~A.}\ \bibnamefont {Constantin}}, \bibinfo {author} {\bibfnamefont {Xiaolan}\ \bibnamefont {Zhou}}, \ and\ \bibinfo {author} {\bibfnamefont {Kieron}\ \bibnamefont {Burke}},\ }\bibfield  {title} {\enquote {\bibinfo {title} {Restoring the density-gradient expansion for exchange in solids and surfaces},}\ }\href {\doibase 10.1103/PhysRevLett.100.136406} {\bibfield  {journal} {\bibinfo  {journal} {Physical Review Letters}\ }\textbf {\bibinfo {volume} {100}},\ \bibinfo {pages} {136406} (\bibinfo {year} {2008})}\BibitemShut {NoStop}%
\bibitem [{\citenamefont {Kresse}\ and\ \citenamefont {Hafner}(1993)}]{kresse1993}%
  \BibitemOpen
  \bibfield  {author} {\bibinfo {author} {\bibfnamefont {G.}~\bibnamefont {Kresse}}\ and\ \bibinfo {author} {\bibfnamefont {J.}~\bibnamefont {Hafner}},\ }\bibfield  {title} {\enquote {\bibinfo {title} {Ab initio molecular dynamics for liquid metals},}\ }\href {\doibase 10.1103/PhysRevB.47.558} {\bibfield  {journal} {\bibinfo  {journal} {Physical Review B}\ }\textbf {\bibinfo {volume} {47}},\ \bibinfo {pages} {558--561} (\bibinfo {year} {1993})}\BibitemShut {NoStop}%
\bibitem [{\citenamefont {Kresse}\ and\ \citenamefont {Hafner}(1994)}]{kresse1994liquid}%
  \BibitemOpen
  \bibfield  {author} {\bibinfo {author} {\bibfnamefont {G.}~\bibnamefont {Kresse}}\ and\ \bibinfo {author} {\bibfnamefont {J.}~\bibnamefont {Hafner}},\ }\bibfield  {title} {\enquote {\bibinfo {title} {Ab initio molecular-dynamics simulation of the liquid-metal--amorphous-semiconductor transition in germanium},}\ }\href {\doibase 10.1103/PhysRevB.49.14251} {\bibfield  {journal} {\bibinfo  {journal} {Physical Review B}\ }\textbf {\bibinfo {volume} {49}},\ \bibinfo {pages} {14251--14269} (\bibinfo {year} {1994})}\BibitemShut {NoStop}%
\bibitem [{\citenamefont {Kresse}\ and\ \citenamefont {Furthm{\"u}ller}(1996{\natexlab{a}})}]{kresse1996efficiency}%
  \BibitemOpen
  \bibfield  {author} {\bibinfo {author} {\bibfnamefont {G.}~\bibnamefont {Kresse}}\ and\ \bibinfo {author} {\bibfnamefont {J.}~\bibnamefont {Furthm{\"u}ller}},\ }\bibfield  {title} {\enquote {\bibinfo {title} {Efficiency of ab-initio total energy calculations for metals and semiconductors using a plane-wave basis set},}\ }\href {\doibase 10.1016/0927-0256(96)00008-0} {\bibfield  {journal} {\bibinfo  {journal} {Computational Materials Science}\ }\textbf {\bibinfo {volume} {6}},\ \bibinfo {pages} {15--50} (\bibinfo {year} {1996}{\natexlab{a}})}\BibitemShut {NoStop}%
\bibitem [{\citenamefont {Kresse}\ and\ \citenamefont {Furthm{\"u}ller}(1996{\natexlab{b}})}]{kresse1996efficient}%
  \BibitemOpen
  \bibfield  {author} {\bibinfo {author} {\bibfnamefont {G.}~\bibnamefont {Kresse}}\ and\ \bibinfo {author} {\bibfnamefont {J.}~\bibnamefont {Furthm{\"u}ller}},\ }\bibfield  {title} {\enquote {\bibinfo {title} {Efficient iterative schemes for ab initio total-energy calculations using a plane-wave basis set},}\ }\href {\doibase 10.1103/PhysRevB.54.11169} {\bibfield  {journal} {\bibinfo  {journal} {Physical Review B}\ }\textbf {\bibinfo {volume} {54}},\ \bibinfo {pages} {11169--11186} (\bibinfo {year} {1996}{\natexlab{b}})}\BibitemShut {NoStop}%
\bibitem [{\citenamefont {Kresse}\ and\ \citenamefont {Joubert}(1999)}]{kresse1999paw}%
  \BibitemOpen
  \bibfield  {author} {\bibinfo {author} {\bibfnamefont {G.}~\bibnamefont {Kresse}}\ and\ \bibinfo {author} {\bibfnamefont {D.}~\bibnamefont {Joubert}},\ }\bibfield  {title} {\enquote {\bibinfo {title} {From ultrasoft pseudopotentials to the projector augmented-wave method},}\ }\href {\doibase 10.1103/PhysRevB.59.1758} {\bibfield  {journal} {\bibinfo  {journal} {Physical Review B}\ }\textbf {\bibinfo {volume} {59}},\ \bibinfo {pages} {1758--1775} (\bibinfo {year} {1999})}\BibitemShut {NoStop}%
\bibitem [{\citenamefont {Togo}\ and\ \citenamefont {Tanaka}(2015)}]{togo2015first-f2e}%
  \BibitemOpen
  \bibfield  {author} {\bibinfo {author} {\bibfnamefont {Atsushi}\ \bibnamefont {Togo}}\ and\ \bibinfo {author} {\bibfnamefont {Isao}\ \bibnamefont {Tanaka}},\ }\bibfield  {title} {\enquote {\bibinfo {title} {First principles phonon calculations in materials science},}\ }\href {\doibase 10.1016/j.scriptamat.2015.07.021} {\bibfield  {journal} {\bibinfo  {journal} {Scripta Materialia}\ }\textbf {\bibinfo {volume} {108}},\ \bibinfo {pages} {1--5} (\bibinfo {year} {2015})}\BibitemShut {NoStop}%
\bibitem [{\citenamefont {Rabe}\ \emph {et~al.}(2007)\citenamefont {Rabe}, \citenamefont {Ahn},\ and\ \citenamefont {Triscone}}]{rabe2007physics}%
  \BibitemOpen
  \bibinfo {editor} {\bibfnamefont {Karin~M.}\ \bibnamefont {Rabe}}, \bibinfo {editor} {\bibfnamefont {Charles~H.}\ \bibnamefont {Ahn}}, \ and\ \bibinfo {editor} {\bibfnamefont {Jean-Marc}\ \bibnamefont {Triscone}},\ eds.,\ \href@noop {} {\emph {\bibinfo {title} {Physics of Ferroelectrics: A Modern Perspective}}}\ (\bibinfo  {publisher} {Springer Berlin Heidelberg},\ \bibinfo {address} {Berlin, Heidelberg},\ \bibinfo {year} {2007})\BibitemShut {NoStop}%
\end{thebibliography}
\end{document}